\documentclass{aa}
\usepackage{natbib}
\usepackage{graphicx}
\usepackage{txfonts}
\usepackage{multirow}
\usepackage{color}
\usepackage{times}
\usepackage{amssymb}
\usepackage{verbatim}
\usepackage{gensymb}
\usepackage{amsmath}
\bibpunct{(}{)}{;}{a}{}{,} 

\begin{document} 
   \title{The Close AGN Reference Survey (CARS)}
   \subtitle{Locating the [\ion{O}{III}] wing component in luminous local Type 1 AGN}
   
   \author{M.~Singha \inst{1}
   \and B.~Husemann \inst{2}
   \and T.~Urrutia \inst{3}
   \and C.~P.~O'Dea \inst{1}
   \and J.~Scharw\"achter \inst{4}
   \and M.~Gaspari \inst{5,6}
   \and F.~Combes \inst{7}
   \and R.~Nevin \inst{8}
   \and B.~A.~Terrazas \inst{8}
   \and M.~P\'erez-Torres \inst{9}
   \and T.~Rose \inst{10}
   \and T.~A.~Davis \inst{11}
   \and G.~R.~Tremblay \inst{8}
   \and J.~Neumann \inst{12}
   \and I.~Smirnova-Pinchukova \inst{2}
   \and S.~A.~Baum \inst{1}
   }

   \institute{Department of Physics \& Astronomy, University of Manitoba, Winnipeg, MB R3T 2N2, Canada\\
    \email{singham4@myumanitoba.ca}
    \and
       Max-Planck-Institut für Astronomie, K\"onigstuhl 17, D-69117 Heidelberg, Germany
    \and 
       Leibniz-Institut f\"ur Astrophysik Potsdam, An der Sternwarte 16, 14482 Potsdam, Germany
    \and
       Gemini Observatory/NSF’s NOIRLab, 670 N. A’ohoku Place, Hilo, Hawai’i, 96720, USA
    \and
       INAF - Osservatorio di Astrofisica e Scienza dello Spazio, via P. Gobetti 93/3, I-40129 Bologna, Italy
    \and 
       Department of Astrophysical Sciences, Princeton University, 4 Ivy Lane, Princeton, NJ 08544-1001, USA
    \and 
       LERMA, Observatoire de Paris, PSL Research Univ., College de France, CNRS, Sorbonne Univ., UPMC, Paris, France
    \and
       Center for Astrophysics $|$ Harvard \& Smithsonian, 60 Garden St., Cambridge, MA 02138, USA
     \and
       Instituto de Astrof\'isica de Andaluc\'ia (IAA-CSIC),
       Glorieta de la Astronom\'ia s/n, E-18008 Granada, Spain
    \and
       Centre for Extragalactic Astronomy, Durham University, DH1 3LE, UK
    \and
       School of Physics \&\ Astronomy, Cardiff University, Queens Buildings, The Parade, Cardiff, CF24 3AA, UK   
    \and
       Institute of Cosmology and Gravitation, University of Portsmouth, Burnaby Road, Portsmouth, PO1 3FX, UK
       }

   \date{Received January 18, 2021; accepted September 21, 2020}

  \abstract
   {The strong asymmetry in the optical [\ion{O}{iii}] $\lambda5007$  emission line is one of the best signatures of AGN-driven warm ($\sim$\,$\mathrm{10}^{4}$\,K) ionized gas outflows on host galaxy scales. While large spectroscopic surveys like SDSS have characterized the kinematics of [\ion{O}{iii}] for large samples of AGN, estimating the associated energetics requires spatially resolving these outflows with, for example, IFU studies.}
   {As part of the Close AGN Reference Survey (CARS) we obtained spatially-resolved IFU spectroscopy for a representative sample of 39 luminous type 1 AGN at $0.01<z<0.06$ with MUSE and VIMOS IFUs at the VLT to infer the spatial location of the ionized gas outflows.}
   {We compare the light distributions of the [\ion{O}{iii}] wing to that of the H$\beta$ broad emission line region, a classical point source (PSF). We then use the PSF to distinguish between the unresolved and resolved [\ion{O}{iii}] wing emission. We further determine its location using spectro-astrometry for the point-like sources.}
   {The [\ion{O}{iii}] wing is spatially unresolved in 23 out of the 36 AGN with $>$80\% of the flux associated with a point-like source. We measure $<$100\,pc offsets in the spatial location of the outflow from the AGN nucleus using the spectro-astrometry technique for these sources. For the other 13 AGN, the [\ion{O}{iii}] wing emission is resolved and possibly extended on kpc scale.}
   {We conclude that [\ion{O}{iii}] wing emission can be compact or extended in an unbiased luminous AGN sample, where both cases are likely to appear. Electron density in the compact [\ion{O}{iii}] wing regions (median $n_\mathrm{e} \sim 1900\ \mathrm{cm}^{-3}$) is nearly a magnitude higher than in the extended ones (median $n_\mathrm{e} \sim 500\ \mathrm{cm}^{-3}$). The presence of spatially extended and compact [\ion{O}{iii}] wing emission is unrelated to the AGN bolometric luminosity and to inclination effects, which means other features such as time delays, or mechanical feedback/radio jets may shape the ionized gas outflow properties.}
   
   
   \maketitle
%

\section{Introduction}

Many massive galaxies tend to host super-massive black holes (SMBH) at their centers \citep[e.g.,][]{Genzel1987,Genzel1996,Eckart2001,Kormendy:2013,Gaspari:2019}. These SMBH usually grow through accreting gaseous material and are visible as "Active Galactic Nuclei"(AGN).
Whether or not these AGN have an important contribution in driving the evolution of their host galaxies had been an unanswered question until AGN feedback was shown to be essential for matching the galaxy luminosity function in models to observations. 

\cite{Benson2003} examined the physics determining the shape of the galaxy luminosity function in a cold dark matter (CDM) universe with a high baryonic content ($\Omega_\mathrm{\text{b}} = 0.02$). They stated that although the cooling and condensation of
gas in a merging hierarchy of dark matter halos could be the dominant process behind galaxy formation, that is not sufficient to explain the sharp cut-off of the galaxy luminosity function and scarcity of highly luminous galaxies. They stated that the sharp cut-off of the galaxy luminosity function can be achieved if the cooling of gas in these massive haloes could be efficiently suppressed by (1) the reheating of cold disk gas; (2) energy injection from supernovae and stellar wind, heating up and expanding the diffuse gas halo; or (3) complete expulsion of cold gas from the disk by highly energetic superwinds. They also added that superwinds can only expel gas if the formation of central SMBHs and generation of the associated energy result in 
limiting the star formation in their host galaxies.
\citet{Bower2006} and \citet{Croton2006} agreed that the observed scarcity of very bright and massive galaxies can be explained by taking into account the energy released from AGN which could effectively prevent the gas from cooling to form the unobserved, highly luminous galaxies.

These results motivated other theoretical studies to analyze the effect of AGN feedback on galaxy growth \citep{Churazov2005,Somerville2008,McCarthy2010,Hopkins2016,Schaye2015,Gaspari:2017}. Most of the galaxy evolution models involved a form of energy injection by the AGN in which the AGN drives galactic scale outflows and hence expels the gas out of their host-galaxies, prohibiting BH growth. This is known as the radiative feedback mode \citep[e.g.,][]{Benson2003,King2003,DiMatteo2005,Hopkins2006,Hopkins2010,Booth2010}. The relativistic jets from the AGN could also heat up the surrounding environment and obstruct star formation. This is known as the mechanical feedback mode \citep[e.g.,][]{McNamara2007,Dubois2010,Birzan2012,Bower2012}. In this feedback mode, the accretion of material onto the SMBH leads to the production of highly relativistic, plasma jets \citep{Best2012}. The mechanical feedback is suspected to be responsible for heating of the intra-cluster medium (ICM) and quenching of cooling flows in the centres of clusters \citep[e.g.,][]{McNamara2007,Gaspari:2020}.
Even though star formation processes such as supernovae and stellar-winds drive galactic scale outflows \citep[e.g.,][]{Heckman1984,Lehnert1996,Swinbank2009,Roberts-borsani2019}, AGN are necessary to drive the outflows of observed velocities $> 1000\ \mathrm{km\ s}^{-1}$ \citep{Benson2003,Mccarthy2011}.

In recent years, X-ray and UV spectroscopy demonstrated that a considerable fraction of highly accreting AGN could drive ultrafast (velocity $\sim$0.1c) outflows (UFOs) close to their accretion disks \citep{Blustin2003,Reeves2003,Ganguly2008,Tombesi:2013,Serafinelli:2019}. 
But, in order to drive the host galaxy's evolution, one key requirement is that the AGN needs to drive galactic scale outflows as mentioned in several galaxy evolution models \citep{Granato2004,Erb2006,Oppenheimer2008,Hopkins2010,Lilly2013}. The optical, highly ionized [\ion{O}{iii}] line is commonly used to trace the signs of AGN ionization over galaxy scales. As [\ion{O}{iii}] is a forbidden line, it cannot trace the dense broad emission line region (BLR) kinematics, but the kinematics in the narrow line region (NLR) which extends from pc to few tens of kpc scale \citep[e.g.,][]{Husemann2016,Husemann2019a}. Therefore, any kinematic disturbance in the NLR would be well reflected in the [\ion{O}{iii}] line shape over the galactic scales.

\citet{Crenshaw2000} suggested that the outflow away from the nucleus of NGC~1068 could follow a biconical kinematic model. They found that the similar amplitudes of the blueshifted and redshifted curves on the northeast side in the position-velocity diagram. This infers that the axis of the bicone is in vicinity to the plane of the sky. The lack of low radial velocities where the curves peak could indicate that that the bicone is evacuated along its axis.

\citet{Bae2016} also found that a simple biconical model of radial outflow is sufficient to explain the general trend of observed outflow velocities. In addition, turbulent motions can also contribute to the broadening of the spectral lines (e.g., \citealt{Tremblay:2018,Simionescu:2019,Gaspari:2020}).

\citet{Mullaney2013} used the SDSS spectroscopic data to study what drives the kinematics of the kilo-parsec scale [\ion{O}{iii}]$\lambda$5007 emitting gas. They analyzed a sample of 24,264 z$<$0.4 optically selected type 1 and type 2 AGN and performed a multi-component fitting to the optical emission-line profiles to estimate the [\ion{O}{iii}]$\lambda$5007 kinematics. For type 2 AGN, the line of sight is towards the central engine, intercepting the equatorial dust. On the other hand, the central engine will be directly visible for a type 1 AGN \citep{Urry1995}. The transition between the type 1 to type 2 usually occurs between inclination angle $45-60\degr$ \citep{Marin2014}. \citet{Mullaney2013} reported that both type 1 and type 2 AGN exhibit asymmetry in [\ion{O}{iii}] line profile, which they interpreted as evidence of outflowing ionized gas. They suggested that the ionized outflows are prevalent in both type 1 and type 2 AGN and are driven by compact radio jets. Previous studies by \citet{Heckman1981,Whittle1992,Blundell1998,Thean2001,Jarvis2019PrevalenceQuasars,Molyneux2019} also concluded that the compact radio sources could be responsible as the launching mechanism of these ionized outflows.
\citet{Woo2016} analysed a sample of $\sim$39,000 type 2 AGN at $z<0.3$ and showed that $\sim 45\%$ of type~2 AGN at $z < 0.3$ show signs of outflows in [\ion{O}{iii}]. They reported an increase in [\ion{O}{iii}] velocity dispersion as AGN luminosity and Eddington ratio increase. \citet{Woo2016} concluded that the radiation pressure from the AGN accretion disk could launch the ionized outflows.

Although the one-dimensional spectroscopic data provide insight into the ionized gas kinematics, they cannot provide any information about the spatial extent or structure of these outflows. This is where spatially resolved spectroscopy has significant advantage. Over the last two decades, both long-slit and integral field unit (IFU) observations of AGN have found evidence of high-velocity and kinematically disturbed ionized gas over several kpc scales \citep[e.g.,][]{McCarthy1996,Colina1999,Swinbank2005,Westmoquette2012,Riffel2013,Diniz2015,Humire2018,Powell2018,Slater2019,Kakkad2020,Scholtz2020}. Previous work \citep{Liu2013,Harrison2014,McElroy2015} found that the AGN activity resulted in high-velocity and disturbed ionized gas (velocity dispersion up to 500\,$\mathrm{km}\,\mathrm{s}^{-1}$) extended to $\sim$10\,kpc in their targets; and the majority (70\%) of their AGN ($z<0.2$) population exhibited highly ionized [\ion{O}{iii}] regions which extended to kpc scales with high velocity ($\sim$510--1100\,$\mathrm{km}\,\mathrm{s}^{-1}$) ionized gas.   

\citet{Villar-Martin2016} investigated the similarity of the [\ion{O}{iii}] light distributions and point-spread functions (PSF)  with VLT-FORS2 observations of type 2 AGN to show that the outflowing ionized [\ion{O}{iii}] intensity distribution follows a point-like profile. \citet{Husemann2016} reported that the bright point-like emission from the center could scatter out and confuse the extended flux. This is known as beam smearing. 
Recent studies have reported sub-kpc scale outflows  \citep[e.g.,][]{Villar-Martin2016,Tadhunter2018,Kawaguchi2018,Bellocchi2019,Baron2019}. Therefore, the spatial extensions of these outflows still remains as an open question. Fortunately, if the [\ion{O}{iii}] wing surface-brightness profile mimics the emission profile of a theoretical point source, it is possible to perform a 2D modelling \citep{Gadotti2008} to estimate the center of the [\ion{O}{iii}] wing emission as well as the location of the central-engine (SMBH). Furthermore, one can achieve the precision of a hundredth of a pixel based on high spatial resolution data using spectro-astrometry \citep[for a discussion, see][]{Bailey1998,Whelan2008,Gnerucci2010,Gnerucci2011a}. Therefore, using type 1 QSOs, we aim to compare the [\ion{O}{iii}] light distribution to that of the PSF (the light distribution of the H$\beta$ broad line region, BLR), and if they are similar, to determine how far off is the region of the maximum [\ion{O}{iii}] wing emission from the SMBH. Similar methods have been applied by \citet{Kakkad2020,Santoro2018}.

Moreover, the energetics associated with the ionized outflow is strongly dependent upon the value of electron densities in the NLR (which can take values from 10--$10^{4}\ \mathrm{cm}^{-3}$ \citep{Nesvadba2006,Liu2013a,Harrison2014}. The electron density is responsible for the largest fraction of systematic uncertainty when estimating the outflow-energetics for warm, ionized gas outflows \citep{Muller-Sanchez2011,Kakkad2016,Perna2017}. It therefore remains unclear how the kinetic power of the outflowing gas couples with AGN properties or the star-formation in the host galaxy.
Hence, it is important to understand how the electron density varies on the different scales in AGN host galaxies and how it is affected by the warm, ionized outflows. \citet{Liu2013a} assumed an electron density of $n_\mathrm{e}\sim 100~\mathrm{cm}^{-3}$ at 1\,kpc away and $n_\mathrm{e} < 10~\mathrm{cm}^{-3}$ at 6 kpc away from the nucleus for the outflowing ionized gas. Studies by \citet{bennert2006,sharp2010,cresci2015,freitas2018} have reported that the electron density in the central region can go up to a few $10^{3} \mathrm{cm}^{-3}$, whereas it drops to $\sim10^{2} \mathrm{cm}^{-3}$ at $\sim$ 1.2 kpc away from the nucleus.
Recent spatially resolved studies by \citet{Kakkad2018,Baron2020} and XSHOOTER observations from the study by \citet{Davies2020} reported 
a similar decrease for the electron densities of the outflowing gas with increasing distance from the central region. A systematic estimation of electron densities is hence required to properly quantify the outflow energetics.

In this paper we explore optical IFU observations of a large number of nearby, luminous type 1 QSOs from the close AGN reference survey \citep[CARS,][; Husemann et al. in prep.]{Husemann2017,Husemann2019}. We focus on characterizing the light distribution for the asymmetric [\ion{O}{iii}] wing component with respect to the PSF. Thereby we can systematically verify whether such putative outflows are unresolved (at the limit of our observations) or possibly extended over kpc scales. In addition, we also measure electron densities associated with the outflowing wing component based on the [\ion{S}{ii}] $\lambda\lambda6717,6731$ doublet to test previous assumptions incorporated into outflow energetics calculations.

Throughout this paper we adopt the standard $\Lambda$CDM cosmology with $H_0 = 70~\text{km s}^{-1}~\text{Mpc}^{-1}$, $\Omega_\mathrm{\text{m}} = 0.3$, and $\Omega_\mathrm{\Lambda} = 0.7$.

\section{The IFU data set}
\label{section:data}
For this study we use the IFU data obtained as part of CARS. CARS represents an ambitious spatially-resolved multi-wavelength survey of a representative sample of the most luminous type 1 AGN at redshifts $0.01 < z < 0.06$.  The CARS targets are drawn from the Hamburg-ESO Survey \citep[HES,][]{wisotzki2000}, which is a purely flux-limited AGN catalog based on $B$-band photometry and slitless spectroscopy. A random sub-sample of HES AGN in this redshift range have follow-up in the cold gas phase \citep{bertram2007,konig2009} which defines the CARS targets. It is important to note that none of the sample selection criteria are related to ionized gas outflow properties in any sense, which makes this sample unbiased and fully representative for the outflow properties of luminous local AGN. All details of the sample can be found in \citet{bertram2007} and Husemann et al. (in prep.).
IFU observations have been obtained with the multi-unit spectroscopic explorer \citep[MUSE,][]{bacon2010,bacon2014} at the very large telescope (VLT) for 36 objects (programs 094.B-0345(A) and 095.B-0015(A), PI: B. Husemann), the VIsible MultiObject Spectrograph \citep[VIMOS,][]{lefevre2003} at the VLT for HE~1310$-$1051 and HE~1338$-$1423 (program 083.B-0801(A), PI: K. Jahnke), and the potsdam multi-aperture spectrophotometer \citep[PMAS,][]{roth2005} at the Calar Alto Observatory for HE~0853$-$0126 and HE~0949$-$0122 (program H18-3.5-010, PI: B. Husemann).

The MUSE data were reduced by the standard ESO pipeline for the instrument \citep{weilbacher2020} while the VIMOS and PMAS data were reduced with the Py3D fiber-fed IFU reduction package developed initially to reduce the data of the CALIFA survey \citep{husemann2013}. For more technical details on the IFUs and the data reduction process we refer the reader to \citet{Husemann2019} and Husemann et. al (in prep.). Apart from the basic CARS IFU data processing as used by \citet{Husemann2019} and explained in Husemann et. al (in prep.), a stellar continuum and emission-line modelling of the 3D IFU data is performed after QSO-subtraction using PYPARADISE \citep{Husemann2016,Husemann2019}. For the analysis presented in this paper we make use of the stellar continuum model datacubes.
The original MUSE and QSO-subtracted data will be available at CARS data release 1 (DR1).
In our analysis we neglect the CARS target HE~0021$-$1810 due to its poor data quality comprising just a single short MUSE exposure, which is not sufficient for robust measurements. Furthermore, PMAS IFU has a sampling size of 1$\arcsec$ which is clearly not as precise as MUSE/VIMOS. Therefore, we excluded the 2 AGN from the PMAS IFU in this paper.

\section{Analysis and Results}
\label{section:analysis}
\subsection{Modelling the [\ion{O}{iii}] line shape}
A primary signature of ionized gas outflows from AGN are the commonly observed  broad and blue-shifted wings in the [\ion{O}{iii}] emission lines. In order to extract the integrated [\ion{O}{iii}] emission line profile  parameters, we co-added all the spectra within a 3\arcsec\ diameter around the the AGN location. As our sample consists of only type 1 AGN, the broad H$\beta$ and \ion{Fe}{ii} $\lambda\lambda$4923,5018 lines from the broad line region (BLR) are blended with the narrow emission lines of H$\beta$ and [\ion{O}{iii}] $\lambda\lambda4959,5007$, respectively. We modelled the spectral region from 4750\AA\ to 5090\AA\ (rest-frame wavelength) simultaneously to deblend all broad and narrow emission lines in this region. This is a standard process for the analysis of type 1 AGN spectra \citep[e.g.,][]{Jin2012,Mullaney2013,Husemann2016,Husemann2019}.

Here, we assumed a simple superposition of Gaussian profiles to model all lines in the spectrum. We adopt 1-2 Gaussian components to model each of the narrow [\ion{O}{iii}] $\lambda\lambda4959,5007$ doublet and the narrow H$\beta$ component. The first Gaussian component corresponds to the core and the second Gaussian component corresponds to the broad, usually blue-shifted, wing of the lines which might be potentially attributed to ionized gas outflows. We further deployed 1 to 2 Gaussian components for each of the BLR H$\beta$ and \ion{Fe}{ii} $\lambda\lambda4923,5018$ emission lines depending upon their complexities. Finally, we approximated the local AGN continuum with a first-order polynomial given that the considered wavelength range is small. 

Additionally, we always assume that the absolute velocity and the intrinsic velocity dispersion of the narrow [\ion{O}{iii}] $\lambda\lambda4959,5007$ and H$\beta$ components are coupled, because they likely originate from the same physical region with similar kinematics. Similar coupling is assumed for the BLR components of the H$\beta$ and \ion{Fe}{ii} $\lambda\lambda4923,5018$ emission lines. We further put two constraints on the line fluxes in addition to our previous assumptions of emission line kinematics. One is that the [\ion{O}{iii}] $\lambda\lambda4959,5007$ has a line flux ratio of 1/3 \citep{Storey2000}. Secondly, we found that the line flux ratio of 
\ion{Fe}{ii} $\lambda 4923$ and \ion{Fe}{ii} $\lambda5018$ was nearly constant across the sample with a mean ratio to 0.81; hence we used 0.81 as \ion{Fe}{ii} $\lambda 4923$/\ion{Fe}{ii} $\lambda5018$ to constrain the line fluxes of the \ion{Fe}{ii} doublet.

\begin{figure}
   \resizebox{\hsize}{!}{\includegraphics{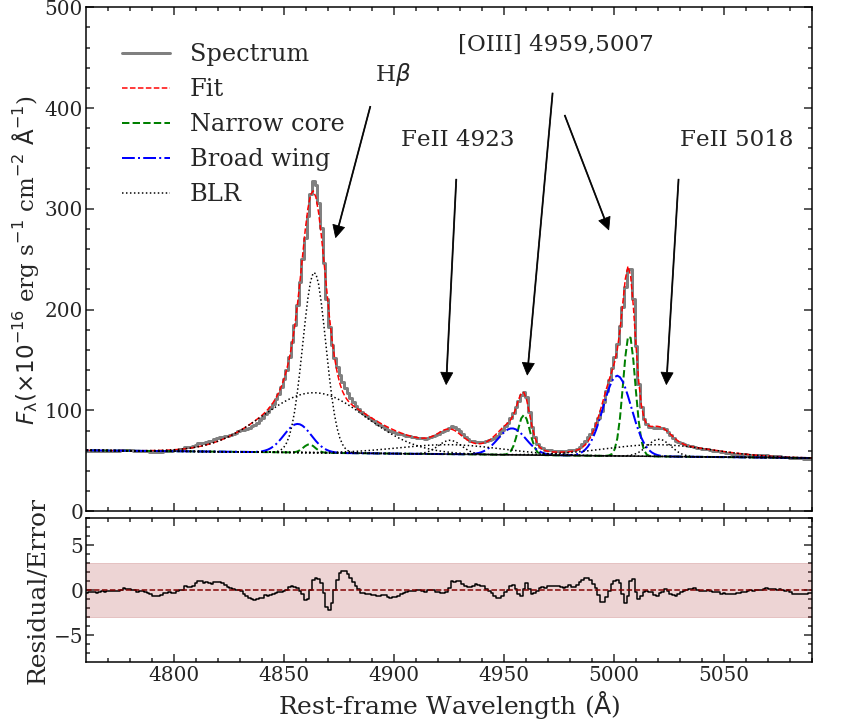}}
      \caption{Example of the multi-component modelling for the spectrum of HE~0345$+$0056. \textit{Upper panel:} The brightest spaxel spectrum of HE~0345$+$0056 and the best-fit spectral model.  The blue green lines refer to the narrow core and the blue lines refer to the broad wing components of H$\beta$ + [\ion{O}{iii}]. The gray lines refer to the individual emission lines of the BLR H$\beta$ and \ion{Fe}{ii} lines.
      The narrow width of the BLR component representing H$\beta$ is due to the fact that HE~0345$+$0056 is a narrow line Seyfert 1 (NLS1), where the broad line region tends to show narrower Balmer lines. These narrow BLR components are common in these objects \citep[e.g.,][]{Mathur2000a,Mullaney2008}.
      \textit{Lower panel:} Residual spectrum normalized by the error spectrum. The dark grey shaded area highlights the $\pm$3$\sigma$ limit and the red dashed line provides the reference to 0.}
         \label{Fig:AGNmodel}
   \end{figure}

We independently fit the brightest AGN spaxel and the integrated 3\arcsec\ spectra using a non-linear Levenberg-Marquardt algorithm. For the error estimation, we create 100 different spectra by fluctuating the original spectrum with the noise spectrum and re-model the spectra to obtain the associated uncertainties for each parameter as the standard deviation of the repeated modelling. For the two Gaussian components of [\ion{O}{iii}], we only consider the second, broad Gaussian component when it has a signal-to-noise ratio of $S/N > 5 $ and the fractional error in the velocity dispersion in each of the Gaussian components is $<1$. All sources in our sample satisfy these criteria.
An example of the best-fit AGN spectral model is shown for HE~1126$-$0407 in Fig.~\ref{Fig:AGNmodel}.

   \begin{figure*}
   \includegraphics[width=\textwidth]{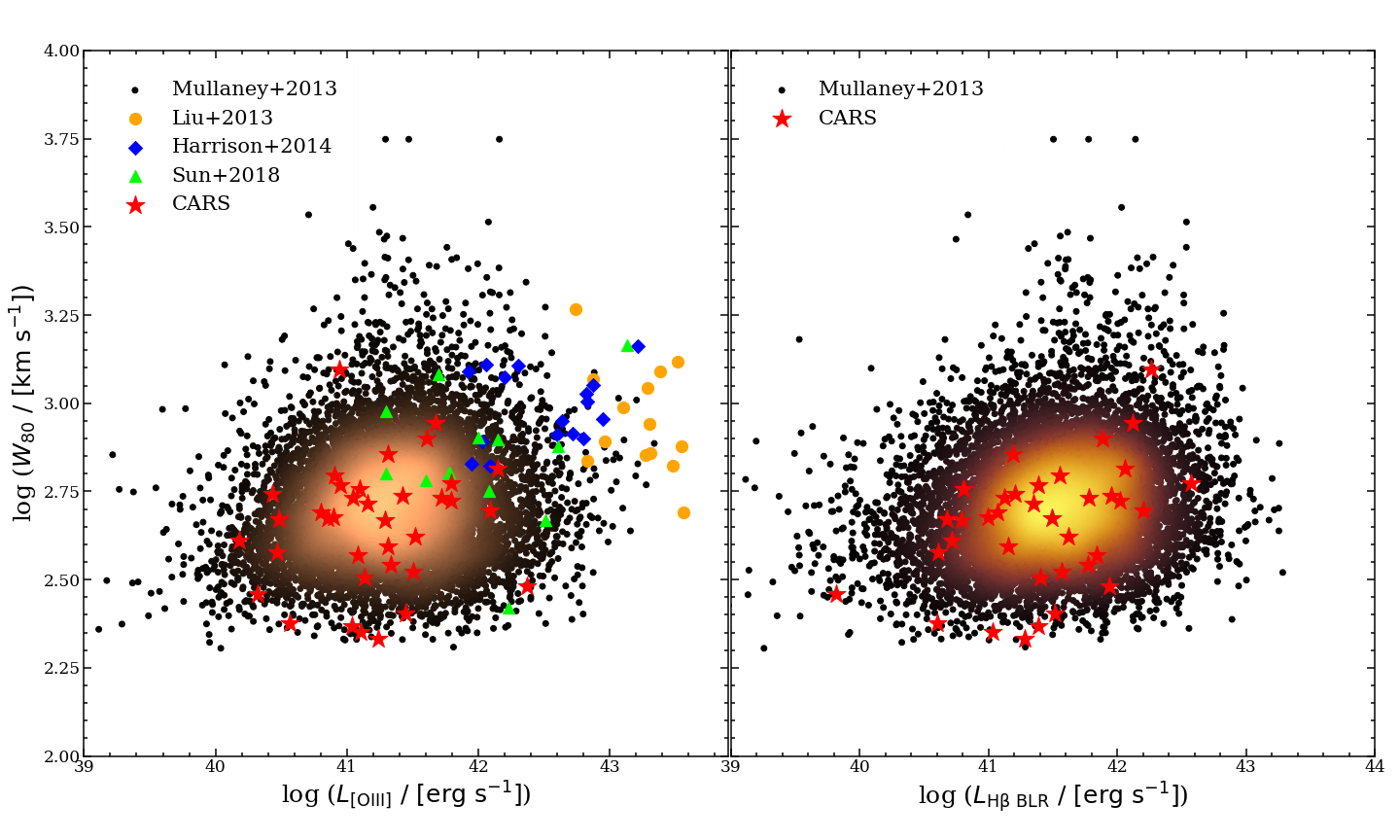}
   \caption{[\ion{O}{iii}] line width described by $W_\mathrm{80}$ against AGN bolometric [\ion{O}{iii}] luminosity (left panel) and BLR H$\beta$ luminosity (right panel). The yellow circles denote the type 2 AGN sample from \citep{Liu2013}, the blue diamonds denote the type 2 AGN sample by \citep{Harrison2014}, the lime triangles denote the type 2 AGN sample by \citep{Sun2017} and the red stars denote our CARS sample of type 1 AGN. The dense scatter plot corresponds  to the overall SDSS sample of both type 1 and type 2 (left panel) and type 1 AGN (right panel) from the ALPAKA library published by \citep{Mullaney2013}.}
         \label{Fig:O3width_sample}
   \end{figure*}
   
From the best-fit emission-line model of the central $3\arcsec \times 3 \arcsec$ spectra, we report the  [\ion{O}{iii}] line fluxes, their velocity difference $\Delta v = v_\mathrm{wing} - v_\mathrm{core}$ and the [\ion{O}{iii}] line width ($W_\mathrm{80}$) in Table~\ref{Table:1}. Any negative $\Delta v$ will represent a blue-wing and a positive $\Delta v$ will denote a red-wing [\ion{O}{iii}] asymmetry. $W_\mathrm{80}$ is defined as the velocity width of the emission-line that contains 80\% of the line flux. If $v_\mathrm{10}$ and $v_\mathrm{90}$ are the velocities at the 10th and 90th percentiles, then $W_\mathrm{80}$ = $v_\mathrm{90} - v_\mathrm{10}$ \citep{Harrison2014}. In addition, we also report the total BLR H$\beta$ flux ($f_\mathrm{\mathrm{H}\beta_\mathrm{BLR}}$).

In Fig.~\ref{Fig:O3width_sample}, we compare the [\ion{O}{iii}] line width described by $W_\mathrm{80}$ with two often used bolometric luminosity indicators, the total [\ion{O}{iii}] luminosity and the broad H$\beta$ luminosity.  The CARS sample is consistent with the bulk population of AGN as probed by the SDSS \citep{Mullaney2013} and is therefore a representative sub-sample of the low-redshift AGN population in this parameter space. Several other IFU studies of SDSS-selected AGN instead focused on more extreme systems.

The asymmetry in the [\ion{O}{iii}] line-shape does not necessarily correspond to an outflow: turbulent motions \citep[e.g.,][]{Gaspari:2018} and simple gravity may also drive the broadening and related red/blue wings in the [\ion{O}{iii}] line-profile towards the center below certain velocity limits. Nevertheless, the velocity offsets are seen at the galaxy center co-located with the AGN, which indicates that this asymmetric feature in the [\ion{O}{iii}] line profile is potentially related to an AGN-driven outflow as the dominant process.
Previous detailed IFU studies of ionized gas outflows \citep[e.g.,][]{Liu2013,Liu2013a,Harrison2014,McElroy2015,Sun2017} targeted rare sub-samples at the edges of the distribution, with significantly higher $W_\mathrm{80}$ and $L_\mathrm{[\ion{O}{iii}]}$ values. Those are therefore not representative for the overall AGN population and their results cannot be necessarily extrapolated to be common properties of the overall AGN population.

\subsection{Spatial distribution of the [\ion{O}{iii}] wing component}

\begin{table*}
\caption{Kinematic Table for CARS sources}
\label{Table:1}
\centering
\begin{tabular}{c c c c c c cc}\hline\hline
Object & z & log$(L_\mathrm{[\ion{O}{iii}]})$ \tablefootmark{a} & log$(L_\mathrm{BLR\: H\beta})$ \tablefootmark{b} & $W_\mathrm{80}$ \tablefootmark{c} & $\Delta v$ \tablefootmark{d} & \multicolumn{2}{c}{$\sigma_\mathrm{[\ion{O}{iii}]}) \tablefootmark{e}$} \\ \smallskip
 & & $(\mathrm{erg\: s}^{-1})$ & $(\mathrm{erg\: s}^{-1})$ & $(\mathrm{km\: s}^{-1})$ & $(\mathrm{km\: s}^{-1})$ & $\hspace{2cm}(\mathrm{km\: s}^{-1})$\\ \hline \smallskip
& & & & & & core & wing\\ \hline \smallskip
HE0021$-$1810 & 0.054 & 41.10 $\pm$ 0.00 & 40.81 $\pm$ 0.02 & 570 $\pm$ 4 & -83 $\pm$ 2 & 102 $\pm$ 2 & 288 $\pm$ 4\\
HE0021$-$1819 & 0.053 & 41.29 $\pm$ 0.00 & 40.79 $\pm$ 0.00 & 465 $\pm$ 1 & -130 $\pm$ 1 & 73 $\pm$ 0 & 296 $\pm$ 1\\
HE0040$-$1105 & 0.042 & 41.31 $\pm$ 0.00 & 41.15 $\pm$ 0.00 & 392 $\pm$ 1 & -98 $\pm$ 1 & 67 $\pm$ 0 & 254 $\pm$ 1\\
HE0108$-$4743 & 0.024 & 40.57 $\pm$ 0.00 & 40.60 $\pm$ 0.00 & 238 $\pm$ 2 & -55 $\pm$ 2 & 41 $\pm$ 2 & 114 $\pm$ 2\\
HE0114$-$0015 & 0.046 & 40.47 $\pm$ 0.00 & 40.61 $\pm$ 0.01 & 377 $\pm$ 5 & 9 $\pm$ 13 & 125 $\pm$ 1 & 402 $\pm$ 21\\
HE0119$-$0118 & 0.054 & 41.72 $\pm$ 0.00 & 41.78 $\pm$ 0.00 & 540 $\pm$ 1 & -210 $\pm$ 1 & 106 $\pm$ 1 & 238 $\pm$ 1\\
HE0212$-$0059 & 0.026 & 40.95 $\pm$ 0.00 & 41.38 $\pm$ 0.03 & 587 $\pm$ 9 & -276 $\pm$ 18 & 172 $\pm$ 1 & 472 $\pm$ 20\\
HE0224$-$2834 & 0.060 & 41.52 $\pm$ 0.00 & 41.62 $\pm$ 0.00 & 418 $\pm$ 1 & -26 $\pm$ 2 & 113 $\pm$ 1 & 276 $\pm$ 3\\
HE0227$-$0913 & 0.016 & 40.43 $\pm$ 0.00 & 41.21 $\pm$ 0.00 & 552 $\pm$ 5 & -452 $\pm$ 6 & 139 $\pm$ 1 & 287 $\pm$ 5\\
HE0232$-$0900 & 0.043 & 42.09 $\pm$ 0.00 & 42.20 $\pm$ 0.00 & 496 $\pm$ 1 & -29 $\pm$ 1 & 145 $\pm$ 0 & 330 $\pm$ 1\\
HE0253$-$1641 & 0.032 & 41.31 $\pm$ 0.00 & 41.19 $\pm$ 0.00 & 716 $\pm$ 3 & -249 $\pm$ 2 & 106 $\pm$ 1 & 367 $\pm$ 2\\
HE0345$+$0056 & 0.031 & 41.67 $\pm$ 0.00 & 42.12 $\pm$ 0.00 & 878 $\pm$ 4 & -335 $\pm$ 5 & 154 $\pm$ 2 & 388 $\pm$ 2\\
HE0351$+$0240 & 0.036 & 41.24 $\pm$ 0.00 & 41.29 $\pm$ 0.00 & 215 $\pm$ 1 & -19 $\pm$ 2 & 70 $\pm$ 0 & 234 $\pm$ 4\\
HE0412$-$0803 & 0.038 & 42.37 $\pm$ 0.01 & 41.94 $\pm$ 0.27 & 304 $\pm$ 23 & -21 $\pm$ 6 & 80 $\pm$ 9 & 250 $\pm$ 68\\
HE0429$-$0247 & 0.042 & 41.14 $\pm$ 0.00 & 41.41 $\pm$ 0.00 & 320 $\pm$ 1 & -52 $\pm$ 1 & 87 $\pm$ 0 & 220 $\pm$ 1\\
HE0433$-$1028 & 0.036 & 41.61 $\pm$ 0.00 & 41.89 $\pm$ 0.00 & 795 $\pm$ 2 & -310 $\pm$ 1 & 126 $\pm$ 1 & 336 $\pm$ 1\\
HE0853$+$0102 & 0.052 & 40.90 $\pm$ 0.01 & 41.00 $\pm$ 0.07 & 474 $\pm$ 51 & -35 $\pm$ 6 & 102 $\pm$ 5 & 303 $\pm$ 45\\
HE0934+0119 & 0.050 & 41.34 $\pm$ 0.00 & 41.77 $\pm$ 0.00 & 349 $\pm$ 1 & -123 $\pm$ 2 & 71 $\pm$ 1 & 171 $\pm$ 1\\
HE1011$-$0403 & 0.058 & 41.42 $\pm$ 0.00 & 41.96 $\pm$ 0.00 & 545 $\pm$ 2 & -212 $\pm$ 2 & 98 $\pm$ 1 & 256 $\pm$ 1\\
HE1017$-$0305 & 0.050 & 41.09 $\pm$ 0.00 & 41.84 $\pm$ 0.00 & 372 $\pm$ 6 & -63 $\pm$ 5 & 60 $\pm$ 2 & 217 $\pm$ 6\\
HE1029$-$1831 & 0.040 & 41.15 $\pm$ 0.00 & 41.35 $\pm$ 0.01 & 521 $\pm$ 2 & -135 $\pm$ 3 & 71 $\pm$ 2 & 217 $\pm$ 1\\
HE1107$-$0813 & 0.058 & 40.94 $\pm$ 0.03 & 42.26 $\pm$ 0.03 & 1248 $\pm$ 226 & -70 $\pm$ 28 & 78 $\pm$ 8 & 598 $\pm$ 133\\
HE1108$-$2813 & 0.024 & 40.80 $\pm$ 0.00 & 41.07 $\pm$ 0.00 & 490 $\pm$ 2 & -118 $\pm$ 1 & 68 $\pm$ 1 & 235 $\pm$ 2\\
HE1126$-$0407 & 0.062 & 41.79 $\pm$ 0.00 & 42.57 $\pm$ 0.00 & 594 $\pm$ 5 & -271 $\pm$ 5 & 114 $\pm$ 1 & 344 $\pm$ 4\\
HE1237$-$0504 & 0.009 & 40.49 $\pm$ 0.00 & 40.67 $\pm$ 0.00 & 470 $\pm$ 1 & -181 $\pm$ 2 & 104 $\pm$ 0 & 300 $\pm$ 2\\
HE1248$-$1356 & 0.015 & 40.32 $\pm$ 0.05 & 39.82 $\pm$ 0.01 & 287 $\pm$ 3 & -32 $\pm$ 4 & 57 $\pm$ 1 & 210 $\pm$ 11\\
HE1310$-$1051 & 0.034 & 41.51 $\pm$ 0.00 & 41.57 $\pm$ 0.00 & 333 $\pm$ 1 & -25 $\pm$ 2 & 93 $\pm$ 1 & 214 $\pm$ 3\\
HE1330$-$1013 & 0.022 & 40.18 $\pm$ 0.00 & 40.71 $\pm$ 0.00 & 409 $\pm$ 6 & -145 $\pm$ 9 & 74 $\pm$ 2 & 229 $\pm$ 5\\
HE1338$-$1423 & 0.042 & 41.80 $\pm$ 0.00 & 42.02 $\pm$ 0.00 & 528 $\pm$ 3 & 13 $\pm$ 4 & 115 $\pm$ 4 & 268 $\pm$ 5\\
HE1353$-$1917 & 0.035 & 41.05 $\pm$ 0.00 & 41.13 $\pm$ 0.02 & 541 $\pm$ 4 & -187 $\pm$ 6 & 93 $\pm$ 1 & 360 $\pm$ 3\\
HE1417$-$0909 & 0.044 & 41.44 $\pm$ 0.00 & 41.52 $\pm$ 0.00 & 254 $\pm$ 1 & -33 $\pm$ 1 & 62 $\pm$ 0 & 180 $\pm$ 1\\
HE2128$-$0221 & 0.052 & 41.10 $\pm$ 0.00 & 41.04 $\pm$ 0.00 & 224 $\pm$ 2 & -32 $\pm$ 2 & 51 $\pm$ 2 & 143 $\pm$ 4\\
HE2211$-$3903 & 0.040 & 41.04 $\pm$ 0.00 & 41.38 $\pm$ 0.01 & 234 $\pm$ 3 & -92 $\pm$ 9 & 68 $\pm$ 1 & 205 $\pm$ 9\\
HE2222$-$0026 & 0.059 & 40.85 $\pm$ 0.01 & 41.49 $\pm$ 0.00 & 471 $\pm$ 15 & -164 $\pm$ 17 & 122 $\pm$ 7 & 244 $\pm$ 20\\
HE2233$+$0124 & 0.056 & 40.90 $\pm$ 0.01 & 41.55 $\pm$ 0.02 & 625 $\pm$ 15 & 47 $\pm$ 10 & 164 $\pm$ 4 & 466 $\pm$ 35\\
HE2302$-$0857 & 0.047 & 42.14 $\pm$ 0.03 & 42.06 $\pm$ 0.01 & 651 $\pm$ 17 & 167 $\pm$ 9 & 175 $\pm$ 11 & 452 $\pm$ 29\\
\noalign{\smallskip}\hline
\end{tabular}
\tablefoot{ 
\tablefoottext{a}{\textcolor{black}{The [\ion{O}{iii}]$\lambda5007$ luminosity}}
\tablefoottext{b}{The total broad emission line region (BLR) H$\beta$ luminosity},
\tablefoottext{c}{Velocity width of emission line [\ion{O}{iii}] $\lambda5007$, that contains 80\% of the line flux},
\tablefoottext{d}{Velocity offset between the core and wing components, used to model the of [\ion{O}{iii}] $\lambda5007$ emission-line},
\tablefoottext{e}{Velocity dispersion of the [\ion{O}{iii}] $\lambda5007$ core and wing components.}
}
\end{table*}

The kinematics associated with the [\ion{O}{iii}] broad wings in the ionized gas can be easily obtained from the AGN spectra for large samples as described in the previous section. However, the spatial distribution and the size of the wings can only be determined from spatially-resolved spectroscopy. In recent years, numerous IFU studies  mapped the surface brightness and the kinematics of the ionized gas mainly around type 2 AGN. While several studies reported ionized gas outflows on kpc scales \citep[e.g.,][]{Liu2013,Harrison2014,McElroy2015,Wylezalek2020}, recent work highlighted that ionized gas outflows can also be quite compact on 10--100\,pc scales  \citep[e.g.,][]{Villar-Martin2016,Tadhunter2018,Storchi-Bergmann2019}. Those discrepancies may be caused either by different AGN sample characteristics or in the treatment of the data.

Due to the high gas density and large ionizing flux close to an AGN, the [\ion{O}{iii}] emission is usually centrally concentrated and can outshine the extended [\ion{O}{iii}] core emission from the NLR on larger scale depending on seeing and redshift \citep[e.g.,][]{Husemann2016}. Hence, it is important to characterize the point-spread function (PSF) for an observation to unambiguously separate compact from extended [\ion{O}{iii}] emission. The main advantage of using type 1 AGN is that the broad Balmer emission lines (i.e. H$\alpha$, H$\beta$) allows us to reconstuct the PSF, with the assumption that the BLR emission is intrinsically unresolved (e.g. \citealt{Jahnke2004}.

We take advantage of the type 1 AGN in CARS to accurately characterize the PSF. In the following, we describe the analysis used to test whether the broad wing [\ion{O}{iii}] kinematic components are compact or extended at the limits of our seeing-limited spatial resolution and constrain the actual location using spectroastrometry. Therefore, we start with the hypothesis that the light distribution of the [\ion{O}{iii}] wing component is point like and centrally located. {Our hypothesis will be discarded in case the light distribution is spatially extended implying large-scale ionized gas outflows.

\subsubsection{Constructing the 2D light profiles of emission lines}

Following our hypothesis that the light distribution associated with all the emission lines is compact and that a possible extended component is faint with respect to its central counterpart, we fixed the emission line kinematics and radial velocities to the ones derived from the central brightest pixel modeling as described in Section.~\ref{Fig:AGNmodel}. Only the line fluxes and the continuum shape parameters were allowed to vary within the entire 3\arcsec\ around the AGN position. For VIMOS, we allowed the window to be 6\arcsec\ as the pixel size and spatial resolution is worse than MUSE. Thereafter, the line fluxes for each spaxel were obtained using a nonlinear Levenberg-Marquardt algorithm and we reconstructed the 2D intensity maps for each line separately. We further created 100 different data cubes by fluctuating the main cube with the associated error cubes and repeated the measurements 100 times to obtain proper error maps for the emission line fluxes implicitly taking cross-talk between the lines into account. The resulting emission line maps are exemplarily shown for HE~0040$-$1105 in the top panels of Fig.~\ref{Fig:O3astrometry_HE0040}.

   \begin{figure*}
   \sidecaption
   \includegraphics[width=\textwidth]{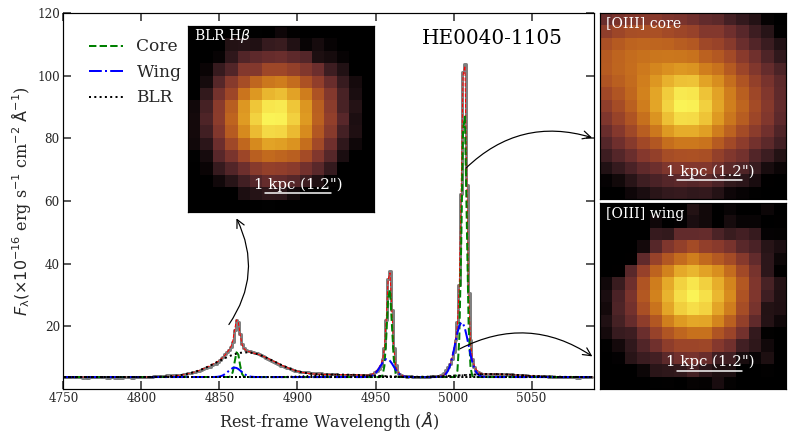}
      \caption{The central 3\arcsec\ intensity maps associated with different spectral lines. The leftmost map describes the light distribution from H$\beta$ broad emission-line region, which is the PSF. The bottom-right map describes the light distribution from blue-shifted ionized [\ion{O}{iii}] wing; and the top-right map describes light-distribution from the bright, narrow core of [\ion{O}{iii}].}
         \label{Fig:O3astrometry_HE0040}
   \end{figure*}

\subsubsection{Modeling the Point-Spread Function}
 
We assume a 2D Moffat function \citep{Moffat1969} to model the PSF following the work by \citet{Gadotti2008}. We use a non-linear Levenberg-Marquardt algorithm to model the observed light profiles. The reference PSF is derived by modelling the BLR H$\beta$ light distribution, where all Moffat model parameters (flux amplitude, centers, PSF shape parameters, and ellipticity) are free parameters.

\subsection{Classification of compact and extended outflows}

If the spatial distribution of the [\ion{O}{iii}] wing is extended, it will not follow a PSF. We assume an extended spatial distribution for the [\ion{O}{iii}] wing component if the deviation from the PSF model is 5$\sigma$. While the [\ion{O}{iii}] emission line profile and gas kinematics can change over kpc scales, such changes would be rather smooth as each kinematic component itself necessarily follows a PSF in the spatial domain. Those kinematic changes would therefore translate into amplitude changes of the wing and core components even at fixed kinematics. Those amplitude changes will manifest themselves as deviations from a pure PSF. Here, we do not consider any [\ion{O}{iii}] kinematic information beyond the central region and distinct line shape components on several kpc-scale would be missed. However, the main purpose of our approach is to characterise the 2D spatial distribution of the [\ion{O}{iii}] wing emission close to the nucleus within the characteristic radius implicitly set by multi-object fibre spectroscopy, like SDSS.

We model the 2D light distributions of the broad wing (and also the narrow core) of [\ion{O}{iii}] by varying the position and amplitude normalization of the initially determined Moffat PSF. All other parameters are kept fixed because the PSF shape should not significantly change within the considered narrow wavelength range. We further fluctuate the 2D intensity maps with the error maps and repeat the entire process 100 times to estimate the uncertainty associated with the measurement of the Moffat parameters. Afterwards, we construct the residual maps normalized by the errors ($R_\mathrm{norm}$) as shown in Fig.~\ref{Fig:full_map}.

   \begin{figure*}
   \includegraphics[width=\textwidth]{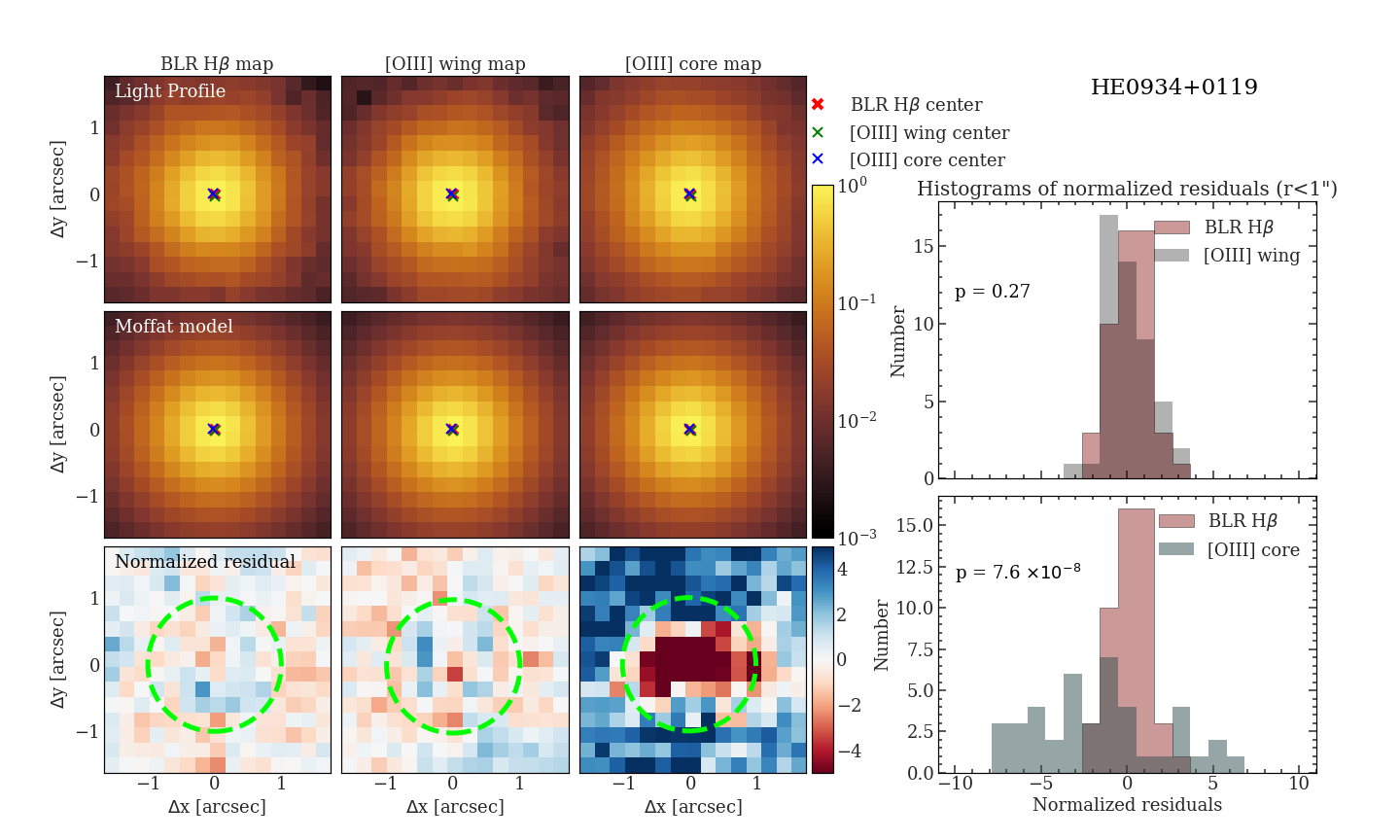}
   \caption{Results of our spectro-astrometric analysis for HE~934$+$0119 from the MUSE IFU observations. We have mapped the surface brightness within the central 3\arcsec\ for all spectral components in the brightest spectrum assuming fixed kinematics. Here we only show the measure light distribution from the BLR H$\beta$ (left), broad [\ion{O}{iii}] wing (middle), and narrow [\ion{O}{iii}] core (right) component. Those 2D light distributions are fitted with a 2D Moffat profile. The red cross indicates the Moffat center for the best-fit elliptical Moffat model to the surface brightness map of the BLR H$\beta$. The green and red crosses indicate the Moffat centers for the best-fit elliptical Moffat models to the surface brightness maps of the [\ion{O}{iii}] wing and core, in which the amplitude and position of the Moffat model were varied only. The histograms show the distributions of the normalized residuals (residual/error) for both the core and wing components in comparison to the BLR H$\beta$ one.}
         \label{Fig:full_map}
   \end{figure*}
   
We distinguish between unresolved and resolved [\ion{O}{iii}] wing emission based on the normalized residuals of the PSF fit. 
As the unresolved emission will be dominant close to the nucleus, we focus on the central 2\arcsec\ region around the AGN for the MUSE sources, and central 4.5\arcsec\ region around the AGN for the VIMOS sources given the larger PSF and poor sampling. In order to quantify the similarity between the normalized residual distributions of BLR H$\beta$ and [\ion{O}{iii}] wing, we perform a Levene's test \citep{Levene1960} within the aforementioned region. The Levene's test checks only the difference between two distributions, but not their relative widths. The $p$-value from Levene's test alone is not necessarily sufficient to conclude whether the [\ion{O}{iii}] wing emission is extended or not. Consequently, we further calculate the standard deviations of the normalized residuals distributions and compare $\sigma_\mathrm{[\ion{O}{iii}]}$ and $\sigma_\mathrm{H\beta}$ within the central region around the nucleus. When $\sigma_\mathrm{[\ion{O}{iii}:H\beta]}$ = $\sigma_\mathrm{[\ion{O}{iii}]}$/$\sigma_\mathrm{H\beta}<1$, we consider the [\ion{O}{iii}] wing to be unresolved, because the residual distribution for the [\ion{O}{iii}] wing is consistent with the PSF at the given noise level. This assumption only holds if the 2D Moffat PSF model is a good model for the empirical PSF as inferred from the BLR H$\beta$ 2D light distribution. In combination with the Levene's test we define two necessary criteria for unresolved [\ion{O}{iii}] wing emission as follows:}
\begin{enumerate}
    \item $p>0.05$
    \item $\sigma_\mathrm{[\ion{O}{iii}]:\mathrm{H}\beta}<1$ if $p\leq0.05$
\end{enumerate}
    
In Fig.\ref{Fig:full_map}, we show the surface brightness profiles for BLR H$\beta$, [\ion{O}{iii}] wing and cores for HE~09340$+$0119, along with their Moffat models and the normalized residuals. We find that the distribution of the normalized residuals for the [\ion{O}{iii}] wing are fully consistent with that of the  BLR H$\beta$ given a $p$-value of 0.27 for the Levene's test. It confirms that the outflow appears to be point-like at the limit of the MUSE data. Therefore, the BLR H$\beta$ and [\ion{O}{iii}] wing components are consistent with a point source with a slight offset in position by $\sim27$\,pc. Comparing the normalized residuals of the [\ion{O}{iii}] core component to that of the BLR H$\beta$ we report a p-value $\sim 10^{-7}$, indicating a completely different distribution than a point source.

\begin{figure*}
   \includegraphics[width=\textwidth]{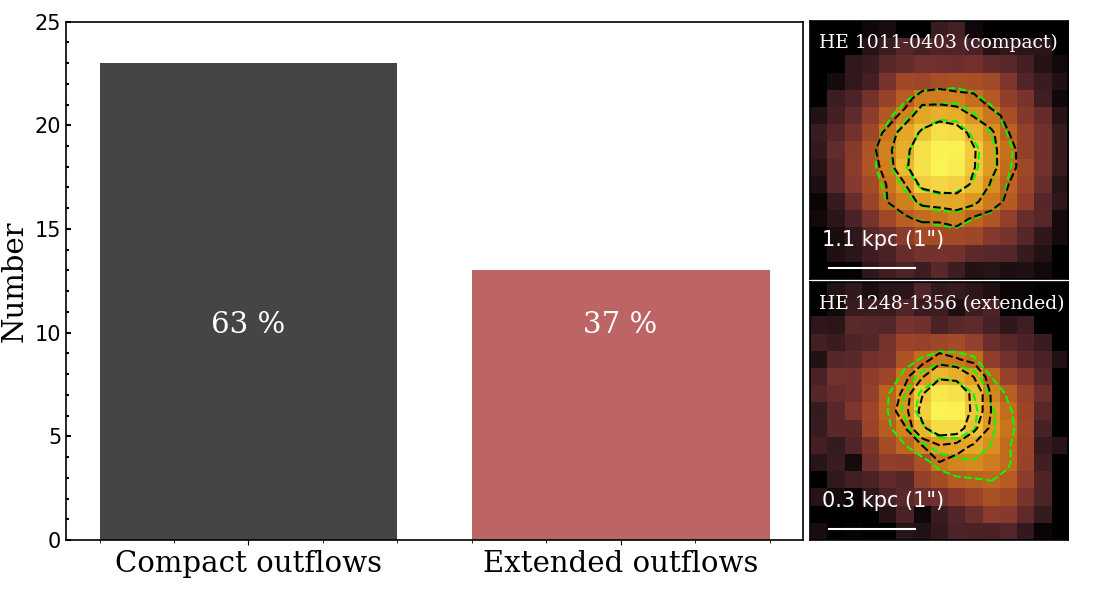}
   \caption{Results from our analysis: Out of 36 CARS targets, 23 agree with a point-like broad [\ion{O}{iii}] wing emission profile, whereas the other 13 targets have broad [\ion{O}{iii}] wing emission profile that is not point-like. We also show one source from each of compact (HE~1011$-$0403) and extended (HE~1248$-$1356) [\ion{O}{iii}] wing emitting sources. We further plot the green contours showing the [\ion{O}{iii}] wing emission-profiles, along with the black contours of the BLR H$\beta$ (PSF) emission profile.}
         \label{Fig:outflow char}
   \end{figure*}

We report 23 sources with compact [\ion{O}{iii}] wing (unresolved or point-like) and 13 sources with extended [\ion{O}{iii}] wing emission possibly extended to kpc scale (see Fig.~\ref{Fig:outflow char}). Furthermore, we calculate the total [\ion{O}{iii}] wing flux from its light distributions or surface brightness map ($F_\mathrm{aperture}$) and the Moffat model ($F_\mathrm{Moffat}$) within 2\arcsec\ region around the Moffat center. $F_\mathrm{aperture}$ gives us an idea how much emission from the broad wing component is contained within the central 2\arcsec\ region; whereas, $F_\mathrm{Moffat}$ demonstrates how much emission would be contained if the emission was entirely due to unresolved or point-like emission. We define the flux ratio of the [\ion{O}{iii}] wing emission within the central 2\arcsec\ to the [\ion{O}{iii}] wing point-like emission as $r_\mathrm{[\ion{O}{iii}]}=F_\mathrm{aperture}/F_\mathrm{Moffat}$.
The advantage of using this parameter is that it gives an indication of how significant the emission is in a region compared to a pure point source assumption, where $r_\mathrm{[\ion{O}{iii}]}=1$ for an ideal point-like source. For these 23 AGN, we calculate how far is the region of the region of maximum [\ion{O}{iii}] wing emission from the AGN central engine ($S_\mathrm{[\ion{O}{III}]}$). 
The distance is a lower limit of the size of the outflowing region, while we cannot exclude that the outflow extends beyond this distance.
We present the characterization of the [\ion{O}{iii}] wing emission for all CARS sources in Table~\ref{Table:2}.

\begin{figure}
   \sidecaption
   \includegraphics[width=9cm]{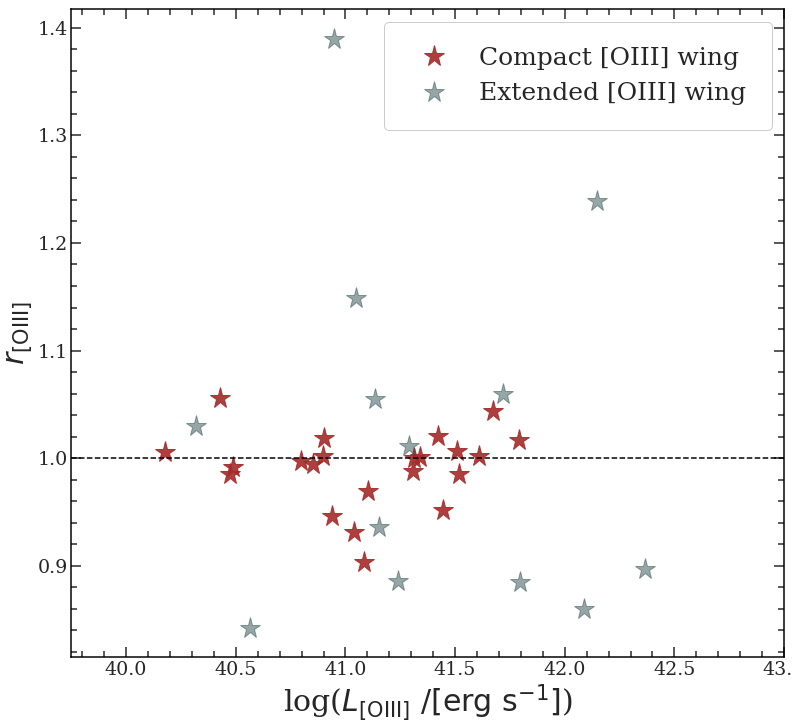}
      \caption{We plot $r_\mathrm{[\ion{O}{iii}]}$ against [\ion{O}{iii}] luminosity for the sources with compact and extended [\ion{O}{iii}] wing emission. The dashed line represent the region where $F_\mathrm{aperture}$ = $F_\mathrm{Moffat}$.
              }
         \label{Fig:Fluxratio}
   \end{figure}

In Fig.~\ref{Fig:Fluxratio}, we plot the [\ion{O}{iii}] luminosity against the flux ratio, $r_\mathrm{[\ion{O}{iii}]}$. For the 23 sources with unresolved [\ion{O}{iii}] wing emission, the ratio is between 0.9--1.05. 3 sources (HE~1107$-$0813, HE~2128$-$0221 and HE~2211$-$3903) appear to have less emission within their surface brightness map than the Moffat model (see Appendix~\ref{section:maps}). This is caused by the fact that the surface brightness in every region does not mimic a 2D Moffat profile, which introduces the difference in total flux values. But the surface brightness profile is so close to the Moffat profile that the residual emission fluctuates within the noise range. 

The 13 sources with extended [\ion{O}{iii}] wing emission show diverse emission profiles. HE~0212$-$0059 and HE~1353$-$1917 have around 15--40\% more flux, clearly showing the extended [\ion{O}{iii}] emission that cannot be described by the Moffat model. 3 sources (HE~0021$-$1819, HE~0119$-$0118 and HE~0429$-$0247) with extended wing emission seem to have $<$10\% [\ion{O}{iii}] flux within the 2\arcsec\ diameter around their peak emission. We note that these sources have the [\ion{O}{iii}] emission towards a certain direction in contrast to a purely spherically symmetric emission profile. While estimating the total flux for these sources, the excess [\ion{O}{iii}] wing emission compared to the Moffat PSF in some regions is often compensated by the other region which has less flux than the Moffat PSF. Therefore, the total flux from the [\ion{O}{iii}] wing emission is within 10\% of a pure point-like emission.

On the other hand, the 4 out of 13 sources with extended [\ion{O}{iii}] wing emission have up to 10\% less flux in the central 2\arcsec\ compared to a purely unresolved or point-like emission profile. This feature is most prominently seen in HE~1338$-$1423, which exhibits a large bi-polar outflow \citep{Husemann2014IntegralMetallicities}. The outflow in HE~1338$-$1423 is extended in the south-east and north-west direction (See Appendix~\ref{section:maps}). Although, the central 3\arcsec\ diameter region has less flux than a pure point source (which explains the flux deficiency), there is an excess [\ion{O}{iii}] wing emission beyond the central 3\arcsec\ visible till the edge of the central 4\arcsec\ window suggesting that the ionized gas might be extended to several kpc scales. 

\begin{table*}
\caption{Intensity Modelling Table for CARS sources}
\label{Table:2}
\centering
\begin{tabular}{c c c c c c c c c}\hline\hline
Object & $FWHM_\mathrm{max}$ \tablefootmark{a} & $e$ \tablefootmark{b} & \textcolor{black}{Scale} \tablefootmark{c} & $S_\mathrm{[O~III]}$ \tablefootmark{d} & $r_\mathrm{[O~III]}$ \tablefootmark{e} & p-value \tablefootmark{f} & $\sigma_\mathrm{[O~III]:H\beta}$ \tablefootmark{g} &\textcolor{black}{Type} \tablefootmark{h}\\ \smallskip
& (mas) & & \textcolor{black}{(kpc/")} & (pc) & & & & \\ \hline \smallskip
HE~0021$-$1810 & 1001 $\pm$ 69 & 0.36 $\pm$ 0.10 & \textcolor{black}{0.845} & 40 $\pm$ 29 $\pm$ 100 & 1.01 $\pm$ 0.01 & $0.01$ & 1.47 & \textcolor{black}{Unresolved}\\
HE~0021$-$1819 & 791 $\pm$ 7 & 0.37 $\pm$ 0.02 & \textcolor{black}{1.051} & 44 $\pm$ 4 $\pm$ 11 & 1.01 $\pm$ 0.01 & $4.1\times10^{-3}$ & 1.97 & \textcolor{black}{Resolved}\\
HE~0040$-$1105 & 741 $\pm$ 3 & 0.33 $\pm$ 0.01 & \textcolor{black}{1.099} & 92 $\pm$ 2 $\pm$ 8 & 0.99 $\pm$ 0.01 & $0.28$ & 1.21 & \textcolor{black}{Unresolved}\\
HE~0108$-$4743 & 1451 $\pm$ 22 & 0.27 $\pm$ 0.05 & \textcolor{black}{1.591} & 67 $\pm$ 8 $\pm$ 12 & 0.84 $\pm$ 0.02 & $3.5\times10^{-7}$ & 2.99 & \textcolor{black}{Resolved}\\
HE~0114$-$0015 & 633 $\pm$ 11 & 0.30 $\pm$ 0.04 & \textcolor{black}{1.104} & 70 $\pm$ 21 $\pm$ 67 & 0.99 $\pm$ 0.03 & $0.75$ & 0.83 & \textcolor{black}{Unresolved}\\
HE~0119$-$0118 & 748 $\pm$ 11 & 0.38 $\pm$ 0.03 & \textcolor{black}{1.026} & 54 $\pm$ 7 $\pm$ 8 & 1.06 $\pm$ 0.01 & $2.1\times10^{-8}$ & 3.02 & \textcolor{black}{Resolved}\\
HE~0212$-$0059 & 673 $\pm$ 7 & 0.12 $\pm$ 0.04 & \textcolor{black}{0.970} & 186 $\pm$ 8 $\pm$ 4 & 1.39 $\pm$ 0.08 & $6.6\times10^{-5}$ & 2.43 & \textcolor{black}{Resolved}\\
HE~0224$-$2834 & 1661 $\pm$ 7 & 0.16 $\pm$ 0.02 & \textcolor{black}{0.951} & 29 $\pm$ 7 $\pm$ 10 & 0.99 $\pm$ 0.02 & $0.23$ & 0.81 & \textcolor{black}{Unresolved}\\
HE~0227$-$0913 & 1204 $\pm$ 3 & 0.19 $\pm$ 0.01 & \textcolor{black}{1.005} & 16 $\pm$ 3 $\pm$ 14 & 1.06 $\pm$ 0.02 & $0.01$ & 0.62 & \textcolor{black}{Unresolved}\\
HE~0232$-$0900 & 1222 $\pm$ 2 & 0.23 $\pm$ 0.01 & \textcolor{black}{1.994} & 126 $\pm$ 2 $\pm$ 4 & 0.86 $\pm$ 0.02 & $5.2\times10^{-11}$ & 3.41 & \textcolor{black}{Resolved}\\
HE~0253$-$1641 & 954 $\pm$ 10 & 0.28 $\pm$ 0.02 & \textcolor{black}{1.630} & 22 $\pm$ 3 $\pm$ 14 & 1.00 $\pm$ 0.01 & $0.21$ & 1.23 & \textcolor{black}{Unresolved}\\
HE~0345$+$0056 & 921 $\pm$ 3 & 0.18 $\pm$ 0.01 & \textcolor{black}{1.015} & 7 $\pm$ 1 $\pm$ 8 & 1.04 $\pm$ 0.01 & $0.49$ & 1.06 & \textcolor{black}{Unresolved}\\
HE~0351$+$0240 & 794 $\pm$ 3 & 0.29 $\pm$ 0.01 & \textcolor{black}{1.657} & 16 $\pm$ 5 $\pm$ 19 & 0.89 $\pm$ 0.02 & $3.8\times10^{-3}$ & 1.68 & \textcolor{black}{Resolved}\\
HE~0412$-$0803 & 781 $\pm$ 1 & 0.30 $\pm$ 0.00 & \textcolor{black}{1.087} & 26 $\pm$ 1 $\pm$ 4 & 0.90 $\pm$ 0.01 & $1.4\times10^{-5}$ & 1.98 & \textcolor{black}{Resolved}\\
HE~0429$-$0247 & 980 $\pm$ 8 & 0.24 $\pm$ 0.02 & \textcolor{black}{1.850} & 42 $\pm$ 4 $\pm$ 11 & 1.06 $\pm$ 0.01 & $4.5\times10^{-4}$ & 1.76 & \textcolor{black}{Resolved}\\
HE~0433$-$1028 & 606 $\pm$ 4 & 0.12 $\pm$ 0.03 & \textcolor{black}{1.588} & 14 $\pm$ 1 $\pm$ 6 & 1.00 $\pm$ 0.01 & $0.08$ & 1.42 & \textcolor{black}{Unresolved}\\
HE~0853$+$0102 & 628 $\pm$ 8 & 0.20 $\pm$ 0.04 & \textcolor{black}{1.044} & 12 $\pm$ 6 $\pm$ 8 & 1.00 $\pm$ 0.01 & $0.10$ & 1.32 & \textcolor{black}{Unresolved}\\
HE~0934$+$0119 & 744 $\pm$ 6 & 0.07 $\pm$ 0.03 & \textcolor{black}{1.011} & 27 $\pm$ 5 $\pm$ 13 & 1.00 $\pm$ 0.01 & $0.27$ & 1.24 & \textcolor{black}{Unresolved}\\
HE~1011$-$0403 & 820 $\pm$ 8 & 0.17 $\pm$ 0.04 & \textcolor{black}{1.070} & 17 $\pm$ 6 $\pm$ 25 & 1.02 $\pm$ 0.01 & $0.70$ & 1.10 & \textcolor{black}{Unresolved}\\
HE~1017$-$0305 & 806 $\pm$ 3 & 0.28 $\pm$ 0.01 & \textcolor{black}{1.106} & 14 $\pm$ 8 $\pm$ 27 & 0.90 $\pm$ 0.02 & $1.8\times10^{-3}$ & 0.63 & \textcolor{black}{Unresolved}\\
HE~1029$-$1831 & 751 $\pm$ 10 & 0.26 $\pm$ 0.05 & \textcolor{black}{1.080} & 57 $\pm$ 5 $\pm$ 12 & 0.94 $\pm$ 0.02 & $3.3\times10^{-6}$ & 5.52 & \textcolor{black}{Resolved}\\
HE~1107$-$0813 & 697 $\pm$ 8 & 0.39 $\pm$ 0.02 & \textcolor{black}{1.087} & 40 $\pm$ 14 $\pm$ 164 & 0.95 $\pm$ 0.03 & $0.71$ & 0.98 & \textcolor{black}{Unresolved}\\
HE~1108$-$2813 & 510 $\pm$ 2 & 0.27 $\pm$ 0.01 & \textcolor{black}{0.981} & 35 $\pm$ 2 $\pm$ 8 & 1.00 $\pm$ 0.01 & $0.61$ & 1.00 & \textcolor{black}{Unresolved}\\
HE~1126$-$0407 & 811 $\pm$ 3 & 0.29 $\pm$ 0.01 & \textcolor{black}{1.075} & 23 $\pm$ 4 $\pm$ 28 & 1.02 $\pm$ 0.01 & $0.92$ & 0.96 & \textcolor{black}{Unresolved}\\
HE~1237$-$0504 & 614 $\pm$ 1 & 0.23 $\pm$ 0.01 & \textcolor{black}{0.982} & 5 $\pm$ 0 $\pm$ 3 & 0.99 $\pm$ 0.01 & $3.3\times10^{-6}$ & 0.46 & \textcolor{black}{Unresolved}\\
HE~1248$-$1356 & 646 $\pm$ 9 & 0.41 $\pm$ 0.02 & \textcolor{black}{0.960} & 11 $\pm$ 2 $\pm$ 5 & 1.03 $\pm$ 0.02 & $1.8\times10^{-8}$ & 5.17 & \textcolor{black}{Resolved}\\
HE~1310$-$1051 & 536 $\pm$ 1 & 0.60 $\pm$ 0.00 & \textcolor{black}{1.026} & 4 $\pm$ 1 $\pm$ 2 & 1.01 $\pm$ 0.01 & $4.3\times10^{-3}$ & 0.43 & \textcolor{black}{Unresolved}\\
HE~1330$-$1013 & 775 $\pm$ 12 & 0.25 $\pm$ 0.05 & \textcolor{black}{0.817} & 19 $\pm$ 5 $\pm$ 17 & 1.01 $\pm$ 0.02 & $0.09$ & 0.76 & \textcolor{black}{Unresolved}\\
HE~1338$-$1423 & 398 $\pm$ 4 & 0.68 $\pm$ 0.00 & \textcolor{black}{1.105} & 49 $\pm$ 1 $\pm$ 6 & 0.74 $\pm$ 0.01 & $2.0\times10^{-8}$ & 10.67 & \textcolor{black}{Resolved}\\
HE~1353$-$1917 & 811 $\pm$ 9 & 0.26 $\pm$ 0.04 & \textcolor{black}{0.853} & 802 $\pm$ 7 $\pm$ 17 & 1.15 $\pm$ 0.11 & $8.6\times10^{-11}$ & 7.18 & \textcolor{black}{Resolved}\\
HE~1417$-$0909 & 769 $\pm$ 2 & 0.27 $\pm$ 0.01 & \textcolor{black}{0.615} & 33 $\pm$ 3 $\pm$ 10 & 0.95 $\pm$ 0.01 & $0.01$ & 0.68 & \textcolor{black}{Unresolved}\\
HE~2128$-$0221 & 757 $\pm$ 6 & 0.11 $\pm$ 0.03 & \textcolor{black}{0.899} & 38 $\pm$ 8 $\pm$ 25 & 0.97 $\pm$ 0.02 & $0.12$ & 0.73 & \textcolor{black}{Unresolved}\\
HE~2211$-$3903 & 529 $\pm$ 4 & 0.32 $\pm$ 0.02 & \textcolor{black}{0.803} & 78 $\pm$ 9 $\pm$ 15 & 0.93 $\pm$ 0.02 & $0.12$ & 0.69 & \textcolor{black}{Unresolved}\\
HE~2222$-$0026 & 680 $\pm$ 6 & 0.23 $\pm$ 0.03 & \textcolor{black}{0.905} & 45 $\pm$ 14 $\pm$ 44 & 0.99 $\pm$ 0.02 & $5.9\times10^{-4}$ & 0.55 & \textcolor{black}{Unresolved}\\
HE~2233$+$0124 & 927 $\pm$ 15 & 0.33 $\pm$ 0.03 & \textcolor{black}{0.575} & 37 $\pm$ 23 $\pm$ 39 & 1.02 $\pm$ 0.03 & $3.3\times10^{-3}$ & 0.69 & \textcolor{black}{Unresolved}\\
HE~2302$-$0857 & 729 $\pm$ 8 & 0.17 $\pm$ 0.03 & \textcolor{black}{1.131} & 178 $\pm$ 6 $\pm$ 15 & 1.24 $\pm$ 0.03 & $3.2\times10^{-5}$ & 2.14 & \textcolor{black}{Resolved}\\
\noalign{\smallskip}\hline
\end{tabular}
\tablefoot{ 
\tablefoottext{a}{Maximum FWHM estimated from the semi-major axis of the best-fit elliptical Moffat model fitted to the surface brightness map of BLR H$\beta$},
\tablefoottext{b}{Ellipticity of the best-fit elliptical Moffat model},
\tablefoottext{c}{\textcolor{black}{The kiloparsec per arcsecond scale achievable for individual objects}},
\tablefoottext{d}{The offset between the Moffat centers of BLR H$\beta$ and [\ion{O}{iii}] wing. 
It quantifies how the is the region of the region of maximum [\ion{O}{iii}] wing emission from the AGN central engine. It further gives an indication of the lower limit of the size of the outflowing region. The size of the outflowing region could be extended far beyond this distance, but this quantifies that the outflow has already traversed at least distance; and hence it is a lower limit of the size of the outflowing region.
 The first uncertainty denotes the random uncertainty due to measurements; and the second uncertainty denotes the uncertainty in position measurement due to systematics},
\tablefoottext{e}{The flux ratio of the integrated surface brightness profile within central 2-3$\arcsec$ (2$\arcsec$ for MUSE and 3$\arcsec$ for VIMOS) and that of the best-fit Moffat model for [\ion{O}{iii}] wing},
\tablefoottext{f}{The p-value estimated from Levene's test to check the similarity of distributions between the normalized residuals of BLR H$\beta$ and [\ion{O}{iii}] wing},
\tablefoottext{g}{The ratio of standard deviation of the normalized residuals of [\ion{O}{iii}] wing and BLR H$\beta$},
\tablefoottext{h}{\textcolor{black}{The characterization of the spatial profile of the [\ion{O}{iii}] wing- if they are unresolved or resolved.}}
}
\end{table*}

\subsection{Systematic error estimation for offset measurements}
Previously, we reported that 23 AGN in our sample are consistent with spatially unresolved [\ion{O}{iii}] wing emission. The 2D Moffat centroid for the [\ion{O}{iii}] wing emission was offset from the central engine by $<100$\,pc, with a median offset of $\sim27$\,pc. Although we estimated the measurement error for the 2D offset by repeating the process several times with fluctuated data,  we also need to consider the systematic uncertainty of positional measurements.
The systematics include other instrument effects like systematic flat-fielding residuals, geometrical distortion of the CCD, or detector noise, which would artificially change the flux distribution at different wavelength and thereby affect the measured flux-weighted centroids.

In order to estimate the systematic positional uncertainty, we construct an artificial source based on the AGN spectrum model and the best-fit 2D PSF model of the BLR H$\beta$ and add it in random locations within the MUSE FoV. We excluded regions with (a) a star, and (b) the presence of bright AGN continuum and verify if $S/N<5$ in all spaxels within a 3\arcsec window. \newline We put the artificial source on 300 arbitrary locations within IFU FoV following the aforementioned criteria and carried out a spectro-astrometric analysis to determine the angular offset between the known and measured  offset for the outflowing ionized [\ion{O}{iii}]. The standard deviation of the angular offset distribution gives us an indication of systematic error in estimating the offsets, which we present in Table~\ref{Table:2}. 

\subsection{Estimation of the electron densities}
With the sensitivity of MUSE we can robustly detect the optical emission lines [\ion{S}{ii}]$\lambda\lambda$6716,6731 ([\ion{S}{ii}] doublet). 
The doublet lines occur due to two different energy levels with very similar excitation energies.
Their relative fluxes are dependent upon the electron density that occupies these energy levels. In this scenario, any degeneracy that could arise because of different ions is removed \citep{Osterbrock2006}. One key issue is that, the measurement of the electron density is temperature dependent. But, the uncertainty that will be introduced with the assumption of an electron temperature, $T_\mathrm{e} = 10^4$ K, will be lower than the measurement uncertainty of the electron densities \citep{Sanders2015}. Additionally, the [\ion{S}{ii}] doublet is sensitive only to a limited range of electron densities, which is well within the typical range of electron densities observed in the Extended Narrow Line Region (ENLR). The ENLR is extended over a few kpc scales where densities in the diffuse ISM are at the level of $100\ \mathrm{cm}^{-3}$ \citep[e.g.,][]{Harrison2014,Husemann2016}.

\begin{figure}
   \resizebox{\hsize}{!}{\includegraphics{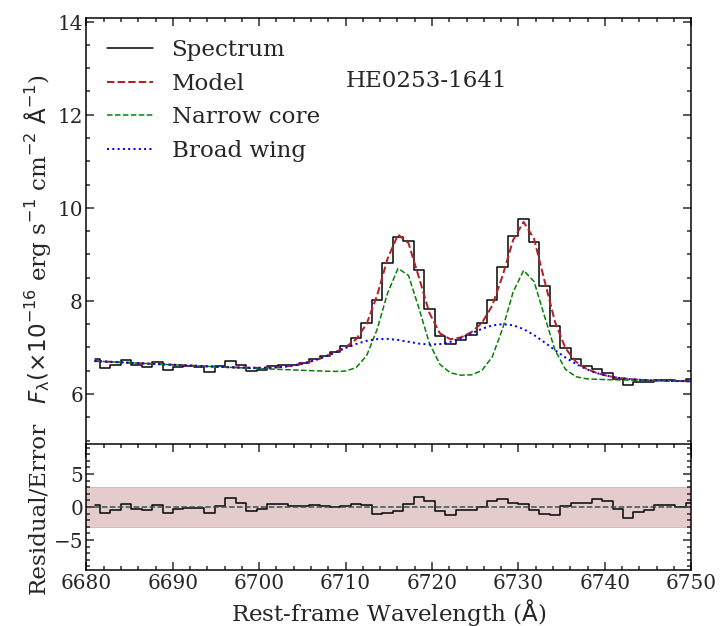}}
      \caption{Example of the multi-component modelling for the spectrum of HE~0253$-$1641. \textit{Upper panel:} Integrated 3 $\times$ 3 spaxel-spectrum of HE~0253$-$1641 and the best-fit spectral model.  The green lines refer to the narrow core and the blue lines refer to the broad wing components of [\ion{S}{ii}]. \textit{Lower panel:} Residual spectrum normalized by the error spectrum. The dark grey shaded area highlights the $\pm 3\sigma$ limit and the red dashed line provides the reference to 0.}
         \label{Fig:SIImodel}
   \end{figure}

To infer the electron densities of the core and wing components we 
deblend the [\ion{S}{ii}] doublet accordingly. We model the doublet within the rest-frame wavelength range from 6680\AA\ to 6750\AA\ which covers any possible blue-shifted component. Because the [\ion{S}{ii}] has a lower $S/N$ than other emission lines (i.e. [\ion{O}{iii}], [\ion{N}{ii}]), it is difficult to obtain enough signal from a single pixel. The majority of the outflows in our sample agree with an unresolved source, so that the majority of the emission is confined within the central 1\arcsec\ diameter around the AGN. Therefore, we co-add all the spectra of 3$\times$3 spaxels around the AGN to improve the $S/N$. Similar to our H$\beta$ + [\ion{O}{iii}] + \ion{Fe}{ii} spectral modelling,  we model the the [\ion{S}{ii}] doublet with kinematically coupled multi-Gaussian components. For the underlying continuum we adopt a 2nd order polynomial to capture the 'tail' of the H$\alpha$ BLR emission.  Uncertainties are again estimated by fitting 100 mock spectra fluctuated within their noise and by calculating the standard deviation of the resulting model parameter distributions. An example of the best-fit AGN spectral model for HE~0253$-$1641 along with the data and the residual is shown in Fig.~\ref{Fig:SIImodel}.

We use the PyNeb package \citep{Luridiana2015} and assume a gas temperature, $\mathrm{T} = 10^{4}\,\mathrm{K}$ for all calculations to convert the [\ion{S}{ii}] doublet flux ratio to electron densities. Since the wing component of the [\ion{S}{ii}] doublet has much lower $S/N$, we can only reliably determine the electron density for 9 sources among which 6 sources have unresolved [\ion{O}{iii}] wing emission and 3 other sources have extended [\ion{O}{iii}] wing emission. We present our finding in Table~\ref{Table:3}.

Following the work of \citet{Osterbrock2006} it is evident that as the electron density decreases, the flux ratio of the [\ion{S}{ii}] doublet approaches the upper limit of 1.45, whereas if the electron density increases, the flux ratio of [\ion{S}{ii}] doublet approaches the lower limit of 0.45. For the core component we find a flux ratio of $\sim 1.0$ indicating an electron density $\sim 500\ \mathrm{cm}^{-3}$.

We derive  a flux ratio of  $\sim 0.7$ for the wing component. Such a flux ratio corresponds to a much higher electron density $> 10^3\ \mathrm{cm}^{-3}$. Hence our results suggest that the electron densities for the wing [\ion{S}{ii}] component are systematically higher than the bright, narrow core component.

\section{Discussion}
\label{section:discussion}
\subsection{Where does the majority of the broad [\ion{O}{iii}] wing emission come from?}

\subsubsection{The range in the extent of turbulent circum-nuclear ionized gas}

Numerous studies used HST/ACS imaging and spatially-resolved spectroscopy to quantify the extent of turbulent ionized gas potentially related to AGN-driven outflows. Whereas some studies report outflows extending over several kpcs \citep[e.g.,][]{Liu2013,Liu2013a,Harrison2014,McElroy2015,Karouzos2016,Kang2018}, others have reported sub-kpc scale outflows \citep[e.g.,][]{Villar-Martin2016,Tadhunter2018,Kawaguchi2018,Bellocchi2019,Baron2019}.
This paper focuses on the location of the [\ion{O}{iii}] wing emission as an indication of turbulent gas motion driven either by outflows \citep[e.g.,][]{Crenshaw2010,Zhang2021} or other processes \citep[e.g.,][]{Tremblay:2018,Gaspari:2018}. We will use the term outflow to describe this turbulent gas region. Although we cannot physically confirm for individual objects that AGN outflows are the main driver of the turbulence, the outflow interpretation is motivated by other studies.
\citet{Crenshaw2010} found that the NLR gas has a velocity offset $>50 \text{km\ s}^{-1}$ in the [\ion{O}{iii}] line-shapes; where 35\% of the quasars exhibit blue wings and 6\% of the quasars show red wings. They concluded that these red and blue wings could be due to the mass outflow in ionized gas and dust extinction in the inner disk. \citet{Venturi2018,Singha2021} stated that the broad [\ion{O}{iii}] wings could indicate outflow and be related to AGN activities.

In order to determine the location of the [\ion{O}{iii}] wing emission, we compare the 2D light distribution of the [\ion{O}{iii}] wing with the empricially constrained PSF from the BLR emission approximated by a 2D Moffat analytic function. When the distribution is consistent with the Moffat PSF, we can determine the spatial locations of [\ion{O}{iii}] wing emission with high precision using spectro-astrometry. The location of this turbulent gas region might be considered as the current base of an outflow. When the [\ion{O}{iii}] wing emission is spatially extended, estimating a centroid is meaningless and requires a careful PSF debelending to recover the instrinsic distribution and kinematics. This is beyond the scope of the paper and will follow-up on this in future work.

For the compact outflows we can compare the results for our unbiased type 1 AGN sample with previous studies. Different methods have been used to quantify outflow size. We study the difference in inferred sizes by comparing our spectro-astrometric results with other methods. Here, we use the approach taken by \citet{Karouzos2016} as an example for various other works based on seeing-limited observations and showcase the order of magnitude difference in outflow size/location that can result for the [\ion{O}{iii}] wing emission when the impact of beam smearing is not properly considered.

\begin{figure*}
   \includegraphics[width=\textwidth]{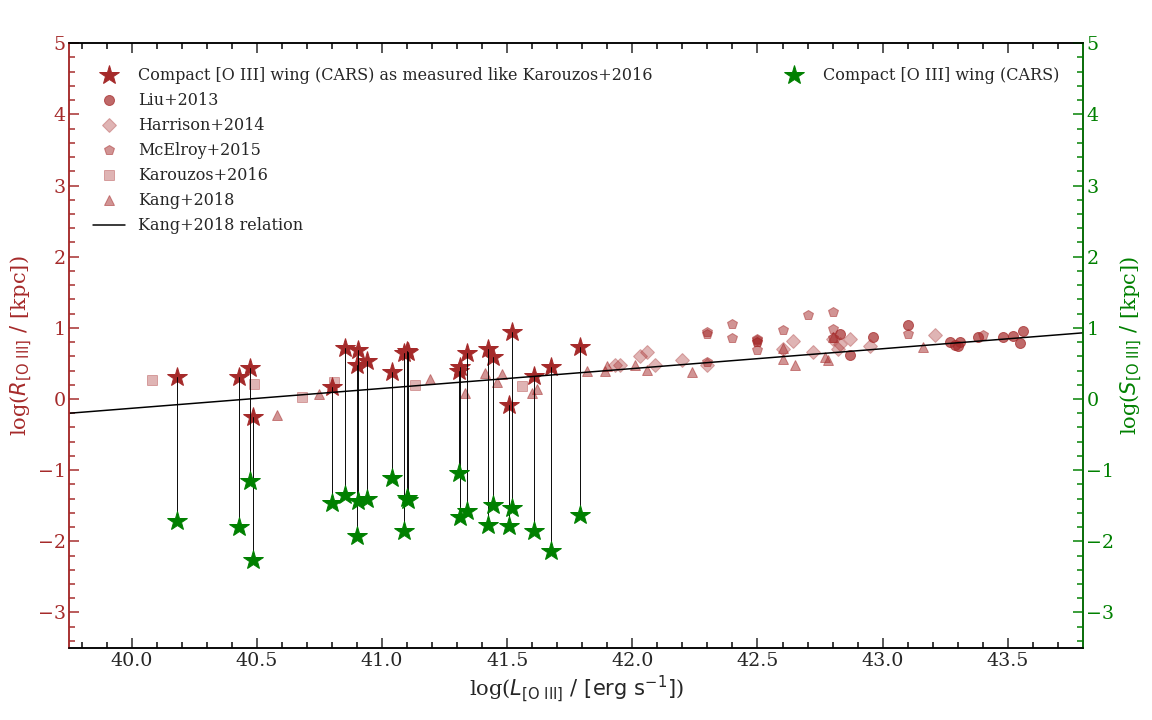}
   \caption{Spatial extension of the [\ion{O}{iii}] wing emission vs [\ion{O}{iii}] luminosity for our samples along with outflow sizes obtained by other studies in the literature at similar or higher [\ion{O}{iii}] luminosity range that find outflows of several kpc scales. For the sources with compact [\ion{O}{iii}] wing emission, we assume the offset sizes obtained by spectro-astrometry to indicate the how far is the peak of the emission from the nucleus. As previous studies considered the [\ion{O}{iii}] wing to indicate outflows \citep[e.g,][]{Crenshaw2010,Zhang2021}, we further plot the outflow sizes measured using the approach taken by \citet{Karouzos2016,Kang2018}. We connect the outflow-sizes obtained using the method of \citet{Karouzos2016} and the offset-sizes using spectro-astrometry for the CARS sources with compact [\ion{O}{iii}] wing emission to quantify the difference in determining spatial extension using these two methods.}
   \label{Fig:outflow_size}
\end{figure*}

In Fig.~\ref{Fig:outflow_size}, we plot the offset between the region of the maximum [\ion{O}{iii}] wing emission ($S_{\text[\ion{O}{iii}]}$) against $L_\mathrm{[\ion{O}{iii}]}$ for the unresolved outflows cases in our sample. In comparison, we also show measurements of the outflow sizes $R_{\text[\ion{O}{iii}]}$ as reported from past works. It is important to note that $S_{\text[\ion{O}{iii}]}$ is not exactly the same parameter as $R_{\text[\ion{O}{iii}]}$ as defined, e,g., in  \citet{Liu2013,Harrison2014,McElroy2015,Karouzos2016,Kang2018} because $S_{\text[\ion{O}{iii}]}$ estimates the distance between the maximum [\ion{O}{iii}] wing emitting region and the nucleus; whereas $R_{\text[\ion{O}{iii}]}$ estimates the extent of the region dominated by outflowing ionized gas. Still, it is useful to compare both quantities as it highlights the impact of beam smearing on ground-based observations. Here, we also compute $R_{\text[\ion{O}{iii}]}$ following the approach of Karouzos et al. (2016) for the compact outflow cases.
The resulting $R_{\text[\ion{O}{iii}]}$ is order of magnitude larger than $S_{\text[\ion{O}{iii}]}$. Intriguingly, the kpc-scale values appear fully consistent with the mild $L_\mathrm{[\ion{O}{iii}]}$  dependence despite the fact that we know the emission is unresolved. An explanation for this could be that past studies did not check for consistency with the PSF and therefore could not distinguish unresolved from resolved emission components. Below we discuss a number of different approaches that have been used to estimate outflow sizes in the literature in the context of the spectro-astrometric method used in this paper.
   
\begin{table*}
\caption{Electron Density Table for CARS sources}
\label{Table:3}
\centering
\begin{tabular}{c cccc}\hline\hline
Object & \multicolumn{2}{c}{$R_{[\ion{S}{ii}]}$ \tablefootmark{a}} & \multicolumn{2}{c}{$n_{e} \tablefootmark{b} \: (cm^{-3}$)} \\ \hline
& core & wing & core & wing\\ \hline \smallskip
HE0040-1105 & \textcolor{black}{1.09 $\pm$ 0.01} & \textcolor{black}{0.79 $\pm$ 0.06} & \textcolor{black}{351 $\pm$ 15} & \textcolor{black}{1165 $\pm$ 276} \\
HE0108-4743 & \textcolor{black}{1.28 $\pm$ 0.09} & \textcolor{black}{0.93 $\pm$ 0.13} & \textcolor{black}{124 $\pm$ 94} & \textcolor{black}{653 $\pm$ 609} \\
HE0119-0118 & \textcolor{black}{0.96 $\pm$ 0.02} & \textcolor{black}{0.88 $\pm$ 0.02} & \textcolor{black}{593 $\pm$ 39} & \textcolor{black}{799 $\pm$ 53} \\
HE0224-2834 & \textcolor{black}{1.13 $\pm$ 0.05} & \textcolor{black}{0.66 $\pm$ 0.13} & \textcolor{black}{285 $\pm$ 68} & \textcolor{black}{2124 $\pm$ 1851} \\
HE0253-1641 & \textcolor{black}{0.99 $\pm$ 0.05} & \textcolor{black}{0.64 $\pm$ 0.04} & \textcolor{black}{510 $\pm$ 93} & \textcolor{black}{2455 $\pm$ 591} \\
HE0433-1028 & \textcolor{black}{1.03 $\pm$ 0.02} & \textcolor{black}{0.66 $\pm$ 0.03} & \textcolor{black}{431 $\pm$ 29} & \textcolor{black}{2192 $\pm$ 334} \\
HE1310-1051 & \textcolor{black}{0.95 $\pm$ 0.04} & \textcolor{black}{0.71 $\pm$ 0.04} & \textcolor{black}{598 $\pm$ 95} & \textcolor{black}{1638 $\pm$ 516} \\
HE1338-1423 & \textcolor{black}{0.98 $\pm$ 0.02} & \textcolor{black}{1.27 $\pm$ 0.04} & \textcolor{black}{531 $\pm$ 42} & \textcolor{black}{132 $\pm$ 37} \\
HE2302-0857 & \textcolor{black}{0.97 $\pm$ 0.02} & \textcolor{black}{1.04 $\pm$ 0.01} & \textcolor{black}{565 $\pm$ 46} & \textcolor{black}{427 $\pm$ 20} \\
\noalign{\smallskip}\hline
\end{tabular}
\tablefoot{ 
\tablefoottext{a}{The ratio of the fluxes of the emission-line [\ion{S}{ii}]$\lambda 6716$ to that of [\ion{S}{ii}]$\lambda 6731$ for both the core and wing components.},
\tablefoottext{b}{The estimated electron density from the [\ion{S}{ii}] flux ratio following \citet{Osterbrock2006}}.
}
\end{table*}


\citet{Liu2013} analyzed a sample of 11 type 2 radio-quiet QSOs at z$\sim$0.5 from the sample of \citet{Reyes2008}. They reported that [\ion{O}{iii}] / H$\beta \sim 10$ in the central region until it reaches a 'break radius' ($R_\mathrm{br}$) where it starts to decrease, reaching the 'matter bounded regime'. \citet{Liu2013a} used $R_\mathrm{br}$ as the outflow radius ($D$) in order to calculate the outflow energetics. They estimated $R_\mathrm{br}$ to be $\sim$4--11\,kpc for all the QSOs.
\citet{Liu2013} obtained a flux-calibrated [\ion{O}{iii}] and H$\beta$ surface brightness map by collapsing the continuum subtracted IFU data cube over the [\ion{O}{iii}] and H$\beta$ wavelength range. The drawback of their approach is that it could contain not only the contribution from the outflowing ionized gas but also from non-outflowing photo-ionized gas on several kpc scales. To visualize the effect of their prescription, we perform a similar analysis on HE~0040$-$1105 -- one of the AGN with clearly compact [\ion{O}{iii}] wing emission. We notice that the [\ion{O}{iii}] / H$\beta \sim$ 10 up to 2\,kpc where it sharply falls of to [\ion{O}{iii}] / H$\beta \sim$ 6 and stays almost constant up to 8 kpc where it falls to even lower value.
Contrary to this kpc scale extension, we show that the emission profile of the [\ion{O}{iii}] wing follows the PSF very closely; and the peak of maximum [\ion{O}{iii}] wing emission is $\sim$100\,pc away from the nucleus which is about a magnitude lower than the 'break radius'.

\citet{Harrison2014} analyzed a sample of 16 $z<0.2$ type 2 AGN. They reported the high velocity ($W_\mathrm{80}>600\,\mathrm{km\,s}^{-1}$) ionized gas are prevalent over a distance, $D_\mathrm{600}$, which they defined as the maximum projected spatial extent of the broad [\ion{O}{iii}] wing emission profile, where the outflows dominate the gas kinematics.
They found that all of their AGN showed signatures of spherical outflows and bi-polar super-bubbles and are extended to$\sim$6--16\,kpc. \citet{Harrison2014} concluded that the ionized outflows are not only ubiquitous in luminous ($L_\mathrm{[\ion{O}{iii}]}>5\times 10^{41} \mathrm{erg\, s}^{-1}$) type 2 AGN, but also are extended over several kpc scales in $\geq$70\% (3$\sigma$ confidence) of cases. However, they selected QSOs with significantly broader [\ion{O}{iii}] line profiles than the bulk of the local AGN population as shown in Fig.~\ref{Fig:O3width_sample}. Hence, their sample is not comparable to our sample given the selection criteria and their respective scopes.

\citet{McElroy2015} studied a sample of 17 type 2 nearby (mean z$\sim$0.08), luminous QSOs (average $L_\mathrm{[\ion{O}{iii}]}\sim 6\times10^{42}\,\mathrm{erg\,s}^{-1}$). They defined the spatial extent of the outflow to be the extent of the region in which two-Gaussian components are needed to fit the spaxels. They reported that the high velocity ($W_{\mathrm{80,max}}\sim 400$--$1600 \mathrm{km s}^{-1}$ with mean $\sim790\pm90 \mathrm{km\,s}^{-1}$) ionized gas is indicative of the AGN-driven outflows. They also reported that these outflows are extended to 3--16\,kpc scale around the nucleus. The downside of their approach is that if the emission profile of [\ion{O}{iii}] closely follows the PSF, then the kinematics will roughly stay constant close (a few kpc) to the nucleus. For HE~0040$-$1105, we find the $W_\mathrm{80}$ stays almost consistently $\sim$ 400\,$\mathrm{km\,s}^{-1}$ in the central 2.5\,kpc (in diameter) region, although the [\ion{O}{iii}] wing emission clearly originates from a region only 100\,pc away from the nucleus. This yields again an order of magnitude overestimation compared to the results obtained from our spectro-astrometric analysis.
\citet{McElroy2015} approximated the outflow size as the extent of the region within which the [\ion{O}{iii}] spectra in all spaxels needed two-Gaussian components to be fitted. Again, if the emission profile closely follows the PSF, observing the spectra at several kpc away from the nucleus does not indicate that the kinematically disturbed gas has already traveled that distance. For example, in HE~0040$-$1105,  the surface brightness maps in Fig.~\ref{Fig:O3astrometry_HE0040} indicates that the emission of both the core and wing components extends $\sim$2.5\,kpc away from nucleus; although the [\ion{O}{iii}] wing emission originates from a region only $\sim$100\,pc away from the nucleus as inferred from the PSF comparison analysis. We find that the outflow sizes obtained for the sources with point-like [\ion{O}{iii}] wing emission is significantly larger than the offsets we obtained using the spectro-astrometric analysis.

\citet{Karouzos2016} analyzed a sample of 6 local (z$<$0.1), luminous (average $L_\mathrm{[\ion{O}{iii}]}>10^{42}\,\mathrm{erg\,s}^{-1}$) type 2 AGN. They reported the wing component to have a large velocity ($<$600\,$\mathrm{km\, s}^{-1}$), reaching velocity dispersion $<$800\,$\mathrm{km \,s}^{-1}$, suggestive of AGN driven outflows. In order to estimate the extents of the outflows, they used the effective radius ($r_\mathrm{eff}$), defined as the radius within which half of the total [\ion{O}{iii}] wing emission flux is contained. They assumed that the outflow sizes are to be the distance at which- (1) the [\ion{O}{iii}] wing component's velocity and velocity dispersion become similar to that of the [\ion{O}{iii}] core component and (2) the extreme kinematics of the [\ion{O}{iii}] wing component is prevalent. Their findings suggest that the outflow sizes are roughly $\sim$2--3 \,$r_\mathrm{eff}$ (1.3--2.1\,kpc). Following the method by \citet{Karouzos2016} we estimate an outflow size of $\sim$2.5\,kpc for HE~0040$-$1105.

\citet{Kang2018} analyzed a sample of 23, $z<0.2$, luminous ($L_\mathrm{[\ion{O}{iii}]}> 3\times10^{40}\,\mathrm{erg\,s}^{-1}$) type 2 AGN. They traced the radial decrease of the [\ion{O}{iii}] velocity dispersion and estimated the size of the warm, ionized gas outflows. \citet{Kang2018} defined the outflow sizes to be the distance at which [\ion{O}{iii}] velocity dispersion becomes same as the host galaxy's stellar velocity dispersion.
In order to take into account beam smearing, they subtracted the seeing sizes from the kinematically derived sizes of the outflowing regions and estimated the size of the outflows to be between 0.60--7.45\,kpc.
Furthermore, they obtained a relation between [\ion{O}{iii}] luminosity and outflow radius as- \begin{multline}
  \log\left(\frac{R_\mathrm{out}}{\text{kpc}}\ \right) = (0.279\pm0.035)\times\log \left(\frac{L_\mathrm{[\ion{O}{iii}]}}{10^{42}\,\text{erg s}^{-1}} \right)\\ +(0.427 \pm0.025)  
\end{multline}

Following the size-luminosity relation from the work of \citet{Kang2018} we estimate an outflow size of $\sim$3\,kpc for HE~0040$-$1105. This overestimation occurs as the emission from the unresolved [\ion{O}{iii}] wing covers $\sim$3\,kpc although the region of emission is only offset by $\sim$100\,pc from the nucleus.
Although they subtracted the seeing from the kinematically derived outflowing sizes, they could not compare the [\ion{O}{iii}] wing emission to the PSF because their samples consist of type 2 AGN. Subtracting the seeing sizes from the total sizes of the region, does not ensure that the beam smearing effect has been completely taken out from the measurements particularly for spatially unresolved emission.

Other work uses analytic biconical models based on the kinematics of double-peaked [\ion{O}{iii}] to directly constrain the geometry and energetics of outflows. \citet{Nevin2018} used follow-up longslit observations to investigate the outflows in sample of 18 AGN from SDSS at $z<0.1$ with $L_\mathrm{bol}\sim10^{43}-10^{45}\ \mathrm{erg\ s}^{-1}$. The used a parameter called the turnover radius, at which the outflowing gas starts to decelerate in the best fit biconical model, as an indicator of the size of the outflows. They found that 2 out of 18 AGN have outflow sizes $<$1 kpc and the other 16 AGN have outflow sizes ranging from 1.1-6.5 kpc, where the outflows retain enough kinetic energy that they could drive a two-staged feedback process in their host galaxies. 

The major drawback of these studies is that the outflow sizes were either calculated from an averaged property of the two-Gaussian models, or from the extent of the emission of the wing component, or from an analytic model based on longslit data. They did not test the [\ion{O}{iii}] emission profile against the PSF as their sample consisted of type 2 AGN. Their results are valid for the AGN in which the [\ion{O}{iii}] emission profile does not follow the PSF. For the AGN, where the [\ion{O}{iii}] emission profile follow the PSF, beam smearing \citep{Husemann2016} could make the emission profiles appear to be extended to several kpc scales, although in reality the region of maximum [\ion{O}{iii}] emission could be just a $\sim$ a few hundred pc away from the nucleus. Therefore, if the emission profile is not tested against that of the PSF, a significant overestimation in outflow sizes could occur. Furthermore, they studied highly luminous quasars as seen from Fig.~\ref{Fig:O3width_sample}. The findings from those studies will not account for the general nearby quasar population as outlier samples of AGN could have extreme properties.

\begin{figure*}
 \resizebox{\hsize}{!}{\includegraphics{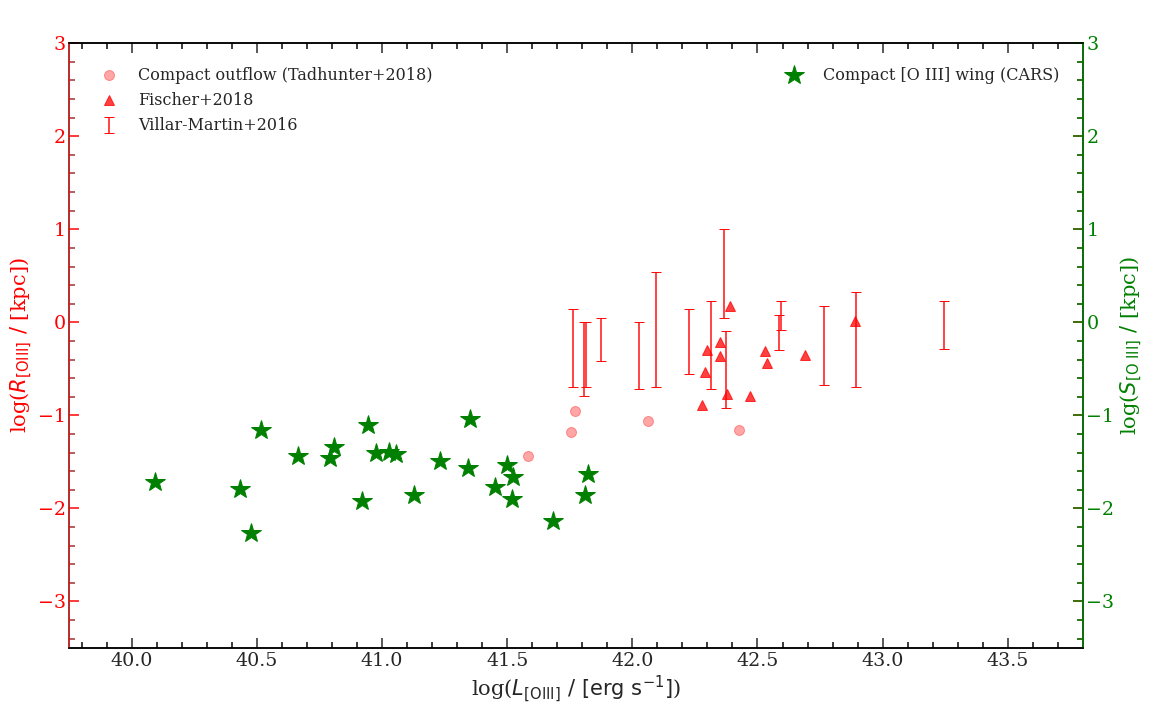}}
   \caption{Spatial extension  of the [\ion{O}{iii}] wing emission vs [\ion{O}{iii}] luminosity for our samples along with outflow sizes obtained by \citet{Villar-Martin2016,Tadhunter2018,fischer2018} at similar or higher [\ion{O}{iii}] luminosity range that find compact outflows. The lower and upper limits of the error bars indicate the lower and upper limits of the outflow sizes obtained by \citet{Villar-Martin2016}.}
     \label{Fig:compact_outflow_sizes}
   \end{figure*}

The situation changes for various studies which either use high-angular resolution HST observations or apply PSF deblending techniques to infer outflow sizes.
In Fig.~\ref{Fig:compact_outflow_sizes}, we plot the offset sizes for the sources in our sample with compact [\ion{O}{iii}] wing components against $L_\mathrm{[\ion{O}{iii}]}$, along with previous studies finding compact outflows, usually $<$ 1 kpc \citep{Tadhunter2018,fischer2018}. These studies used high resolution HST long-slit spectroscopic (STIS) and imaging (ACS and WFCP2) data in order to investigate the AGN driven outflow sizes and gas kinematics. Furthermore, these studies took account of the dominant point like emission from the PSF.
Their outflow sizes are in much better agreement with the sizes reported in this paper and support even the compact outflows found by our approach. Below we briefly discuss some studies individually to discuss their approaches.

\citet{Villar-Martin2016} analyzed 18 luminous type 2 AGNs at $0.3<z<0.6$ using VLT-FORS2 spectroscopic data to investigate the presence of the extended ionized gas outflows. They reported that 15 out of 18 AGN in their sample, show a signature of ionized outflows. They took account of the beam smearing effect and stated that among the 15 AGN with ionized outflows, the [\ion{O}{iii}] wing emission profile follows the emission profile of a standard star very closely, and hence indicate unresolved [\ion{O}{iii}] wing emission. They report a lower limit of outflow sizes $\sim \mathrm{several}\times100$ pc in these AGN. 
In 3 other AGN, where the [\ion{O}{iii}] wing emission did not follow that of the standard star (or the seeing), they report sizes $<1-2 \mathrm{kpc}$. 
They also estimated an upper limit of the outflow sizes to be $\sim$ 2 kpc for the majority of the AGN. The only exceptions were SDSS~J0955$+$03 and SDSS~J0903$+$02 where they reported upper limits of up to 3.5 and 10 kpc respectively.
They did not explore the location of the maximum [\ion{O}{iii}] wing emission, and therefore could not provide an estimation of the offset from the nucleus.

\citet{Tadhunter2018} investigated the AGN-driven outflows in 8 ULIRGs at $z<0.15$ with optical AGN nuclei with narrow-band [\ion{O}{iii}]$\lambda5007$ imaging data from the deep Hubble Space Telescope (HST ACS) observations. They further complimented the HST observations with the ground-based spectroscopic observations from William Herschel Telescope (WHT). \citet{Tadhunter2018} found that the outflowing regions dominated by the warm, ionized gas are usually compact for the majority of the AGN in their sample. They found that the ionized gas outflows are barely resolved at the resolution of HST (65--120\,pc) in 3 objects. HST spatially resolves the outflows in 4 other objects but their flux weighted mean radii ($R_\mathrm{[\ion{O}{iii}]}$) fall in the range $0.65 < R_\mathrm{[\ion{O}{iii}]} < 1.2\mathrm{kpc}$. In one other object (Mrk273), they found conclusive evidence for a large scale outflow, which has a maximum extent of $R_\mathrm{[\ion{O}{iii}]} \sim 5\mathrm{kpc}$. To account for the beam smearing effect, they estimated the FWHM for the star ($FWHM_\mathrm{star}$) and the [\ion{O}{iii}] emission profile ($FWHM_\mathrm{[\ion{O}{iii}]}$) by fitting 2D Gaussians to the HST/ACS images and then subtracted $\mathrm{FWHM}_\mathrm{star}$ from $\mathrm{FWHM}_\mathrm{[\ion{O}{iii}]}$ in quadrature to estimate $R_\mathrm{[\ion{O}{iii}]}$. 
We see that, in almost half of the objects the outflows are unresolved with size of $\sim$100\,pc which is very similar to what we report. 
   
\citet{Mingozzi2019} studied the interstellar medium (ISM) properties of the disc and the outflowing ionized gas in the central regions of 9 nearby ($D_\text{L}<50\text{Mpc}$) Seyfert
galaxies using VLT/MUSE observations. 
They stated that the redshifted and the blueshifted wings of the emission lines are outflows whereas the narrow core represents the disc of the host galaxies. Although they did not take into account the beam smearing effect, which could affect the analysis of 2 objects (NGC~1068 and NGC~1386), the spatial extension of other kpc scale outflows are unlikely to be affected by it.  Based on their [\ion{O}{iii}]$\lambda5007$ flux contours of the red and blue wings, the outflow sizes exhibit a large range from 200\,pc to 3.5\,kpc. However, the velocity dispersion of the wing component is much smaller than usually seen in the AGN spectra so that their results are not fully comparable with our study. It is possible that the turbulence associated with the small velocity dispersion has a different origin than an AGN-driven outflow.
   
\citet{fischer2018} investigated the extent and kinematics of their AGN-ionized gas outflows of 12 nearby ($z < 0.12$), luminous ($L_\mathrm{[\ion{O}{iii}]}>3\times 10^{42}\,\mathrm{erg\,s}^{-1}$) type 2 QSOs with long-slit spectroscopic and imaging data from HST (STIS, ACS and WFCP2). They reported that, the region where the emission-lines exhibit high velocity offsets from the systematic velocities
($> 300 \mathrm{km s}^{-1}$), are likely affected by the AGN driven outflows. They estimated the maximum projected distance of outflowing kinematics/ outflow radius to be 0.15--1.5 kpc (average outflow radius $\sim$ 600 pc). They reported that only 2 of QSOs have outflow radius beyond 1 kpc, while 10 other QSOs have radii $<$ 600 pc. 

\citet{Bellocchi2019} investigated the ionized gas outflows in 6 type 2 QSOs with observations from the optical imager and long-slit spectrograph (OSIRIS) mounted on the 10.4m Gran Telescopio Canarias Spanish telescope (GTC). They identified ionized outflows in 4 out of 6 type 2 QSOs and reported that the ionized outflows are spatially unresolved in 2 QSOs and compact in 1 QSO. The extent of the compact ionized outflow ($R$) is, $R =0.8\pm 0.3\,\mathrm{kpc}$. In only 1 QSO (SDSS~0741$+$3020), \citet{Bellocchi2019} detected a large outflow extending to at least $\sim$ 4 kpc from the central engine.

\citet{Baron2019} investigated 2377 type 2 AGN showing visible [\ion{O}{iii}] wing emission from the parent sample of 24,264 AGN following \citet{Mullaney2013}. They estimated the mass-weighted outflow or wind location according to the SED fitting method \citep{netzer2013} and estimated that the winds are located at 0.1--0.5\,kpc away from the AGN.

Recently, \citet{Zhang2021} analyzed a sample of SDSS spectra of 2621 Type-1 AGN and 1987 Type-2 AGN, with reliable [\ion{O}{iii}] wing component detections. They found different properties of [\ion{O}{iii}] wing components, which could be confirmed between the Type-1 and Type-2 AGN: statistically lower broad [\ion{O}{iii}] luminosities and statistically lower flux ratio of the wing to the core [\ion{O}{iii}] component in the Type-2 AGN, considering necessary effects. Their results indicate stronger obscuration on the [\ion{O}{iii}] wing components in the Type-2 AGN because [\ion{O}{iii}] wing emission regions are nearer to the nucleus, under the framework of the Unified model for AGN. They estimated that the distance between broad [\ion{O}{iii}] emission regions to central BHs [\ion{O}{iii}] wing emitting region is,
$R_\text{B3} \sim  R_\text{sub} \times \tan(\theta/2)$ where $R_\text{sub}$ is the dust sublimation radius and $\theta$ is the opening angle for type 2 AGN. They assumed a global mean value of the opening angle 25$^{\circ}$ and estimated $R_\text{B3} \sim 20 \times R_\text{BLR}$. If we assume $R_\text{BLR} = 0.12$ pc as per \citet{Sturm2018}, the method prescribed by \citet{Zhang2021} will estimate the [\ion{O}{iii}] wing location to be $\sim$ 2 pc away from the nucleus. Additionally, \citet{Zhang2021} stated that the red and blue [\ion{O}{iii}] wings could be robust signs of central outflows. Consequently, \citet{Zhang2021} concluded that their results provide evidence for obscured central outflows in Type-2 AGN. But the drawback of their methods was that they estimated the location of [\ion{O}{iii}] wings using SDSS 1D spectra They did not have the IFU data, that we have in the CARS survey. We can accurately estimate, using spectro-astrometric analysis, how far from the nucleus those maximum [\ion{O}{iii}] wing emitting regions are located, which is impossible with fibre spectroscopy. Therefore, $R_\text{B3}$ might be associated with significant uncertainties.

As per our results, for 23 out of 36 sources we measure a median projected offset of $\sim27$pc between the peak of [\ion{O}{iii}] wing emitting region and the nucleus. Now, the [\ion{O}{iii}] wing emission we have studied in this work could be from highly excited ionized gas clouds. There might be other [\ion{O}{iii}] wing emitting gas clouds, whose emissions could be fainter than the ones we have seen so far. Those fainter [\ion{O}{iii}] wing emitting gas clouds could be extended to kpc scales. The fact that \citet{Zhang2021} found very compact [\ion{O}{iii}] wing emitting regions, itself indicates that the past work showing AGN driven outflows extended to kpc scale, likely was concerned with the fainter [\ion{O}{iii}] wing component in Type 2 AGN, which is not obscured by dust.

\subsubsection{Effect of AGN inclination angle}
As we are only able to measure projected sizes, the classification of compact and extended outflows might be affected by the inclination of the AGN with respect to our line of sight.  We consider the effect of inclination on the outflow-size vs. luminosity relation provided by \citet{Kang2018} for inclination angles of 5$^{\circ}$, 20$^{\circ}$ and 45$^{\circ}$ as shown in Fig.~\ref{Fig:outflow_size}. In these cases, 0\degr\ inclination would indicate that an observer will not be facing the obscuring torus structure and will be straight looking into the pole of the AGN (face-on), whereas 90$^{\circ}$ inclination would indicate that  an observer will be facing the obscuring torus structure and will be straight looking into torus (edge on). As the inclination angle decreases, the projected outflow size also decreases. At an inclination of 5$^{\circ}$, the size-luminosity relation would be matching with our compact [\ion{O}{iii}] wing component offsets. The question is whether a random sample of type 1 AGN would contain so many objects at such low inclinations.

\begin{figure}
 \resizebox{\hsize}{!}{\includegraphics{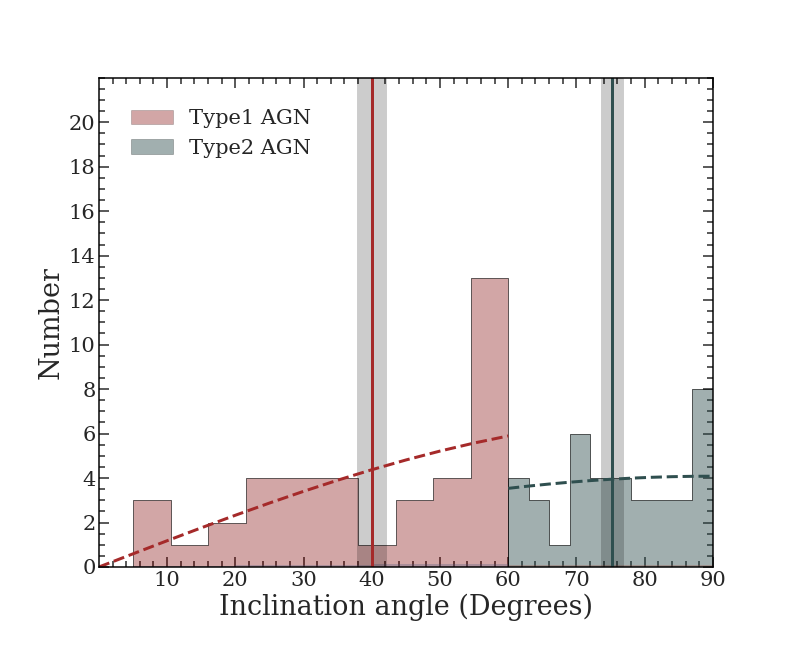}}
   \caption{An example of possible inclination distributions of 39 random samples of AGN with inclination bin-sizes of 5\degr.  We see 39 type 1 AGN randomly distributed between inclination angle between 0--60\degr; and 39 type 2 AGN randomly distributed between inclination angle 60--90\degr. The mean inclination angle for type 1 AGN $\sim(40\pm2)\degr$, and for type 2 AGN it is $\sim(75\pm2)\degr$. We over-plot the probability distribution functions of the inclination angle for type 1 and type 2 AGN as the dashed lines.}
     \label{Fig:inclination}
   \end{figure}

As the CARS sample consists of 39 type 1 AGN, we investigate the inclination distribution of a similar number of AGN to see whether inclination could alter the spatial offsets significantly or not. In Fig.~\ref{Fig:inclination}, we plot the probability distribution function and one possible random sample of 39 type 1 and 39 type 2 AGN assuming an underlying bi-conical geometry. The sample number is equivalent to the number of AGN in the CARS sample. As the bolometric luminosity in our CARS sample is between $6\times10^{41}$--$3\times10^{45}\,\mathrm{erg\,s}^{-1}$ (\citealt{Husemann2017},Husemann et al. in prep), we estimate that the half-opening angle is around 60\degr\ using the work of \citet{Merloni2014}. Hence, anything below 60$^{\circ}$ would be seen as a type 1 AGN and anything beyond will be viewed as type 2. 

We compute the probabilities (see Appendix.~\ref{section:inclination}) for the detailed derivation) of seeing a type 1 and type 2 AGN at any inclination angle ($i$) and find that, the mean inclination angle for a sample is $(40\pm2)\degr$ for a sample of 39 type 1 AGN and $(75\pm2)\degree$ for a sample of 39 type 2 AGN. For our type 1 AGN sample, this mean inclination angle will increase the projected size of the [\ion{O}{iii}] wing component only by a factor of $\sim$1.5. In Fig.~\ref{Fig:inclination}, we notice that the number of AGN with low inclinations rapidly decreases. More specifically, we find that the number of type 1 AGN ($N_\mathrm{type 1}$) with inclination angles between 0--10\degr is at most 3, with 99.5\% confidence. This is in contrast to the 23 compact [\ion{O}{iii}] wing sources in our sample which is much larger than can be predicted by the random type 1 AGN sample of the same sample size.
We already get a mean inclination and an error bar from simulating the distributions of 39 type 1 AGN. The mean inclination is model dependent and could be very different in case the simple ionization cone model is not valid. Therefore, having a larger sample will not alter the mean inclination angle drastically as it is model dependent.

Therefore, we conclude that inclination alone cannot explain the large number of targets with compact [\ion{O}{iii}] wing locations $<$ 100 pc away from the nucleus in our sample. This notion is further supported by other studies that report compact or spatially-unresolved broad [\ion{O}{iii}] on sub-kpc scales even in type 2 AGN \citep[e.g][]{Villar-Martin2016,Tadhunter2018}, which would be even harder to explain by inclination effects.

\subsubsection{Relation to the electron density}
Several previous studies investigating AGN-driven outflows assumed uniform electron densities in the outflowing medium \citep[e.g][]{veilleux2005,canodiaz2012,genzel2014,McElroy2015,Husemann2016,Kakkad2016,leung2017}; but the downside is that the electron density could vary spatially and hence assuming a constant electron density would introduce further systematic uncertainty \citep{Kakkad2018}. \citet{bennert2006} estimated the electron density as a function of radius from the nucleus in a sample of low redshift type 1 and type 2 AGN, and reported that the electron density decreases from $\sim$ $10^{3}\,\mathrm{cm}^{-3}$ in the central region to  $\sim$ $10^{2}\,\mathrm{cm}^{-3}$ at $\sim$ 1.2 kpc away from the nucleus. A similar decrease in electron density from the nucleus starting at $>10^{3}\,\mathrm{cm}^{-3}$ towards the outskirts was also found in AGN with outflows by \citet{sharp2010}, \citet{cresci2015}, and \citet{freitas2018}. Recently, \citet{Kakkad2018}, \citet{Baron2020}, and \citet{Davies2020} measured the electron density for the broad wing component of [\ion{S}{ii}]$\lambda\lambda6716,6731$ and reported radially decreasing electron densities with similar amplitude as previous works. 
Furthermore, other studies involving AGN have clearly demonstrated that the outflowing gas is denser than the ambient, non outflowing gas in luminous AGN \citep[e.g.,][]{Holt2011,Villar-Martin2014,Villar-Martin2016}.
Theoretical/numerical studies (e.g., \citealt{Gaspari2017}) find analogous radial decrease in the electron gas density in simulations of Chaotic Cold Accretion (CCA) feeding and feedback. Thus, this appears to be a natural outcome of galaxy evolution.
Overall, we would expect the electron densities of the compact and extended [\ion{O}{iii}] wing emitting sources to be similar if the broad [\ion{O}{iii}] wing emission intrinsically originates from kpc scales and if our classification of compact and extended is just an observational bias due to projection effects or sensitivity limitations.

\begin{figure}
   \resizebox{\hsize}{!}{\includegraphics{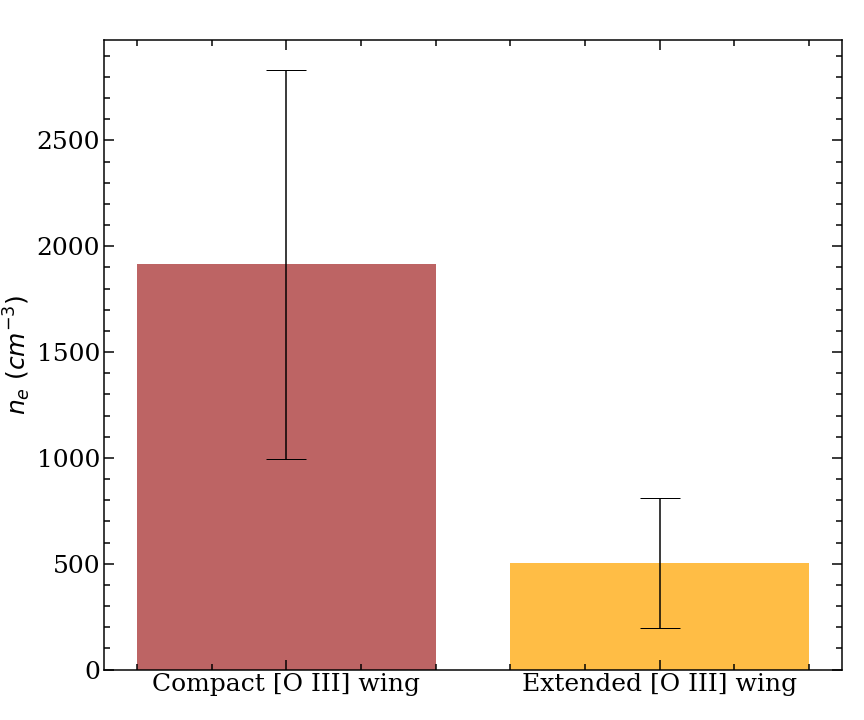}}
   \caption{The bar-graphs of electron-densities for the compact and extended outflows. On the left, the bar represents average electron density for the sources with compact outflows; and on the right, the bar represents average electron density for the sources with extended outflows.}
         \label{Fig:n_e_bar}
   \end{figure}

In Fig.~\ref{Fig:n_e_bar}, we compare the mean electron densities for 7 of our sources with 4 point-like (compact) and 3 extended spatial [\ion{O}{iii}] wing component distributions. We computed the electron densities within the central 0.6\arcsec ($<$500\,pc) for those 7 sources which have sufficient S/N to separate the wing and core component even in the faint [\ion{S}{ii}] line. We find that the median electron density for the sources with a compact [\ion{O}{iii}] wing regions is
$n_\mathrm{e}=1915 \pm 920$\,$\mathrm{cm}^{-3}$. For sources with extended regions [\ion{O}{iii}] wing, the median electron density is $n_\mathrm{e}=503 \pm 307$\,$\mathrm{cm}^{-3}$, which is a factor of 4 lower. A two sample
Kolmogorov-Smirnov test returns a p-value = 0.007, which confirms that the two samples of electron densities are significantly different.
Such a bimodality in electron density would be unexpected if the emission of the wing component originated from similar spatial scales within the galaxy. This dichotomy in electron densities therefore indicates that the  [\ion{O}{iii}] wing emission for the extended sources is physically much more extended than that of the compact sources. In agreement with the observations and simulations quoted above, the significantly higher electron densities support our measurements that the compact wing regions originate much closer to the nucleus than the extended ones. This observation further strengthens our conclusion that projection effects due to different inclination angles cannot explain the measured spatial distribution of [\ion{O}{iii}] and that other parameters need to play an important role. 

\subsection{Is AGN luminosity responsible for [\ion{O}{iii}] wing extent?}
Assuming that the [\ion{O}{iii}] wing is clearly related to an AGN-driven wind, previous studies found a clear trend of increasing outflow size with AGN luminosity.
\citet{Liu2013,Kang2018} reported a correlation between the [\ion{O}{iii}] luminosity and the outflow sizes. As per their finding, the ionized outflow size would increase if the AGN had higher [\ion{O}{iii}] luminosity. \citet{Zakamska2014,Wylezalek2020} reported that the AGN luminosity is primarily responsible for the launching and detection of the outflows.

\begin{figure}
   \resizebox{\hsize}{!}{\includegraphics{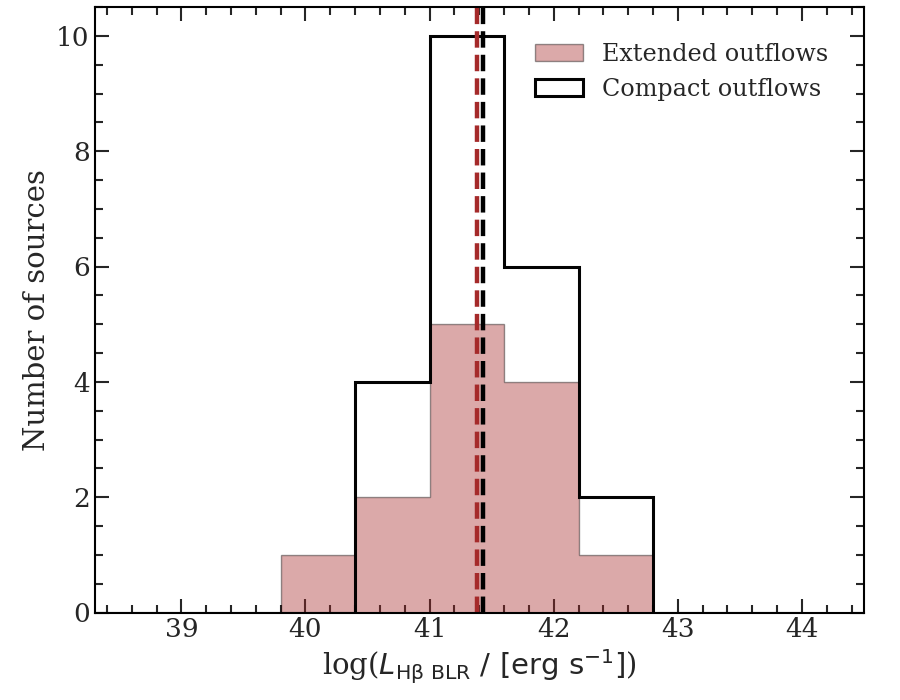}}
   \caption{The distribution of the BLR H$\beta$ luminosity for QSOs with compact and extended ionized outflows. The black vertical dashed lines represent the mean BLR H$\beta$ luminosities for the sources with compact [\ion{O}{iii}] wing emission and the brown vertical dashed lines represent the mean BLR H$\beta$ luminosities for the sources with extended [\ion{O}{iii}] wing emission.}
         \label{Fig:AGNlum_hist}
   \end{figure}

Based on a comparison of the compact versus extended outflows identified for the CARS sources, we find that AGN luminosity is not the main driver of the dichotomity. In Fig.~\ref{Fig:AGNlum_hist}, we show the histograms of the BLR H$\mathrm{\beta}$ ($L_\mathrm{BLR\ H\beta}$) luminosity for the sources with compact and extended outflows. For the sources with compact [\ion{O}{iii}] wing emission, $L_\mathrm{BLR\ H\beta}$ tends to cover the range of $2\times 10^{40}$--$6 \times 10^{42}\,\mathrm{erg\,s}^{-1}$ whereas for the sources with extended outflows the range is wider ($6\times10^{39}$--$6\times10^{42}\,\mathrm{erg\,s}^{-1}$). The mean luminosity for the sources with both compact and extended [\ion{O}{iii}] wing emission is $2.7 \times 10^{41}\,\mathrm{erg s}^{-1}$. This is in agreement with the data from \citet{Mullaney2013} where the highest number of type 1 AGN is located at $L_\mathrm{BLR\ H\beta} = 3 \times 10^{41}\,\mathrm{erg\,s}^{-1}$.

The sample size in Fig.~\ref{Fig:AGNlum_hist} imposes limits on the minimum difference between the samples that can be diagnosed as statistically significant. For the given sample size, we use a two-sample Anderson-Darling (AD) test \citep{Anderson1952}, which results in a p-value of 0.99 indicating that the compact and extended outflow samples in Fig.~\ref{Fig:AGNlum_hist} are consistent with being drawn from the same parent distribution.

However, an offset in AGN bolometric luminosity between the two samples may be hidden due to the systematic uncertainties of the bolometric correction from $L_\mathrm{BLR\ H\beta}$ to $L_\mathrm{bol}$. Here, we assume a systematic uncertainty of 0.3\,dex. This is motivated by the fact that $L_\mathrm{BLR\ H\beta}$ has been reported to be tightly correlated to $L_{5100\AA}$ \citep{Greene2005}, while \citet{Richards2006} estimated the mean and standard deviation in their B-band bolometric correction factor to be 10.4 and 2.5, resulting in $L_\mathrm{bol}\approx(8-12)\times \lambda L_{L5100\AA}$.  We simulate the samples from Gaussian distributions with a standard deviation of 0.3\,dex with a range of offsets between the mean values of the two distributions. The AD test shows that at a separation of $>$0.4\,dex in luminosity we can be sure that the samples are different at a 5\% confidence level. Hence, there could be a small difference in luminosity that we cannot diagnose. On the other hand, if we use the outflow size vs [\ion{O}{iii}] luminosity relation from the work of \citet{Kang2018}, we find that the bolometric luminosity \citep[combining with the work from][]{Heckman2014} difference between AGN with outflow sizes = 100 pc and 1 kpc will be 2.34 dex. This difference in bolometric luminosity is much larger than what we report here.

\begin{figure*}
   \includegraphics[width=\textwidth]{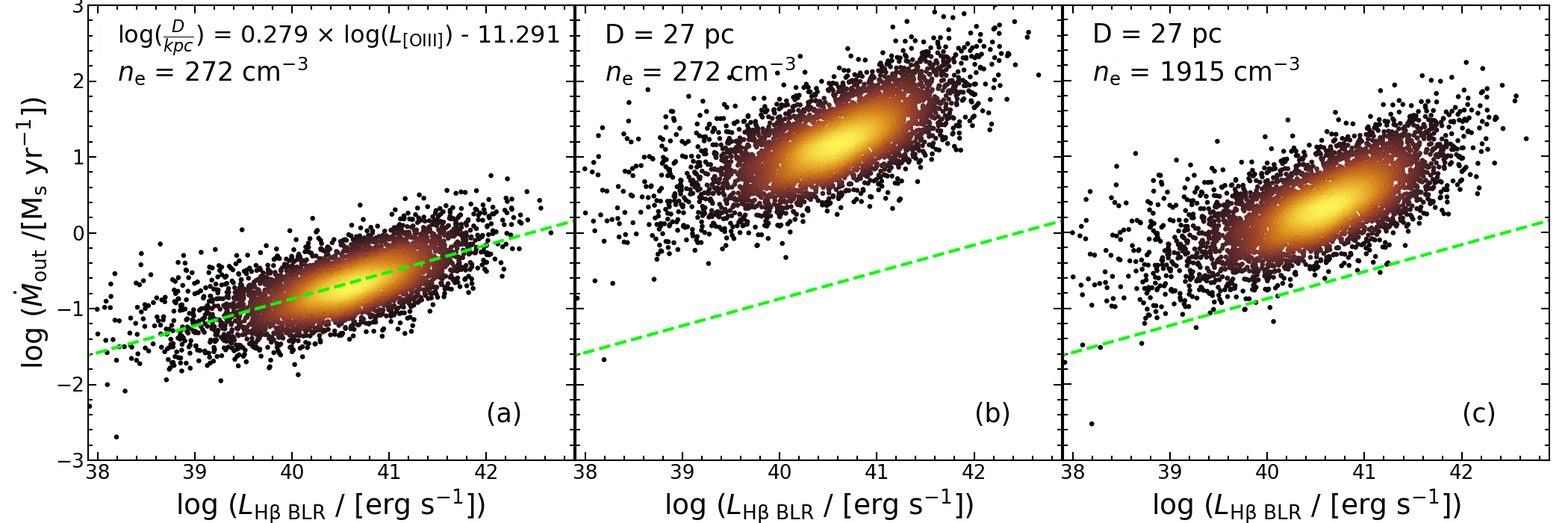}
   \caption{BLR H$\beta$ luminosity against the mass outflow rates for the type 1 AGN from \citet{Mullaney2013}.
   \textit{Left panel:} We use the outflow size vs. [\ion{O}{iii}] luminosity relation from the work of \citet{Kang2018} to calculate the outflow radius (D). We also use the median electron density $n_\mathrm{e} = 272\ \mathrm{cm}^{-3}$ as reported by \citet{Rakshit2018}.The green dashed line indicates the linear regression line of the mass outflow rates and the BLR H$\beta$ luminosity. 
   \textit{Mid panel:} We assume that all of the outflows in this sample of type 1 AGN of \citet{Mullaney2013} to have point like [\ion{O}{iii}] wing emission; in which case the median offset between the AGN nucleus and the [\ion{O}{iii}] wing Moffat center is $\sim$ 27 pc. Therefore, we use 27 pc as the outflow radius in this case. We keep the electron density fixed to $n_\mathrm{e} = 272\ \mathrm{cm}^{-3}$ as reported by \citet{Rakshit2018}.
   \textit{Right panel:} We use $n_\mathrm{e} = 1915\ \mathrm{cm}^{-3}$ which is the median electron density for the AGN with compact [\ion{O}{iii}] wing emission. We use an outflow radius of 27 pc which is about the median offset between the Moffat center of the [\ion{O}{iii}] wing and the nucleus. We over-plot the linear regression line of the mass outflow rates and the BLR H$\beta$ luminosity in all panels to showcase the difference in mass outflow rates in the approach used in all the panels.}
   \label{Fig:energetics_Lbol}
\end{figure*}

\subsection{How do the mass outflow rate results depend on measurements methods?}

In the last decade, several optical IFU studies investigated the warm ionized phase of AGN driven outflows, by tracing the rest-frame optical emission line [\ion{O}{iii}] \citep[e.g.,][]{Liu2013a,Mullaney2013,Harrison2014,Karouzos2016,Karouzos2016a,Woo2016}. But the outflow sizes \citep[e.g.,][]{Husemann2016,Villar-Martin2016,Harrison2018,Tadhunter2018} and the electron densities \citep[e.g.,][]{sharp2010,cresci2015,freitas2018,Kakkad2018,Baron2020} are still uncertain. This imposes a challenge when deriving the outflow energetics, which strongly depends upon of the outflow kinematics, spatial extent and electron density.

A major advantage of CARS is that, the sample consists of nearby ($0.01<z<0.06$) and luminous ($L_\mathrm{[\ion{O}{iii}]} \sim 9\times10^{39}-3\times10^{42}\,\mathrm{erg\,s}^{-1}$) AGN, which well represents the general type 1 AGN population at low redshifts. Because of their proximity, the CARS host galaxies can be studied at sub-kpc scales, while still probing an important part of the local AGN luminosity function (\citealt{Husemann2017,Husemann2019}, Husemann et al. 2020, in prep).

IFU data allows us to directly constrain critical parameters for outflow energetics, such as outflow radius, electron density, outflow velocity and extinction. The issue is that the IFU studies only observe a small sample, whereas fiber spectroscopy covers a large sample. On the other hand, fiber spectroscopy does not probe the outflows radius and the electron density at the outflow radius. Therefore, it is crucial to put our results in perspective with the previous spectroscopic studies of outflow energetics to investigate how much the values of the energetics parameters could change. Assuming a case B recombination \citep{Nesvadba2006,Nesvadba2011}, with an electron temperature, $T_\mathrm{e} = 10^4$ K \citep{Osterbrock2006}, and adopting an intrinsic line ratio of [\ion{O}{iii}]$\lambda$5007\ /\ H$\beta$ = 3.8,

\begin{equation}
    \frac{M_\mathrm{gas}}{4 \times 10^{7}\ M_\mathrm{s}} = \left(\frac{L_\mathrm{\text{[\ion{O}{iii}],\ wing}}}{10^{43}\ \mathrm{erg s}^{-1}}\right) \left(\frac{n_\mathrm{e,\ wing}}{100\ \mathrm{cm}^{-3}}\right)^{-1}
\end{equation}

where $M_\mathrm{s}$ is the solar mass, $L_\mathrm{\text{[\ion{O}{iii}],\ wing}}$ is [\ion{O}{iii}] wing luminosity and $n_\mathrm{e,\ wing}$ is the electron density of the gas emitting [\ion{O}{iii}] wing.

\citet{Rakshit2018} used the SDSS sample to generate a large sample of type 1 AGN. However, estimating the outflow energetics requires the outflow-size within which the energetics is to be calculated. Additionally, outflow energetics values are highly dependent upon the assumption of outflow geometry. 
\citet{Rakshit2018} assumed a bi-conical outflow geometry and 'Case B' recombination \citep{Crenshaw2010,Bae2016}. As their work was confined to a SDSS 1D spectroscopic sample of $\sim$ 5000 type 1 AGNs at $z < 0.3$, they estimated the outflow radius ($D$) using the outflow-size luminosity relation from \citet{Kang2018}. They approximated the outflow velocity as
\begin{equation}
    v_\mathrm{\text{out}} = \left({\sigma_\mathrm{\text{[\ion{O}{iii}]}}}^2\ +\ {V_\mathrm{\text{[\ion{O}{iii}]}}}^2\right)^{1/2}
\end{equation}
where $V_\mathrm{\text{[\ion{O}{iii}]}}$ is the velocity shift, defined as the velocity offset of the first moment of the single/double Gaussian [\ion{O}{iii}]$\lambda 5007$ line profile from the systematic velocity. $\sigma_\mathrm{\text{[\ion{O}{iii}]}}$ is the second moment of the single/double Gaussian [\ion{O}{iii}]$\lambda 5007$ line profile. \citet{Rakshit2018} reported that 1526 objects in their sample overlap with the sample of \citet{Rakshit2017} with a median $n_\mathrm{\text{e}} = 272 \mathrm{cm}^{-3}$. Following the work of \citet{Maiolino2012,Bae2017}, \citet{Rakshit2018} estimated the mass outflow rates to be: 

\begin{equation}
    \dot{M}_\mathrm{\text{out}} = 3\ M_\mathrm{\text{out}}\ \frac{v_\mathrm{\text{out}}}{D}
\end{equation}

We investigate how the energetics will change if the outflows are predominantly compact and unrelated to AGN luminosity.

Although, the measurement of the [\ion{O}{iii}] line shape are very accurate, their interpretation to determine the correct outflow velocity is difficult as different studies use different parameters as outflow velocities. Therefore, we aim to focus on the dependence of mass outflow rates on outflow radius ($D$) and electron densities $n_\mathrm{e}$. 
The energy and momentum outflow rates could be obtained by multiplying the outflow velocity term to the mass outflow rates. Therefore, we entirely focus on the mass outflow rates for an initial comparison of assumptions.


In Fig.~\ref{Fig:energetics_Lbol}, we plot the mass outflow rates against the BLR H$\beta$ luminosity ($L_\mathrm{\text{H$\beta$\ BLR}}$) for the type 1 AGN of the samples of \citet{Mullaney2013}. We only choose AGN with $L_\mathrm{\text{H$\beta$ BLR}} > 4\times10^{37} \text{erg s}^{-1}$. In the left panel, we calculate the outflow radius, $D$ from their respective [\ion{O}{iii}] luminosities \citep{Kang2018} and use $n_\mathrm{\text{e}} = 272\ \mathrm{cm}^{-3}$ \citep{Rakshit2017,Rakshit2018}. We also use the scikitlearn package \citep{Pedregosa2011} to perform a linear regression between $\dot{M}_\mathrm{\text{out}}$ and $L_\mathrm{\text{H$\beta$\ BLR}}$. We obtain --
\begin{equation}
    \text{log}(\dot{M}_\mathrm{\text{out}}) = 0.35\ \text{log}(L_\mathrm{\text{H$\beta$\ BLR}}) - 15.07
\end{equation}
In the middle panel, we assume that all the AGN in this particular sub-sample have spatially-unresolved [\ion{O}{iii}] wing emission. In our CARS sample, the median offset of the Moffat center of the [\ion{O}{iii}] wing from the nucleus is $\sim$ 27 pc. Therefore, we assume the outflow radius, $D$ = 27 pc. We keep the electron density fixed at 272 $\mathrm{cm}^{-3}$ to highlight the sole impact of a small outflow size. We report 30 times increase  in mass outflow rate on average compared to the left panel.

In the right panel, we assume the outflow radius, $D$ = 27 pc similar to the middle panel. This time we also change the electron density to see the cumulative effect of changing the $D$ and $n_e$, compared to the left panel. The AGN with compact [\ion{O}{iii}] wing emission in our CARS sample have an median electron density $n_\mathrm{e}=1915 \pm 920\ \mathrm{cm}^{-3}$. Therefore, we use $n_\mathrm{\text{e}} \sim 1900\ \mathrm{cm}^{-3}$ to see how the mass outflow rate changes. We report an average $\sim$4 times increase in mass-outflow rate; which is about 9 times less than the middle panel. Hence, we see a net increase in mass outflow rate compared to the method by \citet{Rakshit2018} when assuming compact outflows, which should be frequent in this luminosity range according to our studies. While we cannot reconstruct the actual mass outflow rate distribution for the SDSS sample, we predict that many local AGN have more energetic outflows than previously measured if they are also more centrally concentrated.

AGN driven outflows have been observed in various gas phases- from highly ionized X-ray and UV absorption lines \citep[e.g.,][]{Blustin2003,Reeves2003}, ionized optical emission lines \citep[e.g.,][]{Mullaney2013,Harrison2014,Husemann2016,Perna2017}, to atomic and molecular emission and absorption lines \citep[e.g.,][]{Feruglio2010,Rupke2013TheMergers,Veilleux2013}. Some theoretical studies have also confirmed the presence of the multi-phase nature of these outflows \citep{Zubovas2012ClearingGalaxy,Gaspari:2017}. The nature of these multi-phase outflows still remains largely unconstrained; as different gas phases have been found to show different outflow velocities, mass and covering factors \citep[e.g.,][]{Fiore2017,Cicone2018,Veilleux2020}.
Additionally, the warm ionized gas phase traced by the optical emission lines only represents a small fraction of the mass of the multi-phase outflows. \citet{Rupke2013TheMergers} observed a higher mass in the neutral gas outflows than in the ionized gas outflows from ULIRGs. Therefore, other gas phases of these multi-phase outflows need to be investigated in order to get a detailed view of their energetics.

\section{Summary and conclusions}
We used optical IFU observations from CARS, covering the [\ion{O}{iii}] $\lambda\lambda$4959,5007 and H$\beta$ emission lines to investigate the spatial location of the [\ion{O}{iii}] blue wing component. Our CARS targets are selected from the Hamburg-ESO survey, a sample of luminous QSOs with 12.5$<$B$<$17.5 \citep{wisotzki2000} and representative of the bulk of the local AGN population ($10^{40}\,\mathrm{erg\,s}^{-1}< L_\mathrm{[\ion{O}{iii}]} < 3\times10^{42}\,\mathrm{erg\,s}^{-1})$ at z$<$0.06.
All CARS sources show asymmetries in [\ion{O}{iii}]]$\lambda\lambda4959,5007$ as a potential sign of ionized outflows.
In this paper we use a spectro-astrometric analysis to characterize the [\ion{O}{iii}] wing emission. Our main results are:\smallskip

\begin{itemize}
    \item The [\ion{O}{iii}] wing emission profiles in 23 out of 36 AGN are consistent with the PSF profiles with peak positions offset from the nucleus by $<$100\,pc. The other 13 sources display a clearly extended [\ion{O}{iii}] wing spatial emission profile. Therefore, the spatial morphology of the [\ion{O}{iii}] wing emission is diverse and can be compact or extended in an unbiased luminous AGN sample, where both are likely to occur.\smallskip
    
    \item The electron density in the compact [\ion{O}{iii}] wing regions ($n_\mathrm{e}\sim1900\,\mathrm{cm}^{-3}$) is nearly a magnitude higher than the one in the extended [\ion{O}{iii}] wing regions ($n_\mathrm{e}\sim500\,\mathrm{cm}^{-3}$).\smallskip
    
    \item The sources with compact and extended [\ion{O}{iii}] wings follow the same distribution in terms of AGN (BLR H$\beta$) luminosity.
    The inclination correction that would be required to interpret the compact outflows as extended kpc size outflows is inconsistent with the distribution of inclinations expected for our sample.
    
    \item Inclination angle and AGN bolometric luminosity alone cannot explain the large number of compact [\ion{O}{iii}] wing emission in CARS sources.
    
    \item The mass outflow rates for type 1 AGN calculated over kpc scales for
    $L_\mathrm{BLR\ H\beta} = 
    10^{38}\,\mathrm{erg\,s}^{-1}$ to $3\times10^{42}\,\mathrm{erg\,s}^{-1}$ 
    are 
    $\dot{M}_\mathrm{\text{out}} = 0.01 - 1.6 M_\odot$/yr.
    
    \item For a given electron density, the mass outflow rates increase up to 30 times if estimated near the nucleus ($D$ = 27\,pc) compared to the mass outflow rates calculated over kpc scales.
    
    \item The mass outflow rates increase only marginally if estimated near the nucleus ($D$ = 27\,pc) assuming with high electron density ($n_\mathrm{e}\sim1900\,\mathrm{cm}^{-3}$), compared to the mass outflow rates calculated over kpc scales, assuming low electron density ($n_\mathrm{e}\sim272\,\mathrm{cm}^{-3}$).

\end{itemize}

In this paper, we were able to recover compact sub-kpc scale out to extended kpc scale ionized outflows among a homogeneous sample of 36 luminous AGN observed with seeing-limited MUSE and VIMOS IFU observations. Strikingly, AGN luminosity is not the key parameter for the size of circum-nuclear outflow regions as previously proposed from various seeing-limited IFU studies \citep{Liu2013a,Kang2018,Wylezalek2020}.
Therefore, future studies should investigate not only correlations with AGN luminosity, but also with other properties/mechanisms, such as radio jets, that could potentially drive these AGN outflows. Indeed, mechanical (and not radiative) AGN feedback has been shown to be the leading process in shaping the evolution of galaxies over the long term (e.g., \citealt{Santoro2018,Husemann2019,Gaspari:2020} for reviews).
Generally, other factors such as the inclination and duty cycle could play a potential role in setting the apparent size and statistics of these outflows.

Therefore, a lot of things need to be explored in the future with CARS. Clearly, we will focus on measuring the sizes for the sources with extended [\ion{O}{iii}] wing emission, and to properly obtain their outflow energetics. Additionally, we will accurately estimate the electron density associated with these outflows by using the additional spectra from newly acquired, deep VLT/X-SHOOTER observations where MUSE data is currently not deep enough. Moreover, we will investigate the impact of radio jets on outflows by comparing the radio jet sizes and properties from dedicated VLA observations; and systematically check the presence of any obvious relation between the compact outflows and the compact radio jets in radio-quiet QSOs. Lastly, we will also explore the multi-phase nature of such outflows by combining cold, neutral and ionized gas outflows observations \citep[e.g.,][]{Husemann2019} for a large fraction of CARS targets. 

\section*{Acknowledgements}
We thank the referee, Dr. Montserrat Villar Martin, for her valuable and thoughtful comments which significantly improved the quality of the manuscript.
The work of MS was supported in part by the University of Manitoba Faculty of Science Graduate Fellowship (Cangene Award), and by the University of Manitoba Graduate Enhancement of Tri-Council Stipends (GETS) program. 
BH is greatful to  financial support by the DFG grant GE625/17-1 and DLR grant 50OR1911. BH and ISP are additionally supported by the DAAD travel grant 57509925.
TU acknowledges funding by the Competitive Fund of the Leibniz Association through grants SAW-2013-AIP-4, SAW-2015-AIP-2 and SAW-2016-IPHT-2.
JS is supported by the international Gemini Observatory, a program of NSF’s NOIRLab, which is managed by the Association of Universities for Research in Astronomy (AURA) under a cooperative agreement with the National Science Foundation, on behalf of the Gemini partnership of Argentina, Brazil, Canada, Chile, the Republic of Korea, and the United States of America.
MS, CO and SB acknowledge partial support from the Natural Sciences and Engineering Research Council (NSERC) of Canada.
MG acknowledges partial support by NASA Chandra GO8-19104X/GO9-20114X and HST GO-15890.020-A. 
TR is supported by the Science and Technology Facilities Council (STFC) through grant ST/R504725/1. TAD acknowledges support from the UK Science and Technology Facilities Council through grant ST/S00033X/1. 
BAT was supported by the Harvard Future Faculty Leaders Postdoctoral Fellowship. Based on observations collected at the European Southern Observatory under ESO programme(s) 083.B-0801, 094.B-0345(A) and 095.B-0015(A).
GRT acknowledges support from NASA through grant numbers HST-GO-15411 and HST-GO-15440 from the Space Telescope Science Institute, which is operated by AURA, Inc., under NASA contract NAS 5-26555.
MPT acknowledges financial support from the State Agency for Research of the Spanish MCIU through the "Center of Excellence Severo Ochoa" award to the Instituto de Astrofísica de Andalucía (SEV-2017-0709) and through grant PGC2018-098915-B-C21 (MCI/AEI/FEDER, UE). The Science, Technology and Facilities Council is acknowledged by JN for support through the Consolidated Grant Cosmology and Astrophysics at Portsmouth, ST/S000550/1.
Based on observations collected at the Centro Astronómico Hispano-Alemán (CAHA) at Calar Alto, operated jointly by Junta de Andalucía and Consejo Superior de Investigaciones Científicas (IAA-CSIC). This research made use of Astropy, a community-developed core Python package for Astronomy \citep{Astropy2013, Astropy2018}, SciPy \citep{Scipy2020}, NumPy \citep{Numpy2011}; and the plotting packages Matplotlib \citep{Hunter:2007}, CMasher \citep{vandervelden2020}.
\bibliographystyle{aa}

\bibliography{extrabiblio.bib}


\begin{appendix} 

\section{Moffat model formulation}
\label{section:moffat model}

We assumed a 2D Moffat function to model the PSF following the work by \citet{Gadotti2008}
\begin{equation}
      f(r)=a\Big[1+\Big(\frac{r}{\alpha}\Big)^2\Big]^{-\beta}
\end{equation}
where $a$ is the amplitude of the PSF, $\alpha$ denotes the radius of the PSF, and $\beta$ controls the radial shape of the intensity profile. Because the PSF might not be exactly circular, we used an elliptical Moffat model  which is mathematically described as
\begin{equation}
      f(x,y)=a\Big[1+A(x-x_0)^2+B(y-y_0)^2+C(x-x_0)(y-y_0)\Big]^{-\beta}
\end{equation}

with A, B and C as follows
\begin{equation}
    A = \left(\frac{\cos\phi}{\alpha_1}\right)^2+\left(\frac{\sin\phi}{\alpha_2}\right)^2
\end{equation}
\begin{equation}
      B = \left(\frac{\sin\phi}{\alpha_1}\right)^2+\left(\frac{\cos\phi}{\alpha_2}\right)^2
\end{equation}
\begin{equation}
    C = 2 \sin\phi \cos\phi\left[\left(\frac{1}{\alpha_1}\right)^2-\left(\frac{1}{\alpha_2}\right)^2\right]
\end{equation}

Where $\alpha_1$ and $\alpha_2$ are the minor and major axes and the $\phi$ is the position angle. The eccentricity ($e$) of the ellipse could be defined as,

\begin{equation}
    e = \sqrt{1-\frac{\alpha_1^2}{\alpha_2^2}}
\end{equation}

Like an ellipse, there are two axes eventuating two FWHMs which depend on $\phi$. In this case, the FWHM changes due to its ellipticity.

\begin{equation}
    \mathrm{FWHM}_1 = 2\alpha_1\sqrt{2^{1/\beta}-1}
\end{equation}
\begin{equation}
    \mathrm{FWHM}_2 = 2\alpha_2\sqrt{2^{1/\beta}-1}
\end{equation}

\section{The inclination angle - AGN probability relation}
\label{section:inclination}

We assume a differential area of a hemisphere which is 
\begin{equation}
    dS = r^2\ sin(\theta) d\theta\ d\phi
\end{equation}

Now, the azimuthal angle $\phi$ goes from 0 to 2$\pi$ and $\theta$ goes from $\theta_\mathrm{\text{max}}$ to $\theta_\mathrm{\text{min}}$.

Therefore, integrating the area of subtended by the part of hemisphere is,

\begin{equation}
    S = r^2 \int_\mathrm{\theta_\mathrm{\text{min}}}^{\theta_\mathrm{\text{max}}}\text{sin}(\theta)\, d\theta \int_\mathrm{0}^{2\pi}\,d\phi
      = 2\pi r^2[\text{cos}(\theta_\mathrm{\text{min}}) - \text{cos}(\theta_\mathrm{\text{max}})]
\end{equation}

As the area of a hemisphere is $2\pi r^2$, therefore the probability of an AGN to be within angle $\theta_\mathrm{\text{min}}$ to $\theta_\mathrm{\text{max}}$ is,

\begin{equation}
    P = \frac{2\pi r^2[\text{cos}(\theta_\mathrm{\text{min}}) - \text{cos}(\theta_\mathrm{\text{max}})]}{2\pi r^2}
      = \text{cos}(\theta_\mathrm{\text{min}}) - \text{cos}(\theta_\mathrm{\text{max}})
\end{equation}

For type 1 AGN, the inclination angle spans the range 0 to 60 degrees with a inclination bin of 5 degrees. Therefore, the total probability of an AGN to be within these inclination angles should be = 1.

If the normalizing constant is A

\begin{equation}
    A\ \times\ \text{cos}(0\degree) - \text{cos}(60\degree) = 1\\
\end{equation} 

Which yields, A = 2

Therefore, for type 1 AGN , therefore the probability of an AGN to be within angle $\theta_\mathrm{\text{min}}$ to $\theta_\mathrm{\text{max}}$ is,

\begin{equation}
    P = 2\ [\text{cos}(\theta_\mathrm{\text{min}}) - \text{cos}(\theta_\mathrm{\text{max}})]
\end{equation}

Similarly, for type 2 AGN , therefore the probability of an AGN to be within angle $\theta_\mathrm{\text{min}}$ to $\theta_\mathrm{\text{max}}$ is,

\begin{equation}
    P = 2\ [\text{cos}(\theta_\mathrm{\text{min}}) - \text{cos}(\theta_\mathrm{\text{max}})]
\end{equation}

\section{Spectral Modelling}
\label{section:spectra}

In this section, we list all the spectral modeling of the AGN in our analysis. We first present the H$\beta$ + [\ion{O}{iii}]+ \ion{Fe}{ii} modeling for the integrated spectra from all the spaxels in the central 3\arcsec region (comparable to SDSS). Next we present the [\ion{S}{ii}] modeling for 9 AGN where detected the visible broad wing (6 from MUSE and 2 from VIMOS).

\begin{figure*}
    \centering
    \includegraphics[width=0.49\textwidth]{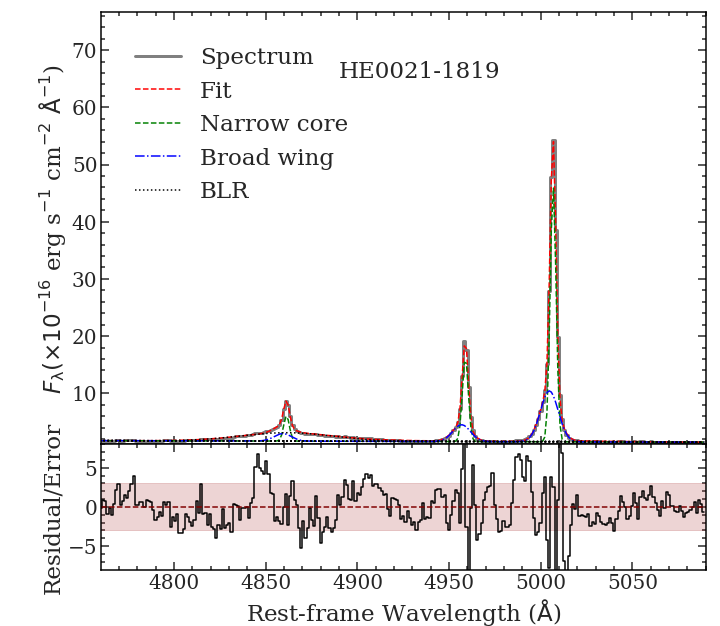}
    \includegraphics[width=0.49\textwidth]{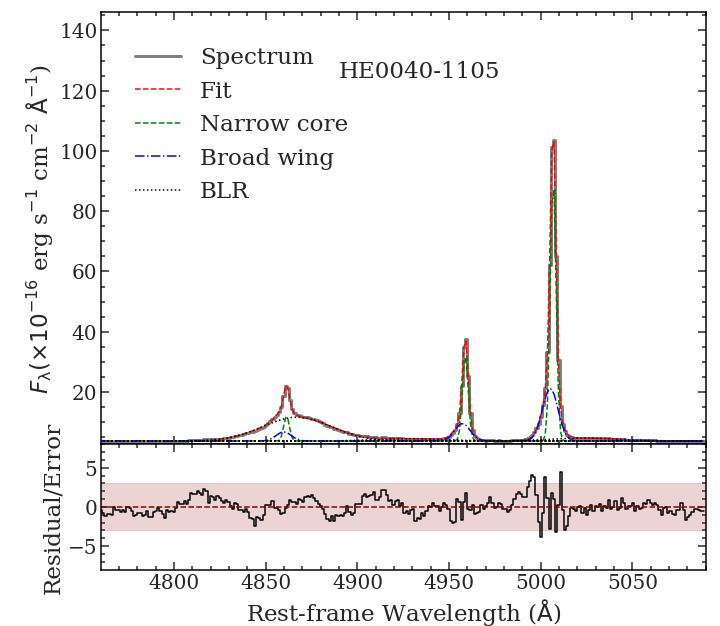}
    \\
\end{figure*}

\begin{figure*}
    \centering
    \includegraphics[width=0.49\textwidth]{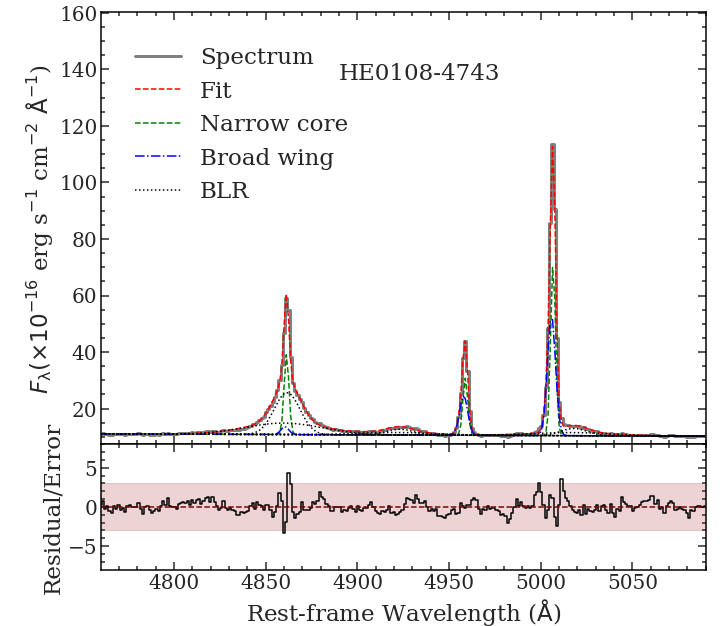}
    \includegraphics[width=0.49\textwidth]{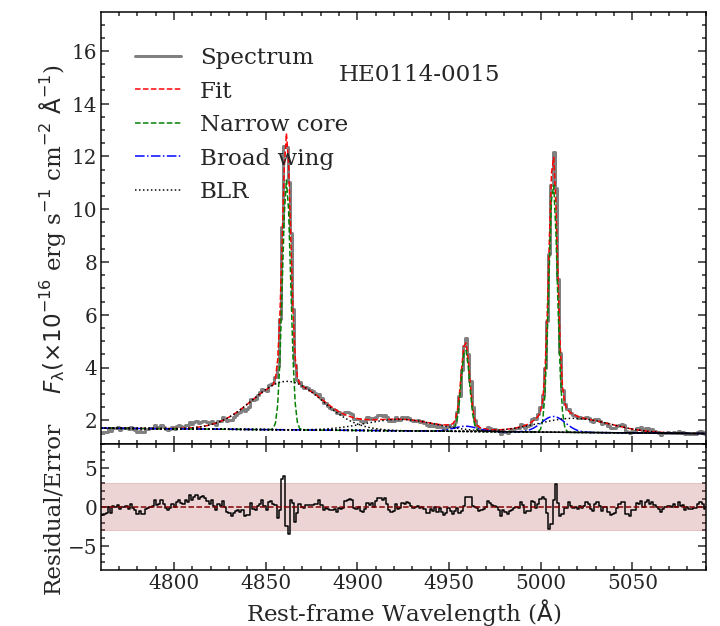}
    \\
\end{figure*}

\begin{figure*}
    \centering
    \includegraphics[width=0.49\textwidth]{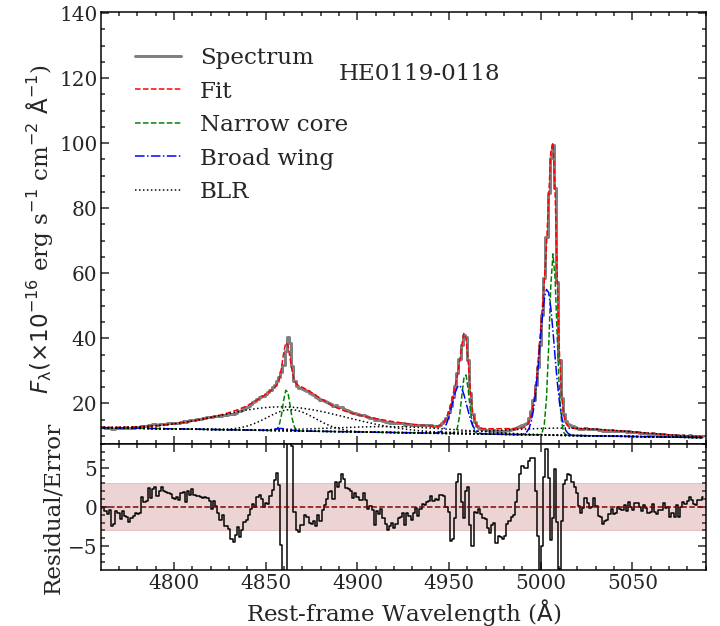}
    \includegraphics[width=0.49\textwidth]{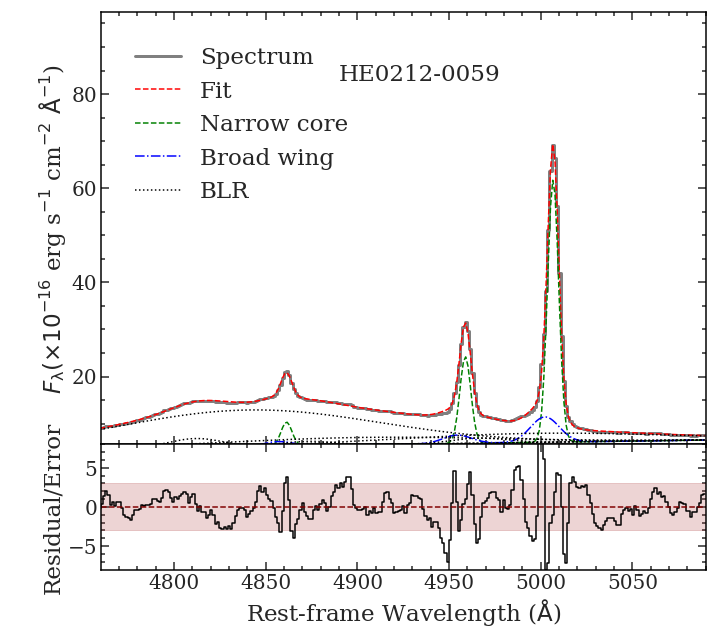}
    \\
\end{figure*}

\begin{figure*}
    \centering
    \includegraphics[width=0.49\textwidth]{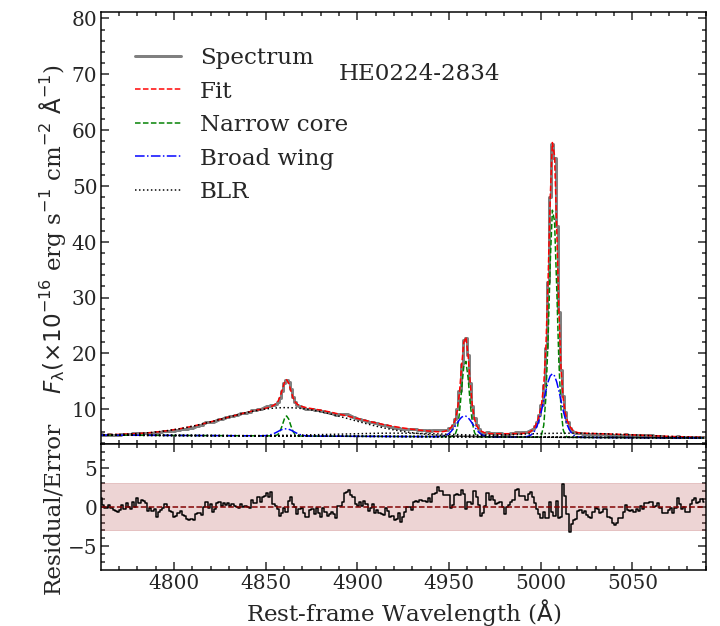}
    \includegraphics[width=0.49\textwidth]{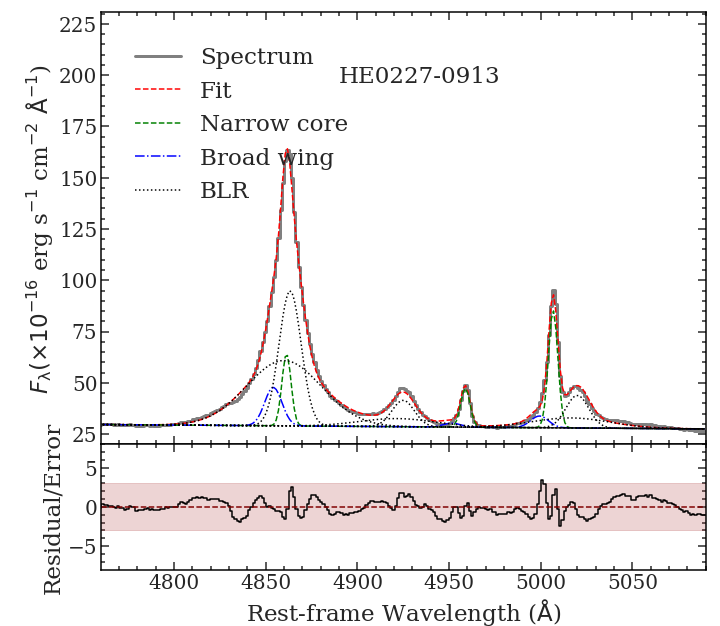}
    \\
\end{figure*}

\begin{figure*}
    \centering
    \includegraphics[width=0.49\textwidth]{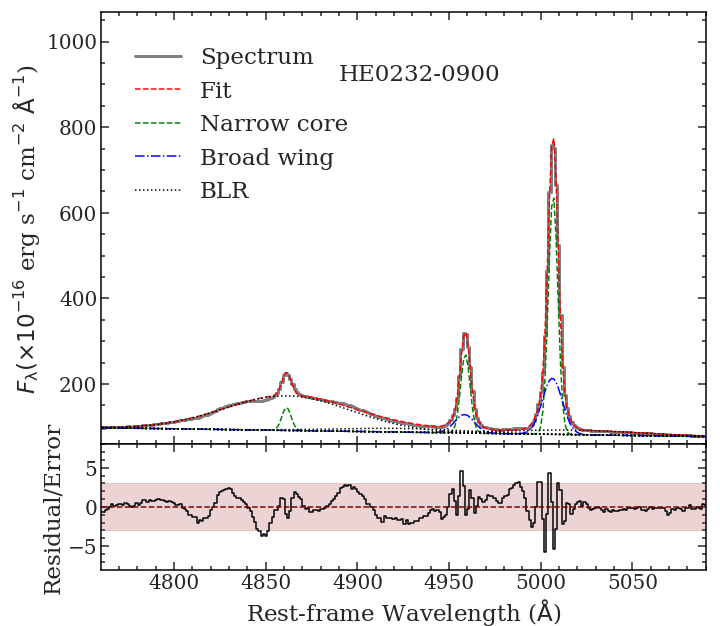}
    \includegraphics[width=0.49\textwidth]{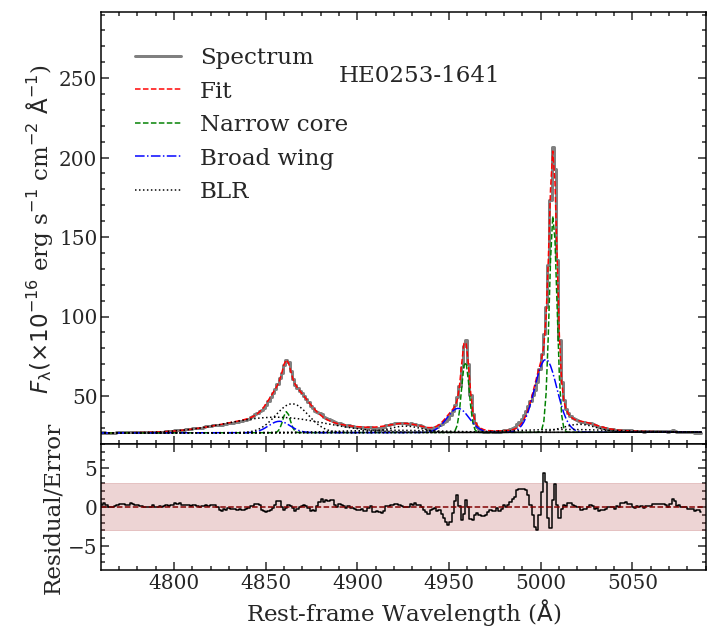}
    \\
\end{figure*}

\begin{figure*}
    \centering
    \includegraphics[width=0.49\textwidth]{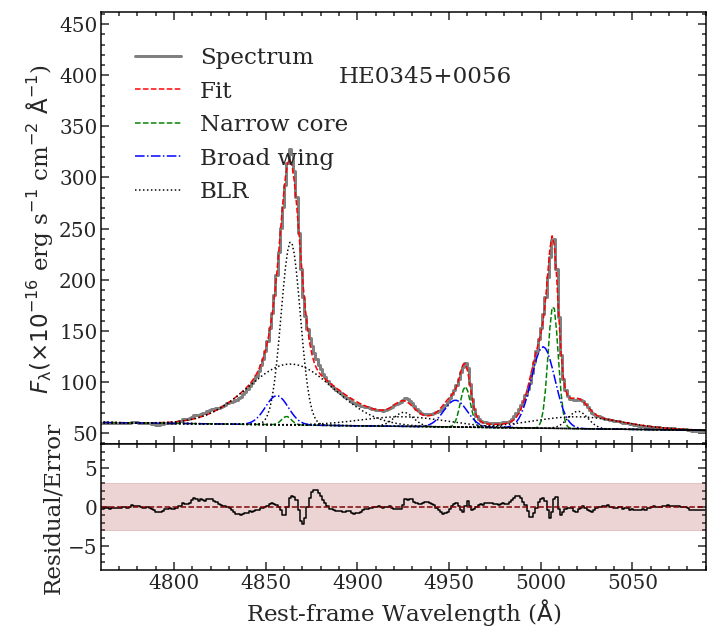}
    \includegraphics[width=0.49\textwidth]{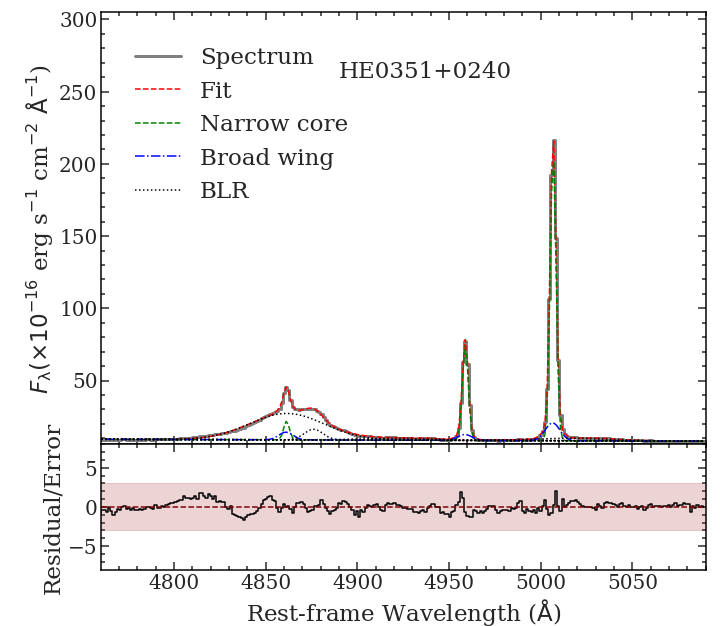}
    \\
\end{figure*}

\begin{figure*}
    \centering
    \includegraphics[width=0.49\textwidth]{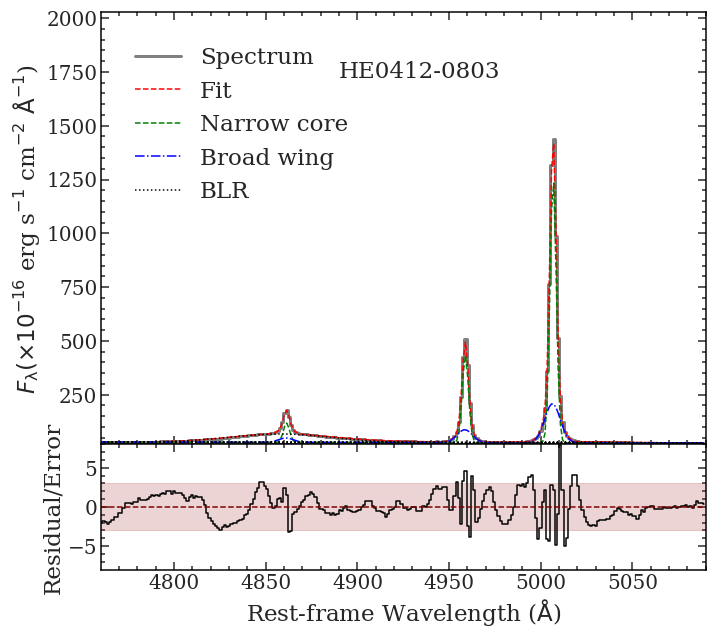}
    \includegraphics[width=0.49\textwidth]{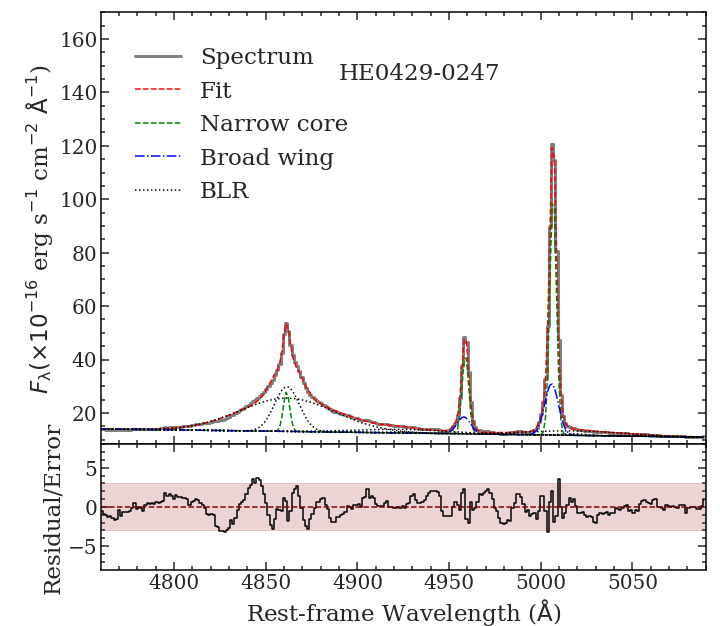}
    \\
\end{figure*}

\begin{figure*}
    \centering
    \includegraphics[width=0.49\textwidth]{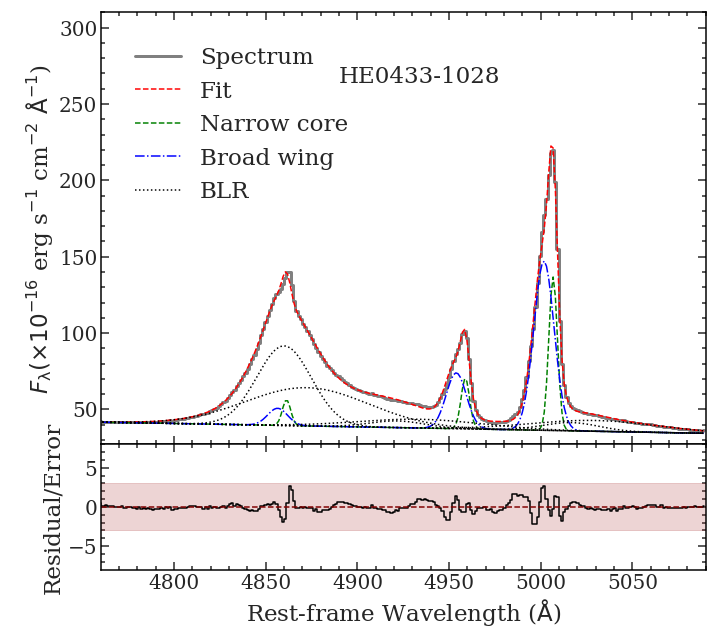}
    \includegraphics[width=0.49\textwidth]{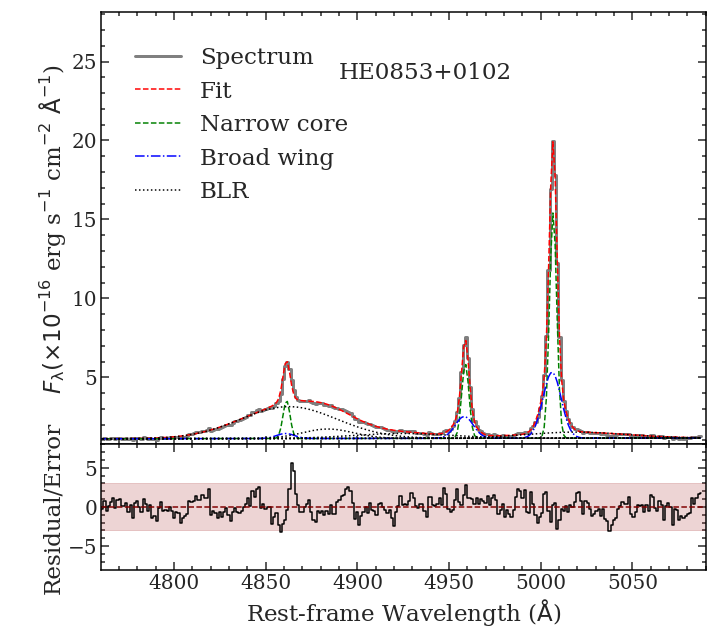}
    \\
\end{figure*}

\begin{figure*}
    \centering
    \includegraphics[width=0.49\textwidth]{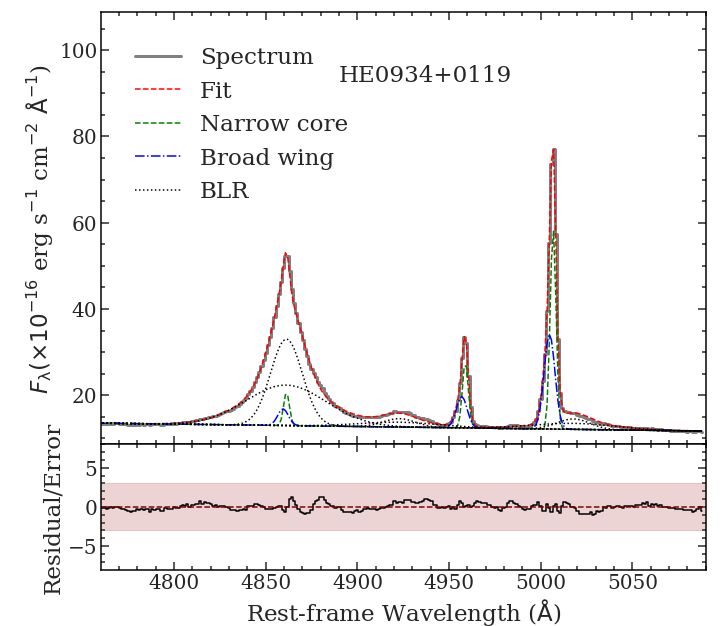}
    \includegraphics[width=0.49\textwidth]{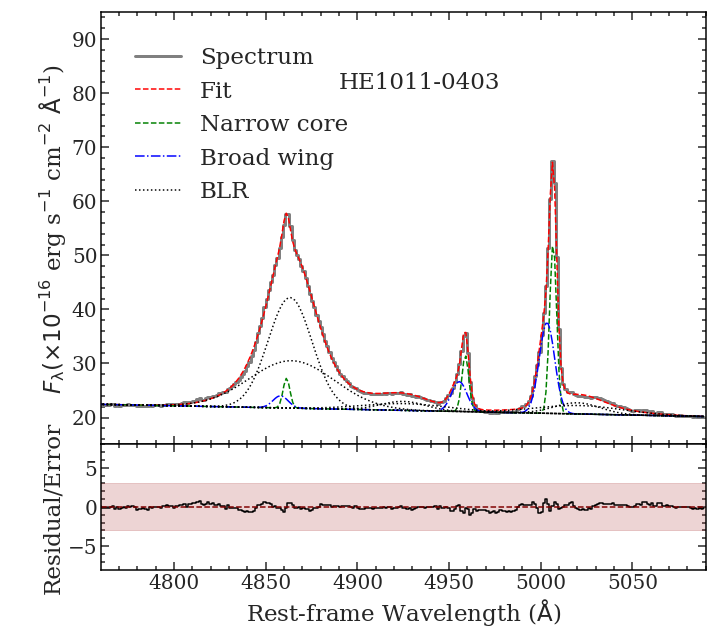}
    \\
\end{figure*}

\begin{figure*}
    \centering
    \includegraphics[width=0.49\textwidth]{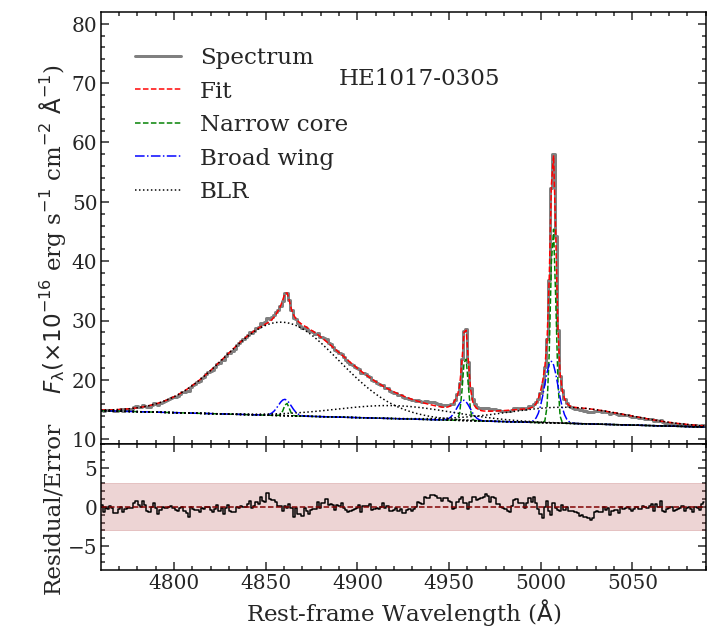}
    \includegraphics[width=0.49\textwidth]{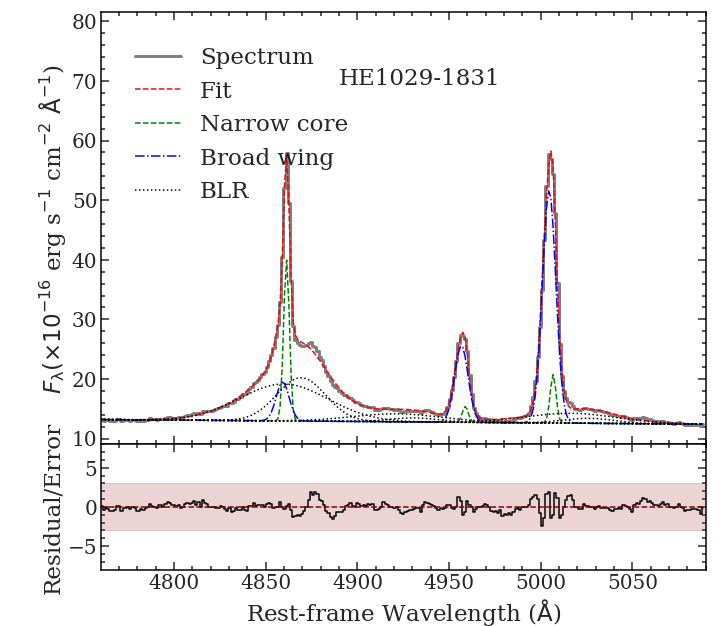}
    \\
\end{figure*}

\begin{figure*}
    \centering
    \includegraphics[width=0.49\textwidth]{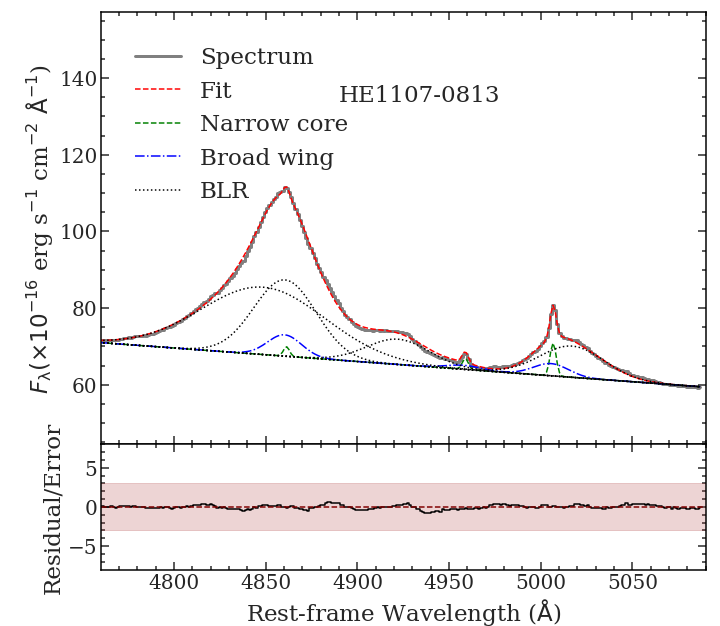}
    \includegraphics[width=0.49\textwidth]{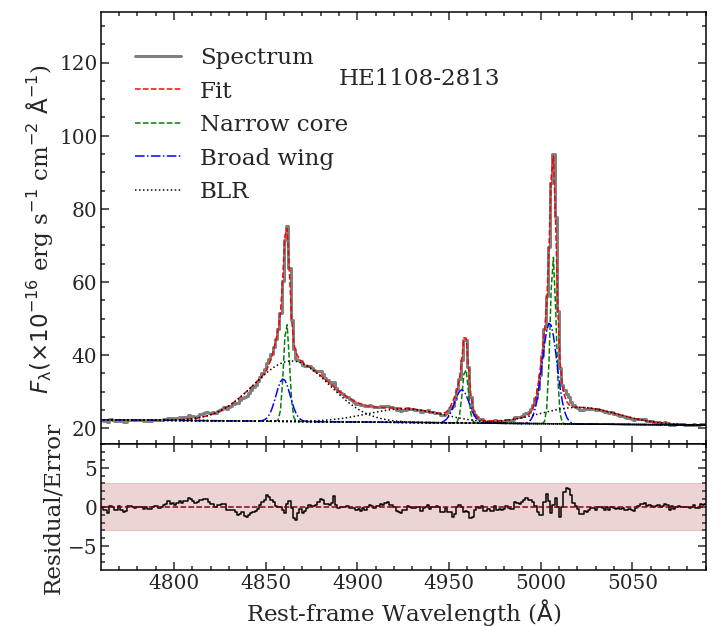}
    \\
\end{figure*}

\begin{figure*}
    \centering
    \includegraphics[width=0.49\textwidth]{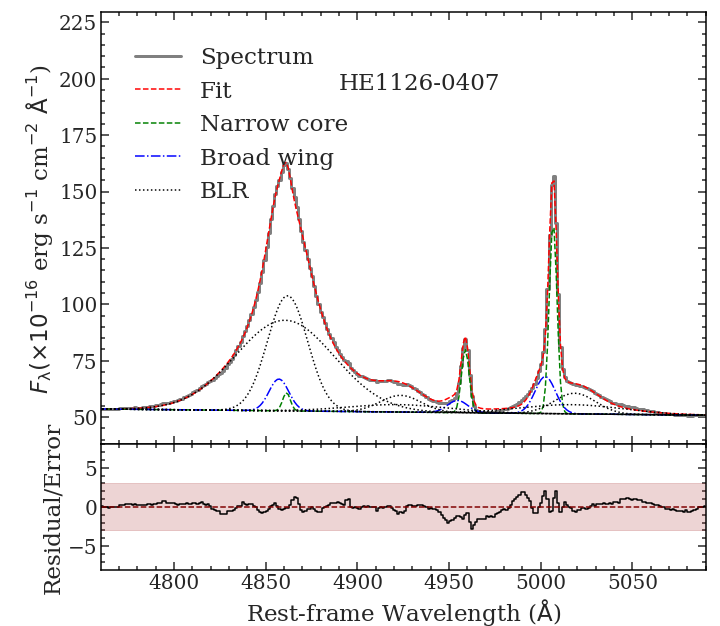}
    \includegraphics[width=0.49\textwidth]{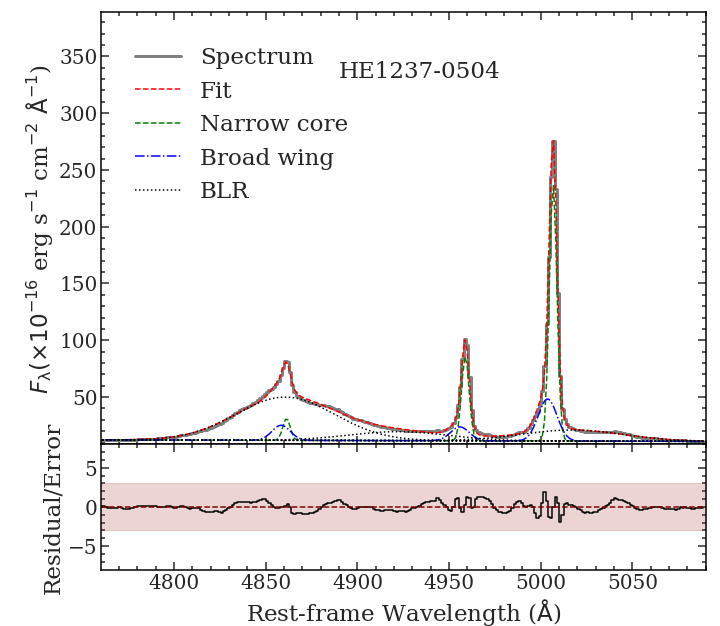}
    \\
\end{figure*}

\begin{figure*}
    \centering
    \includegraphics[width=0.49\textwidth]{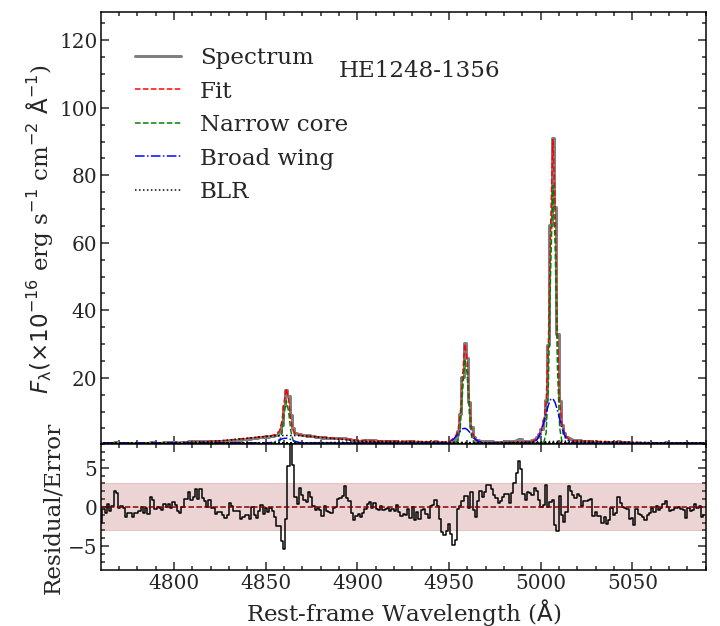}
    \includegraphics[width=0.49\textwidth]{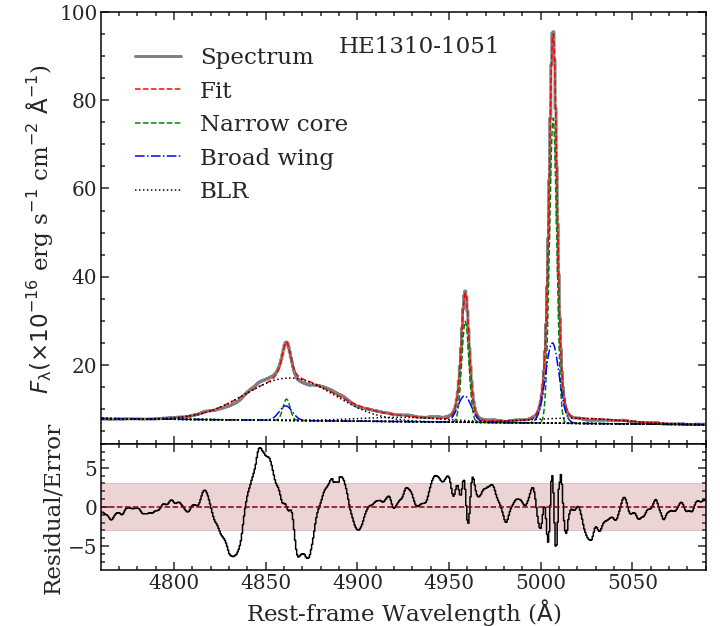}
    \\
\end{figure*}

\begin{figure*}
    \centering
    \includegraphics[width=0.49\textwidth]{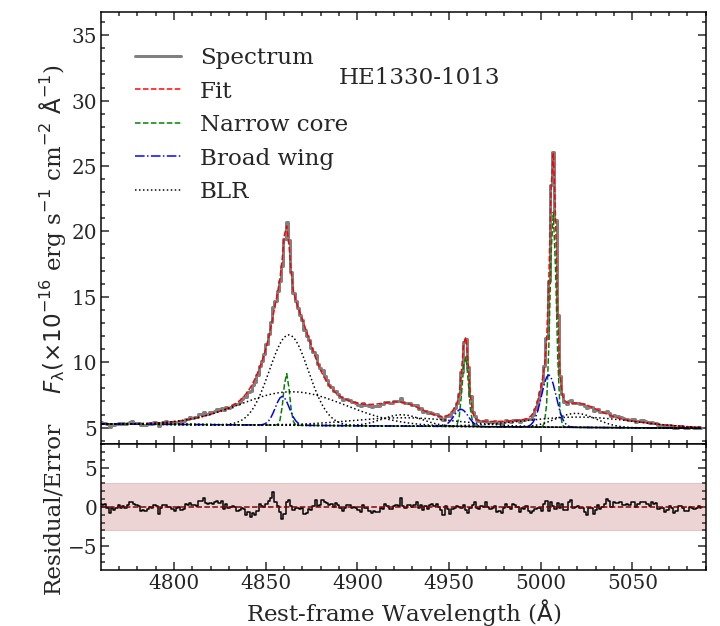}
    \includegraphics[width=0.49\textwidth]{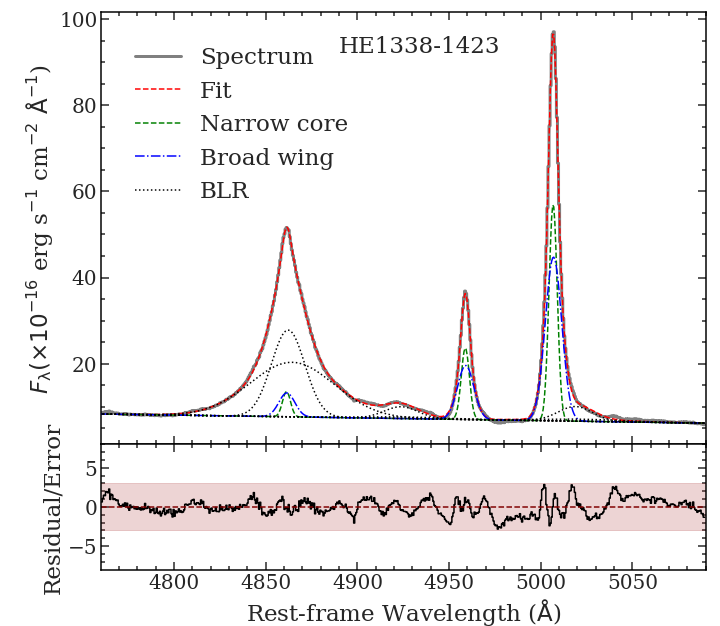}
    \\
\end{figure*}

\begin{figure*}
    \centering
    \includegraphics[width=0.49\textwidth]{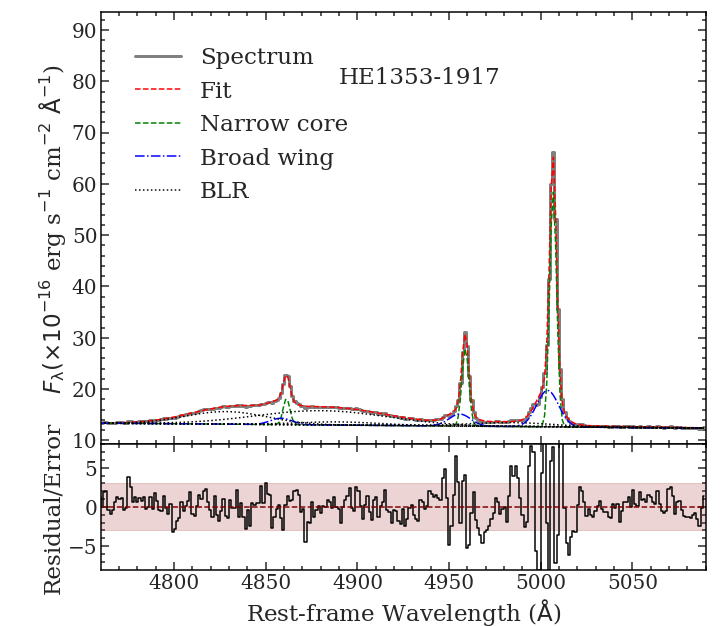}
    \includegraphics[width=0.49\textwidth]{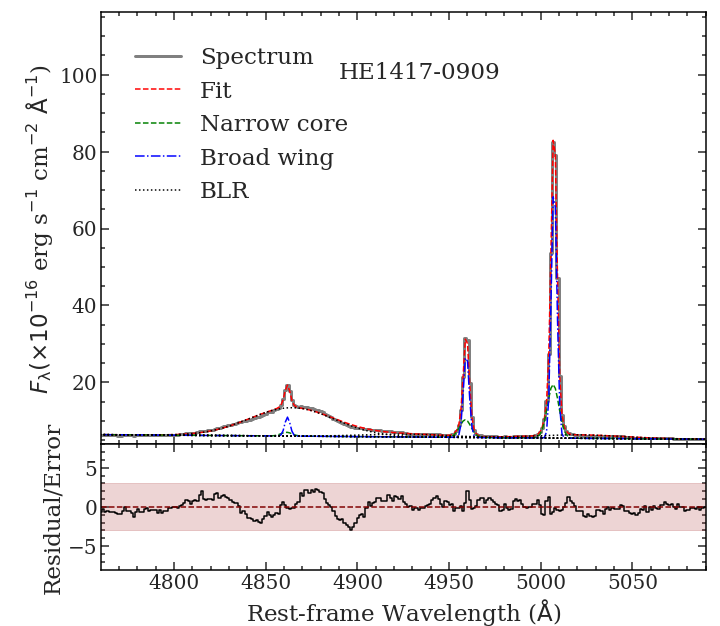}
    \\
\end{figure*}

\begin{figure*}
    \centering
    \includegraphics[width=0.49\textwidth]{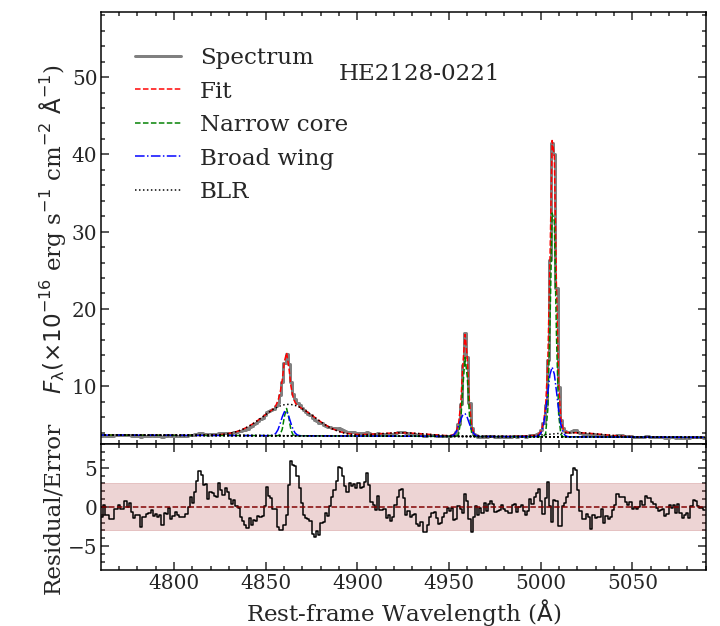}
    \includegraphics[width=0.49\textwidth]{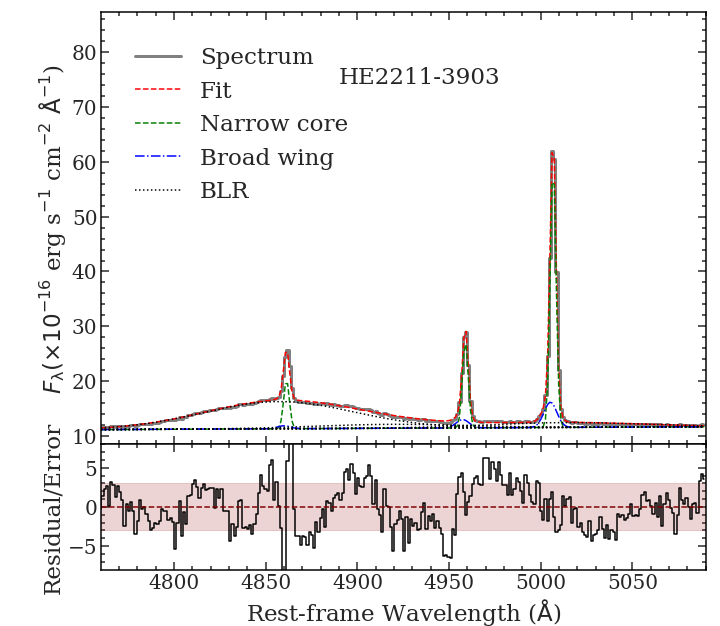}
    \\
\end{figure*}

\begin{figure*}
    \centering
    \includegraphics[width=0.49\textwidth]{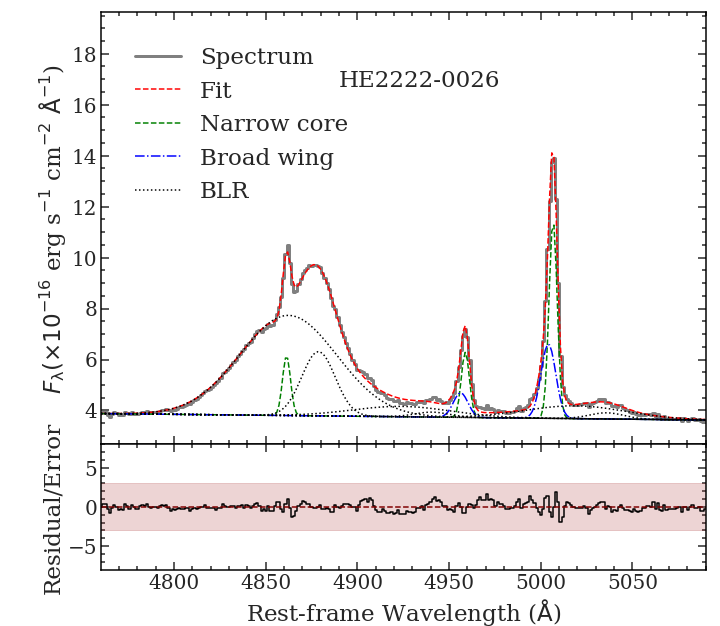}
    \includegraphics[width=0.49\textwidth]{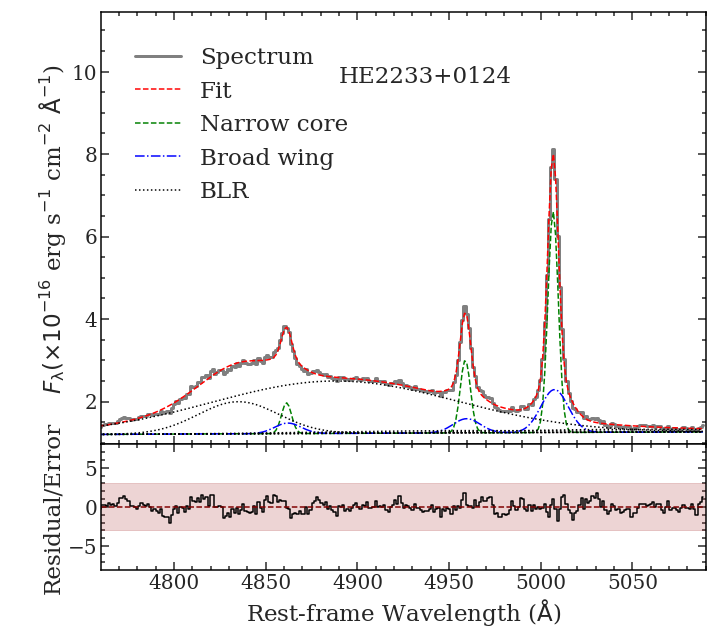}
    \\
\end{figure*}

\begin{figure*}
    \centering
    \includegraphics[width=0.49\textwidth]{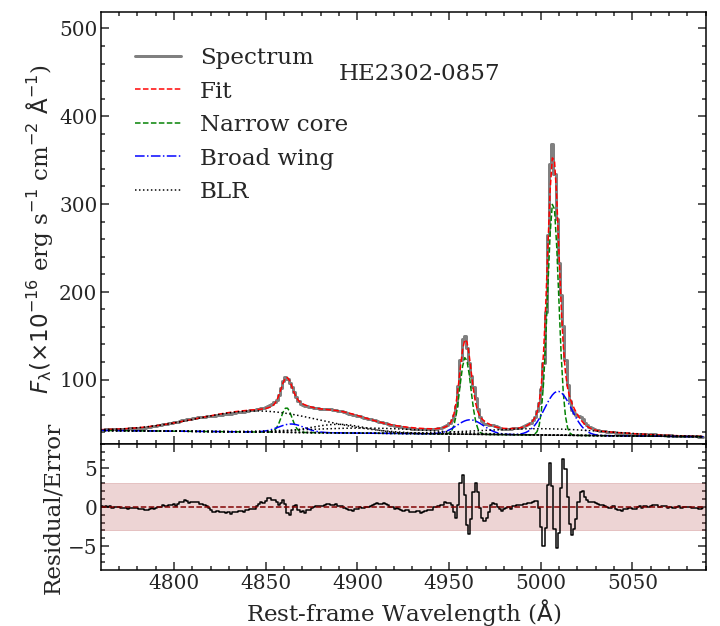}
    \includegraphics[width=0.49\textwidth]{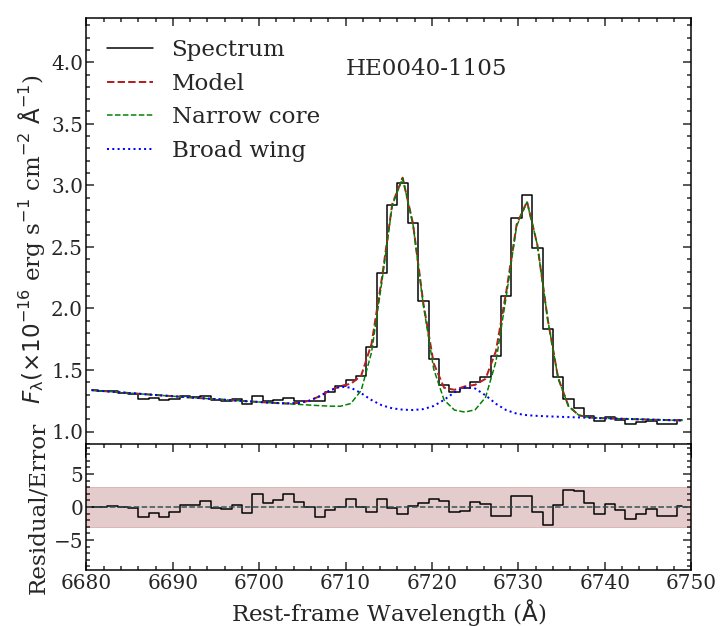}
    \\
\end{figure*}

\begin{figure*}
    \centering
    \includegraphics[width=0.49\textwidth]{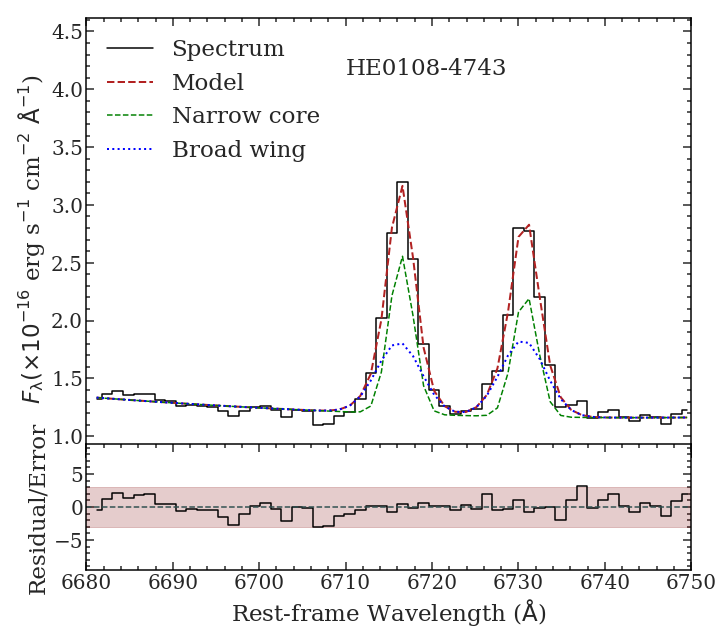}
    \includegraphics[width=0.49\textwidth]{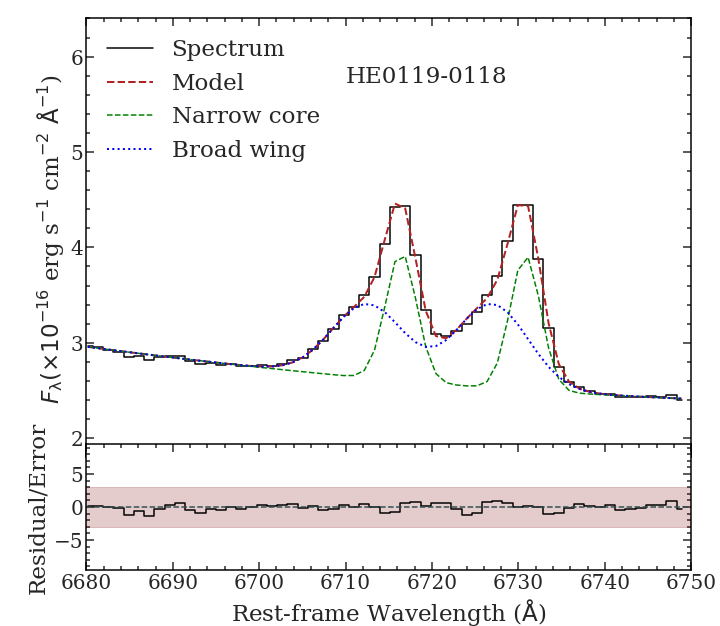}
    \\
\end{figure*}

\begin{figure*}
    \centering
    \includegraphics[width=0.49\textwidth]{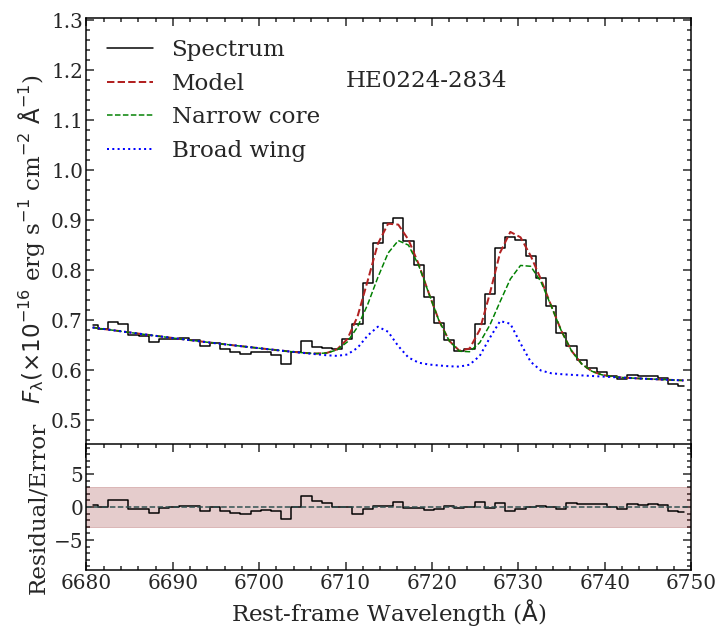}
    \includegraphics[width=0.49\textwidth]{SII_spectra/HE0253-1641_SII_modelling.png}
    \\
\end{figure*}

\begin{figure*}
    \centering
    \includegraphics[width=0.49\textwidth]{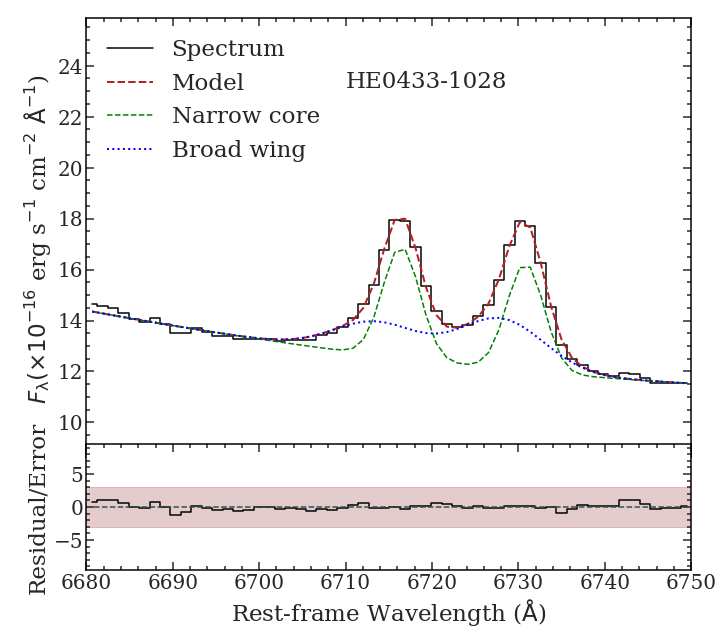}
    \includegraphics[width=0.49\textwidth]{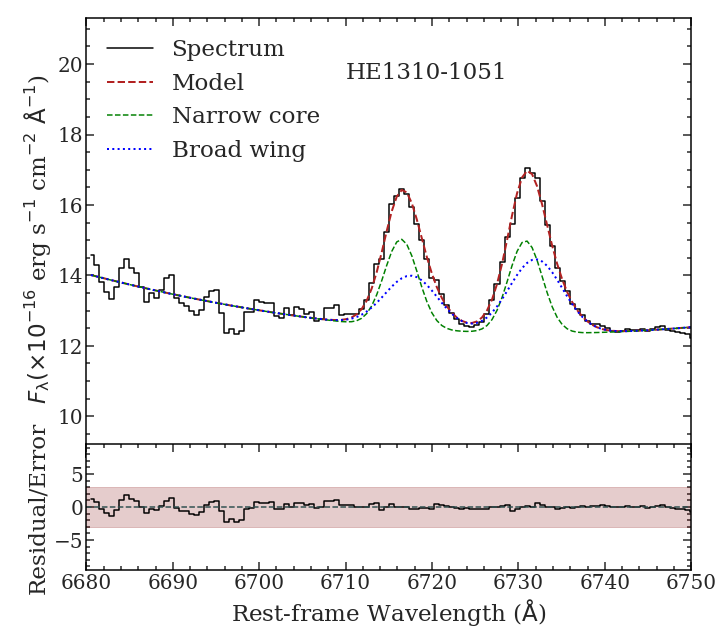}
    \\
\end{figure*}

\begin{figure*}
    \centering
    \includegraphics[width=0.49\textwidth]{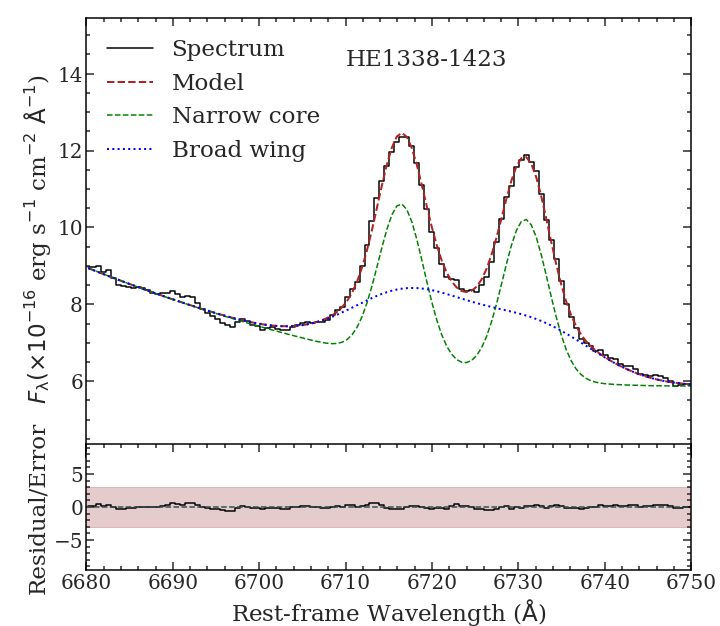}
    \includegraphics[width=0.49\textwidth]{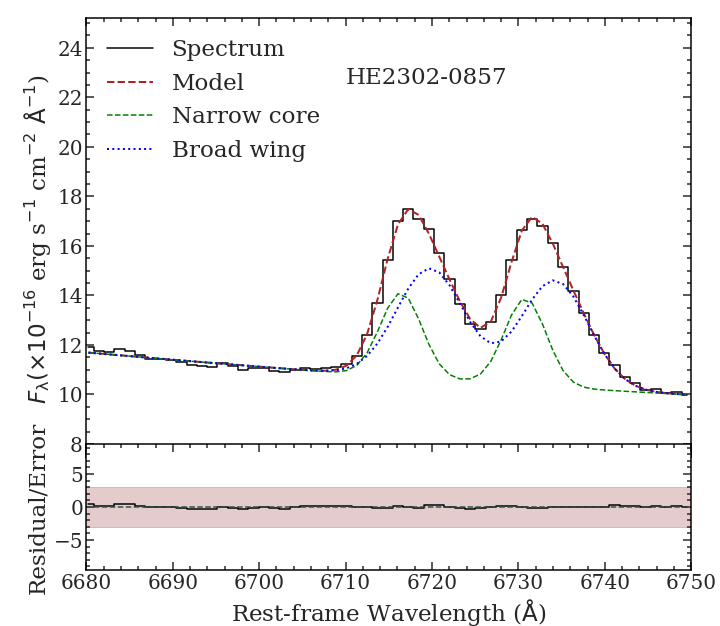}
    \\
\end{figure*}

\section{Flux maps}
\label{section:maps}

In this section, we list all the 2D surface brightness maps H$\beta$ + [\ion{O}{iii}], their 2D Moffat modelling and the normalized residuals. For the MUSE AGN, the 2D maps were performed in the central 3\arcsec\ region; whereas, for the VIMOS sources, it covers central 6\arcsec\ region.

\begin{figure*}
   \resizebox{\hsize}{!}{\includegraphics{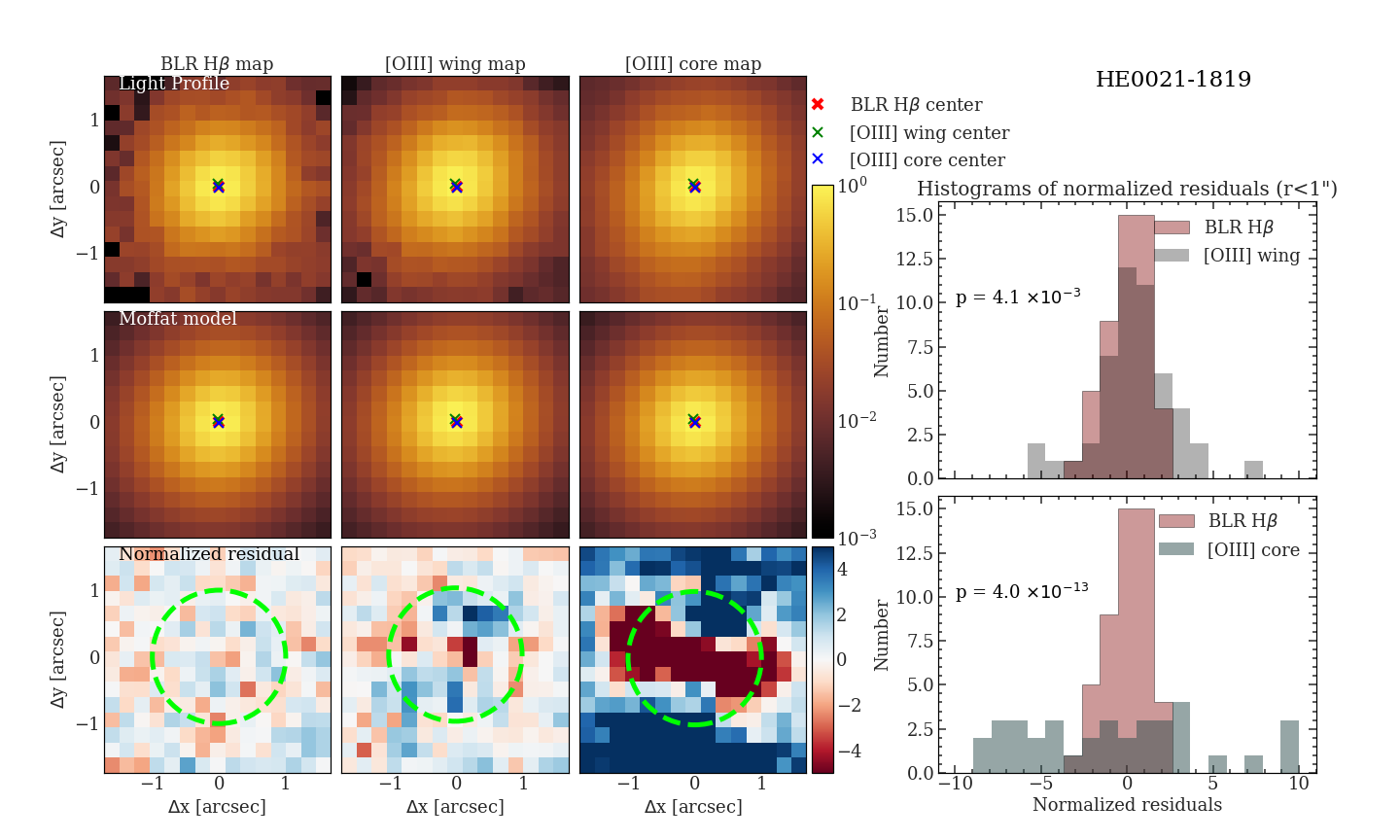}}
   \resizebox{\hsize}{!}{\includegraphics{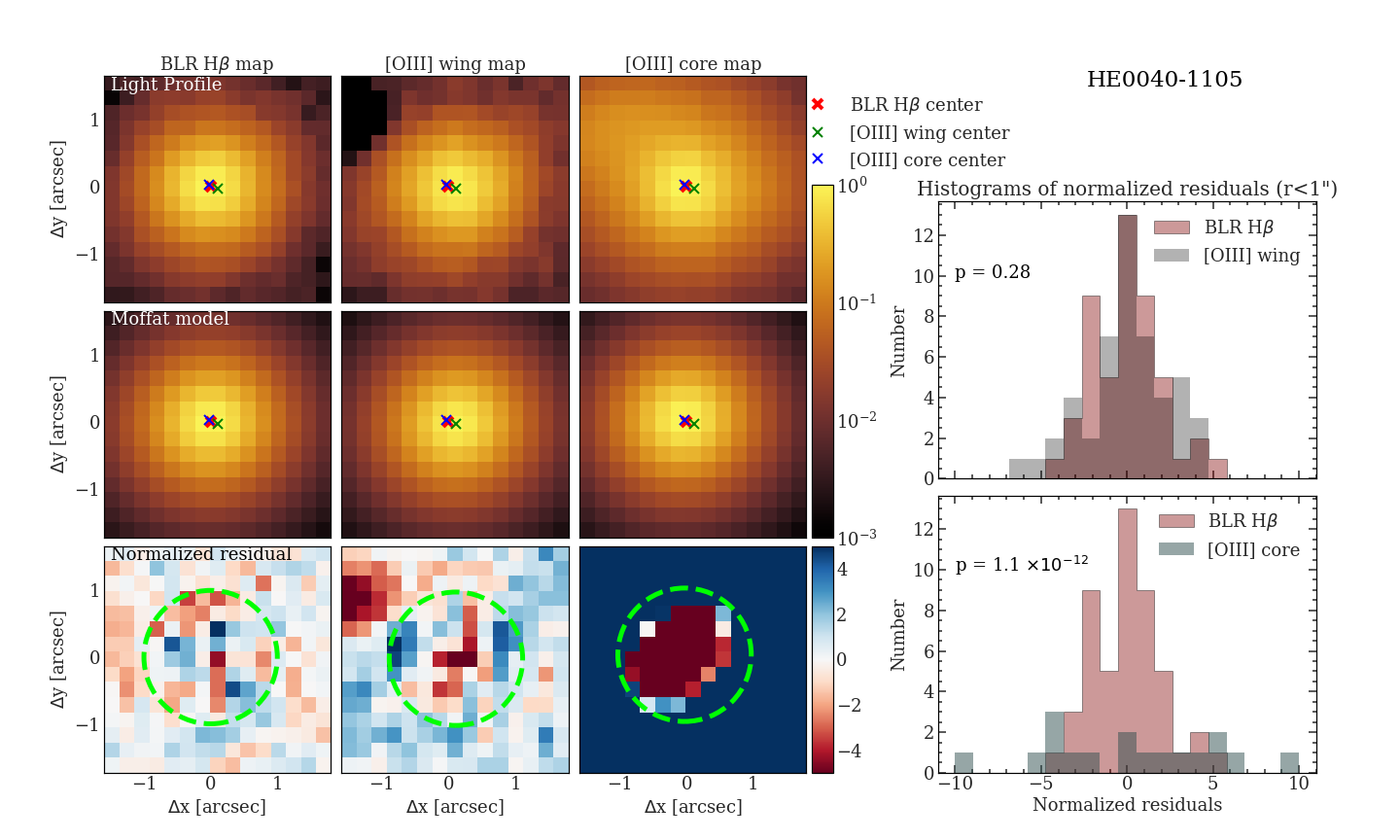}}
\end{figure*}

\begin{figure*}
   \resizebox{\hsize}{!}{\includegraphics{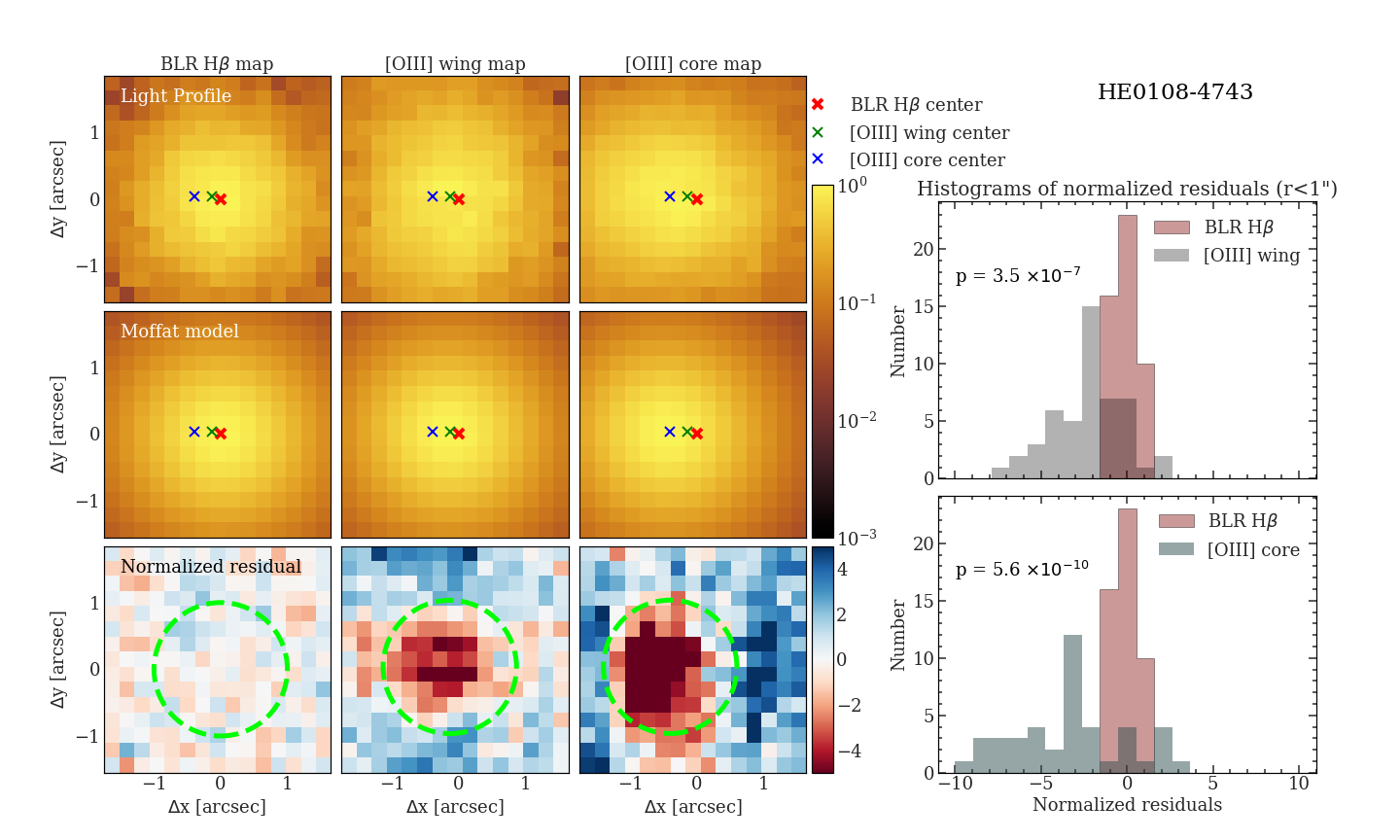}}
   \resizebox{\hsize}{!}{\includegraphics{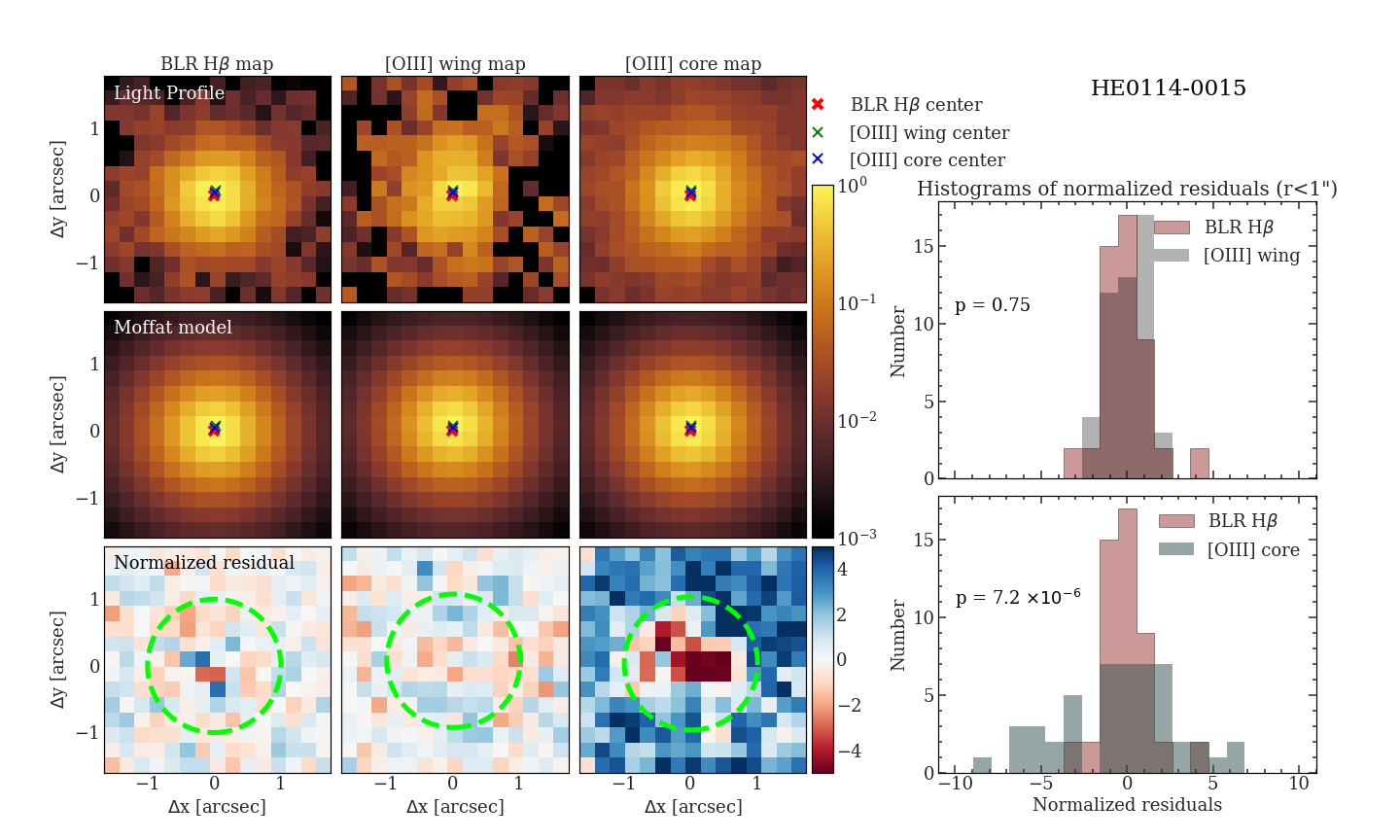}}
\end{figure*}

\begin{figure*}
   \resizebox{\hsize}{!}{\includegraphics{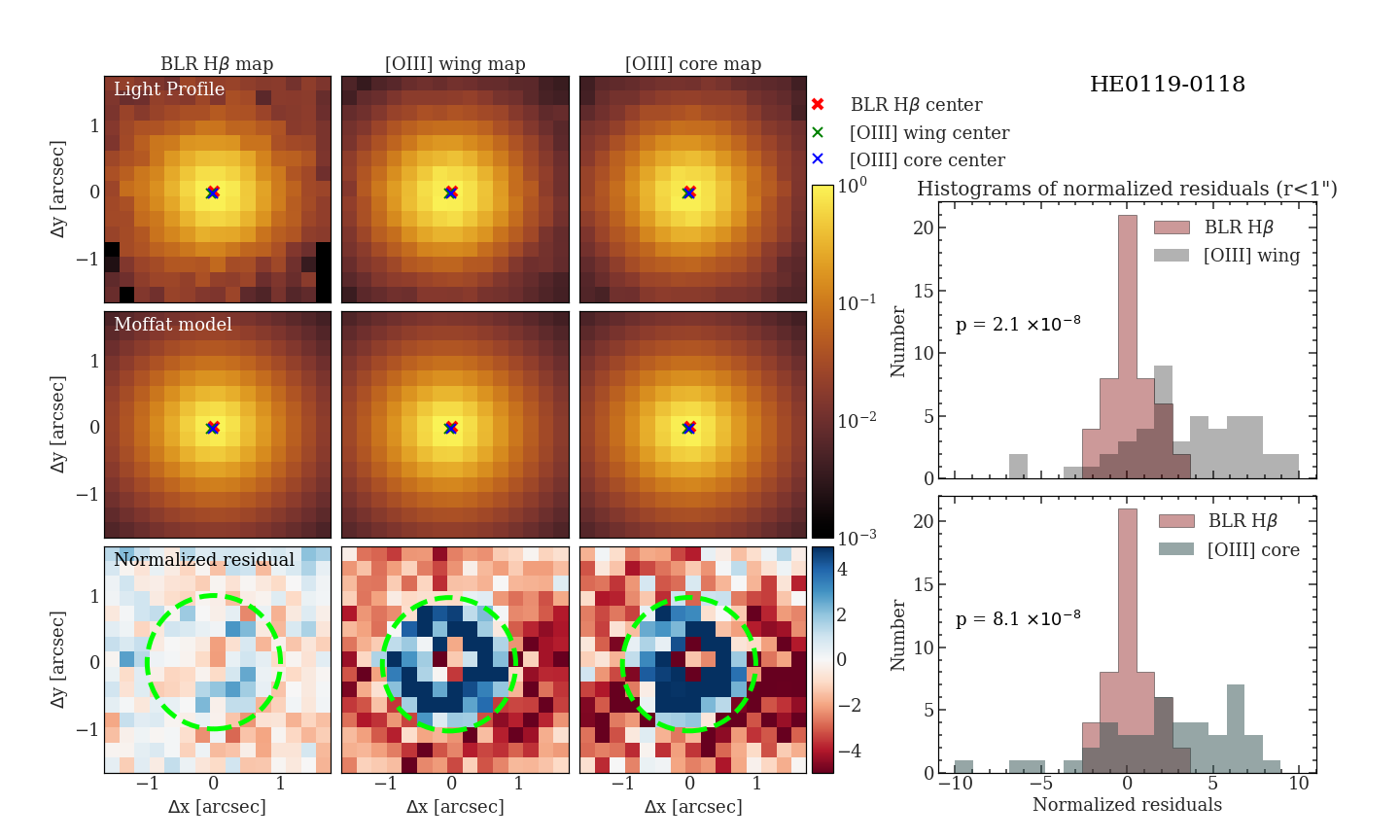}}
   \resizebox{\hsize}{!}{\includegraphics{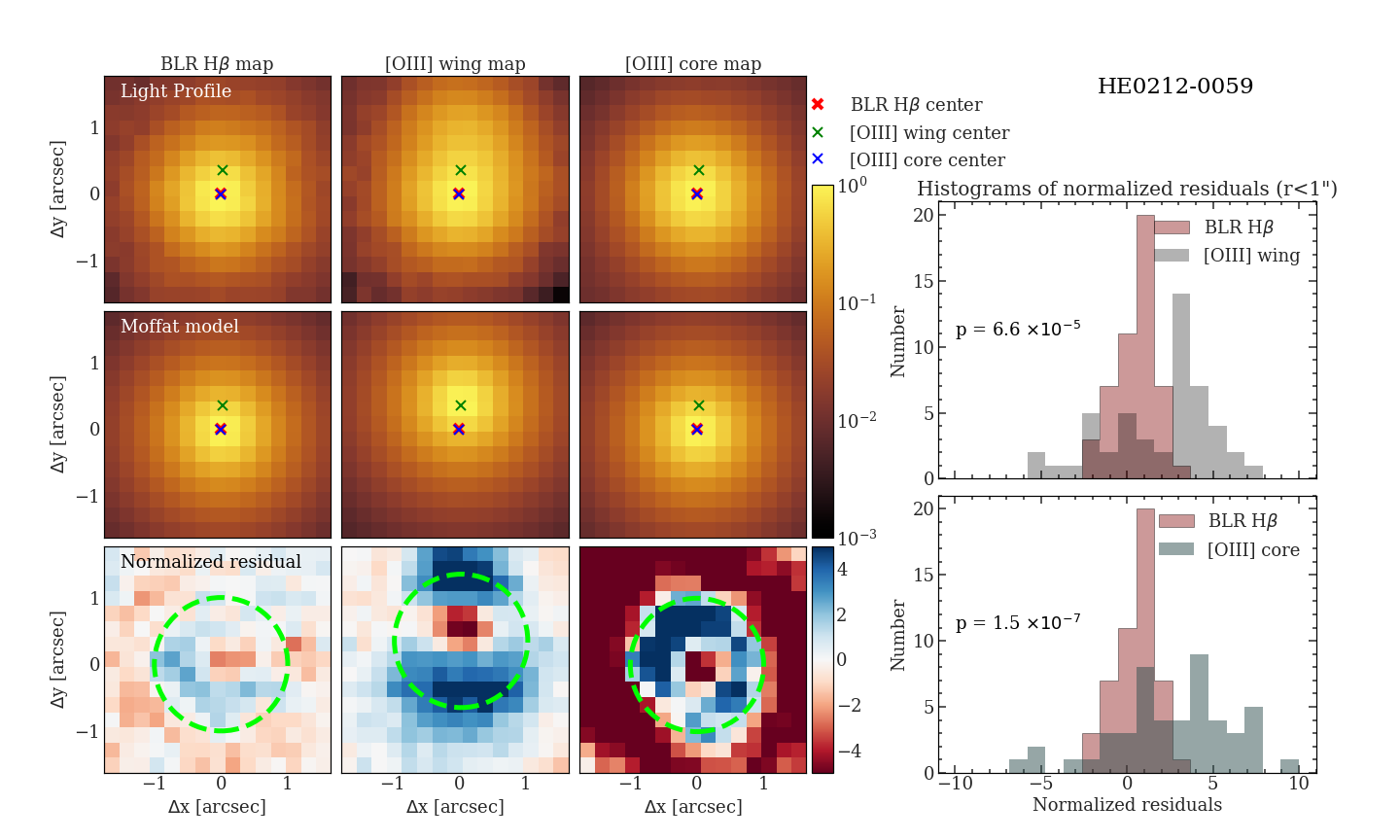}}
\end{figure*}

\begin{figure*}
   \resizebox{\hsize}{!}{\includegraphics{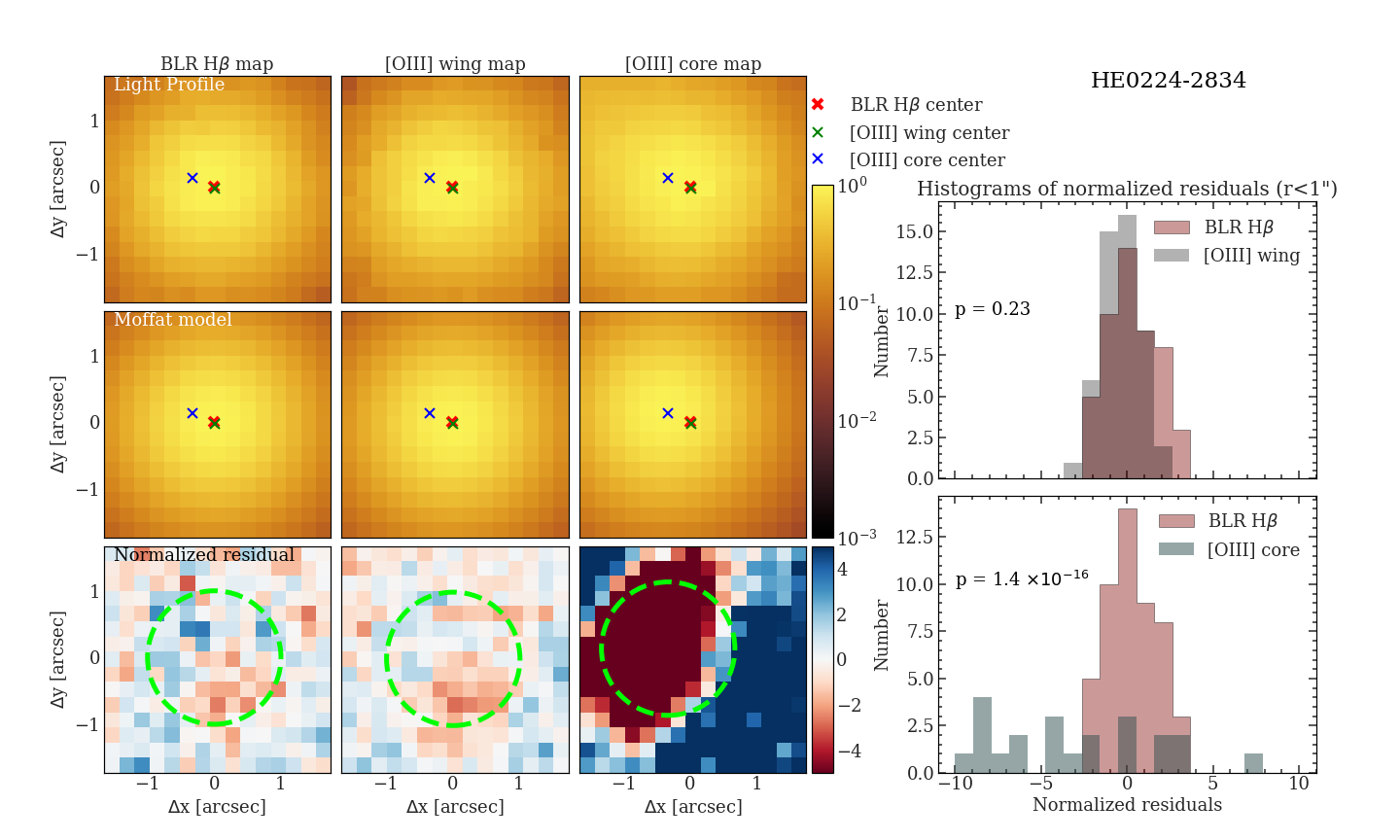}}
   \resizebox{\hsize}{!}{\includegraphics{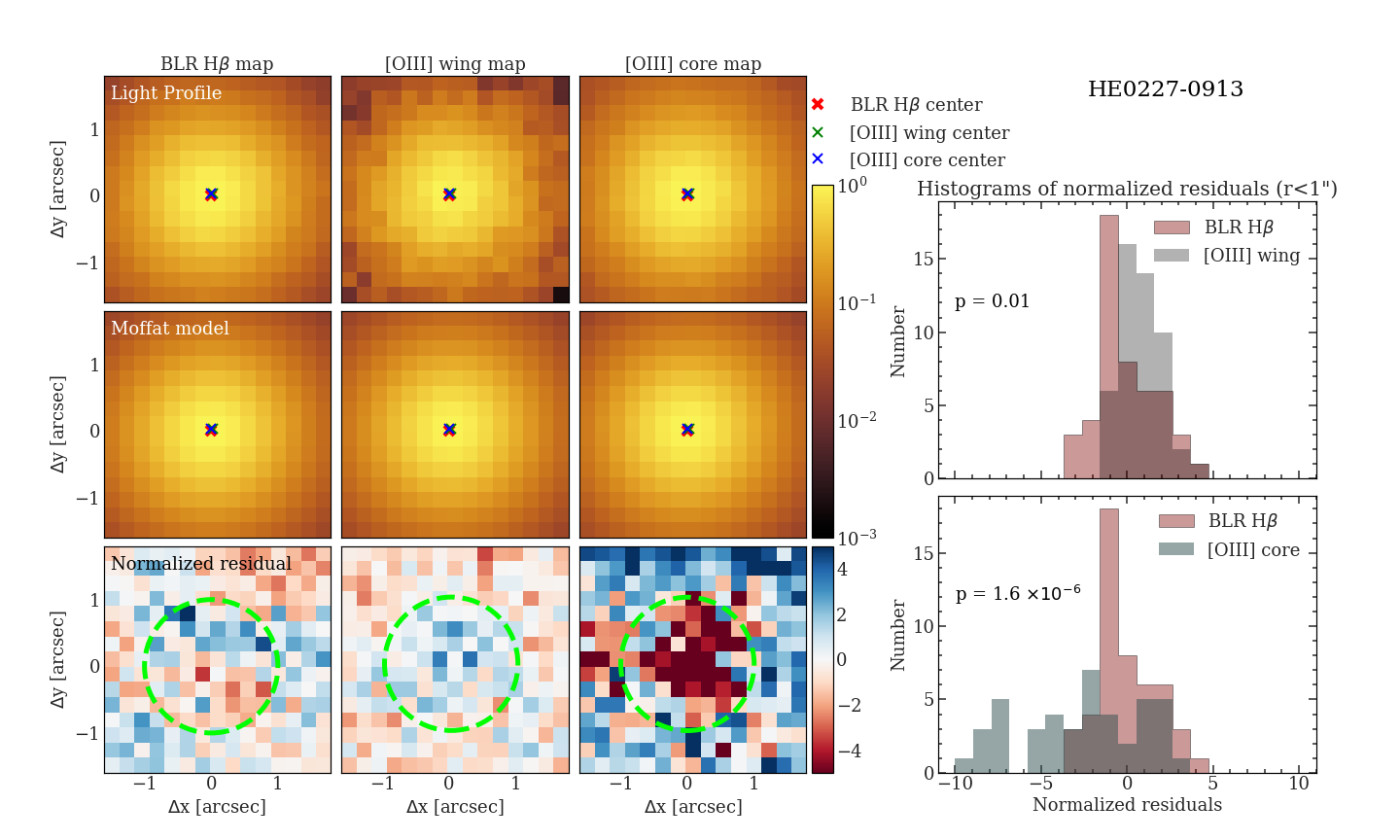}}
\end{figure*}

\begin{figure*}
    \resizebox{\hsize}{!}{\includegraphics{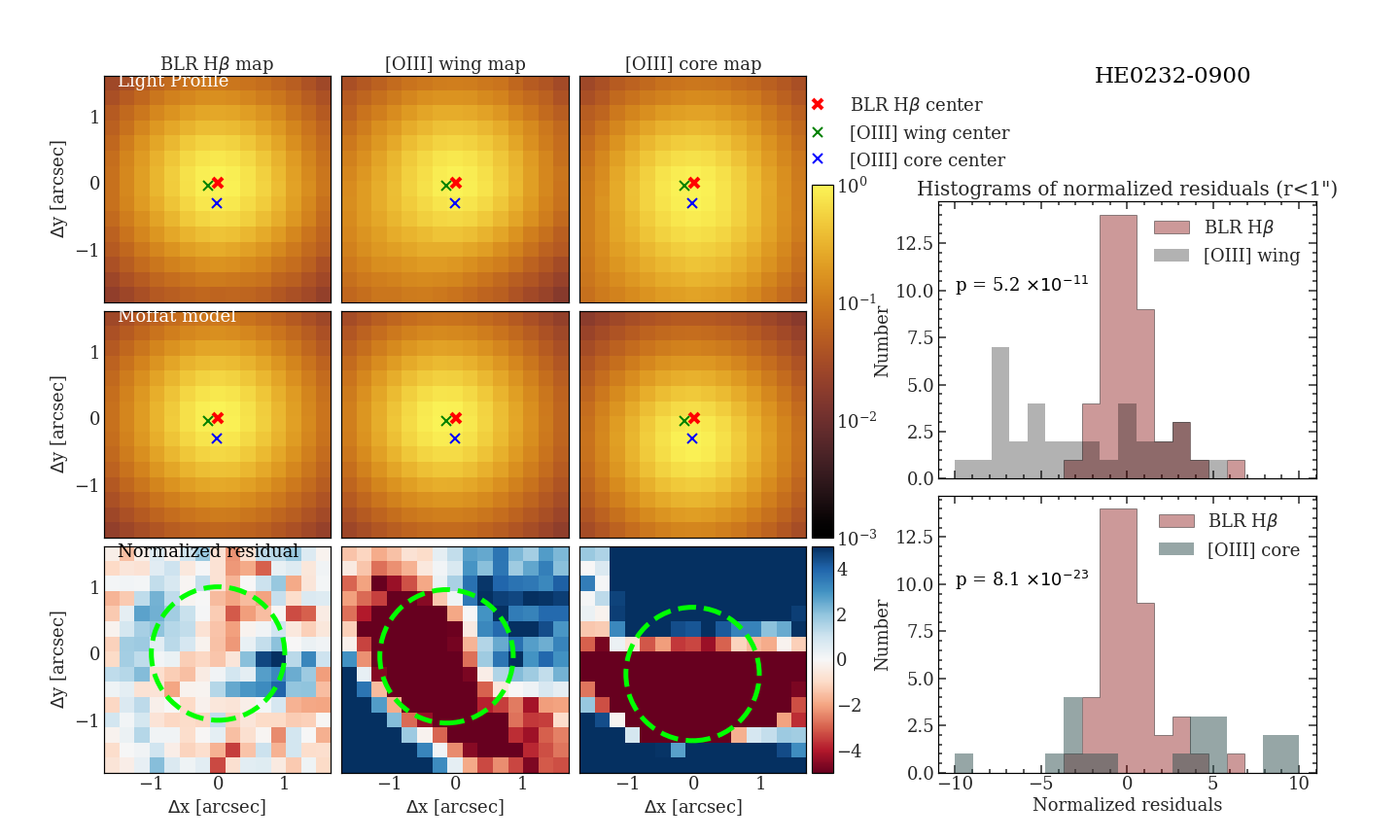}}
    \resizebox{\hsize}{!}{\includegraphics{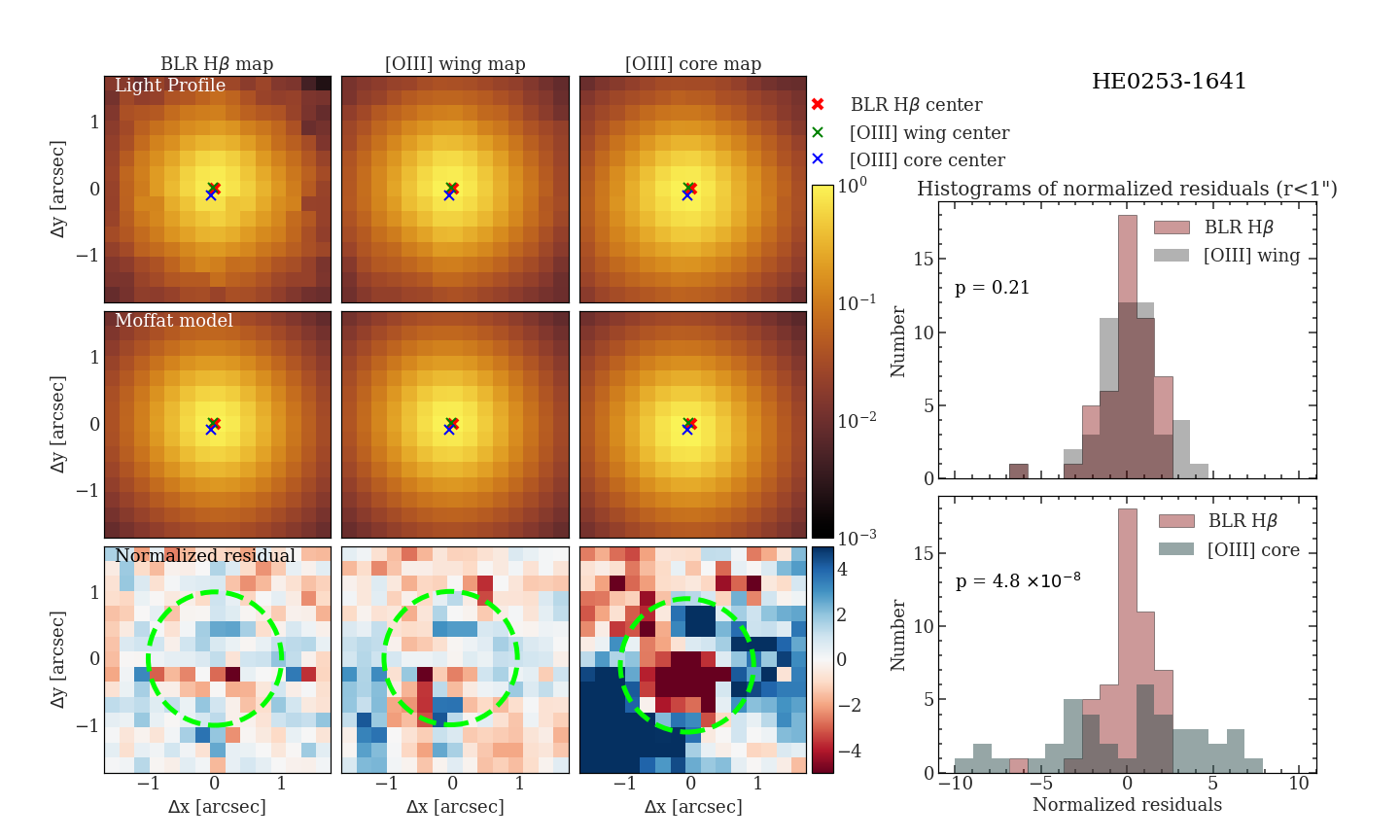}}
\end{figure*}

\begin{figure*}
   \resizebox{\hsize}{!}{\includegraphics{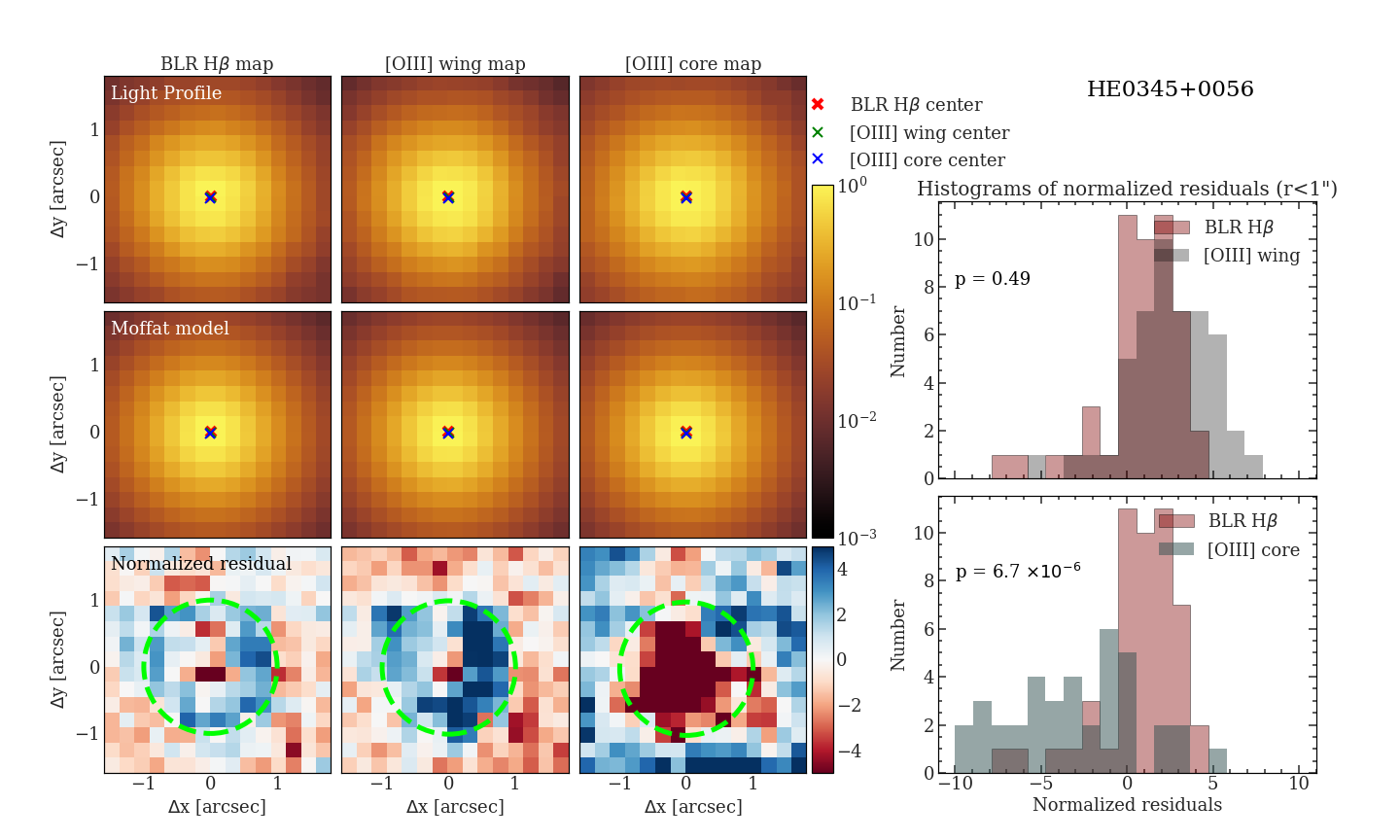}}
   \resizebox{\hsize}{!}{\includegraphics{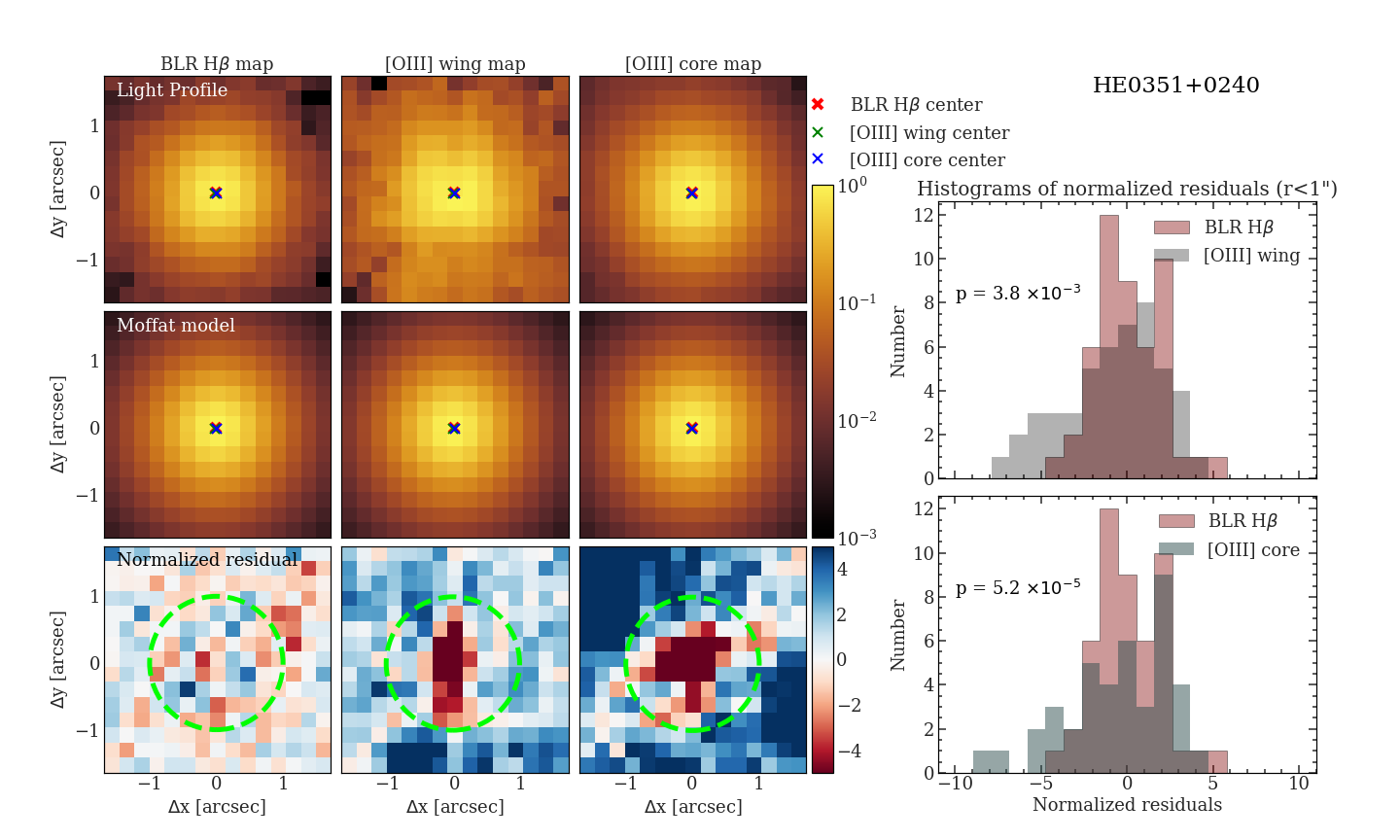}}
\end{figure*}

\begin{figure*}
   \resizebox{\hsize}{!}{\includegraphics{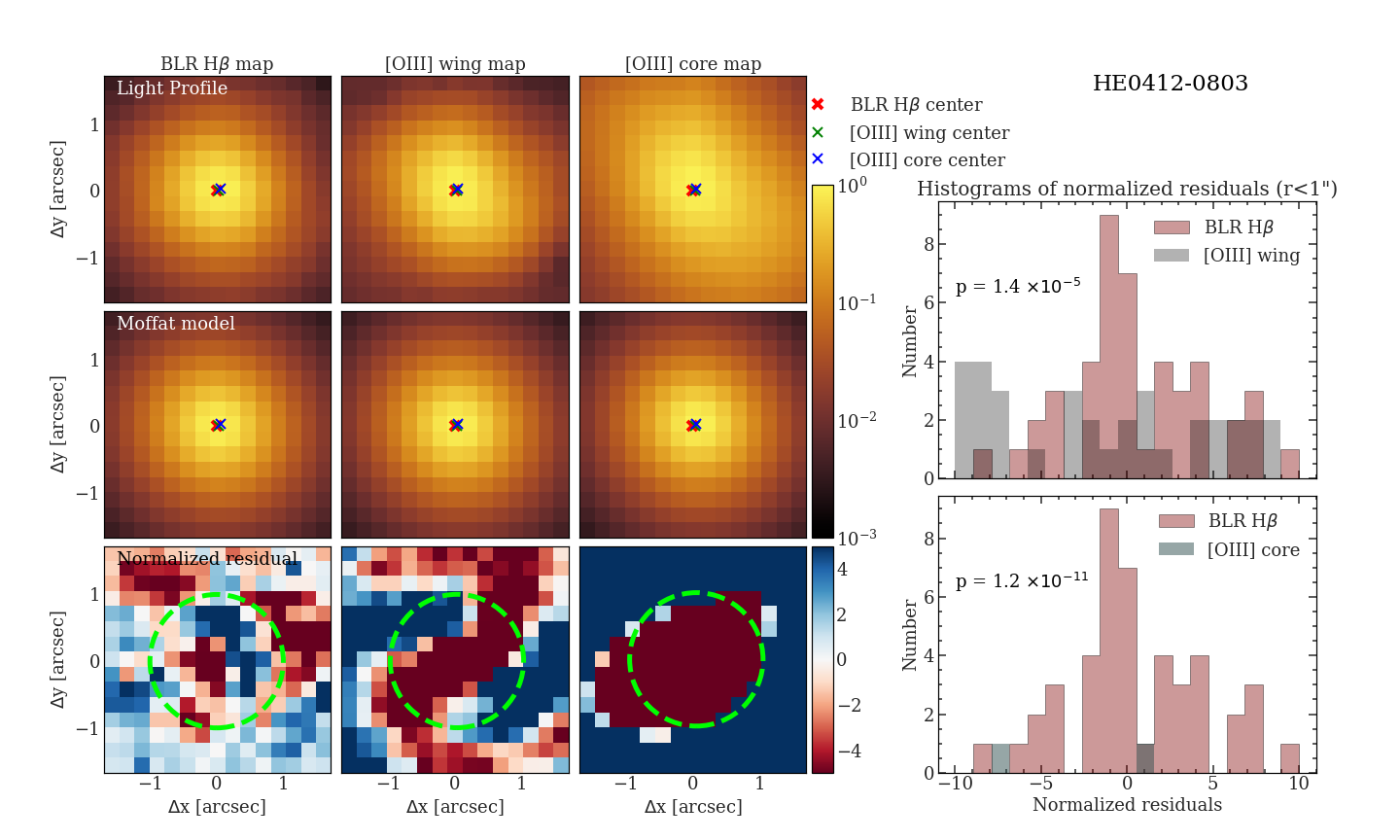}}
   \resizebox{\hsize}{!}{\includegraphics{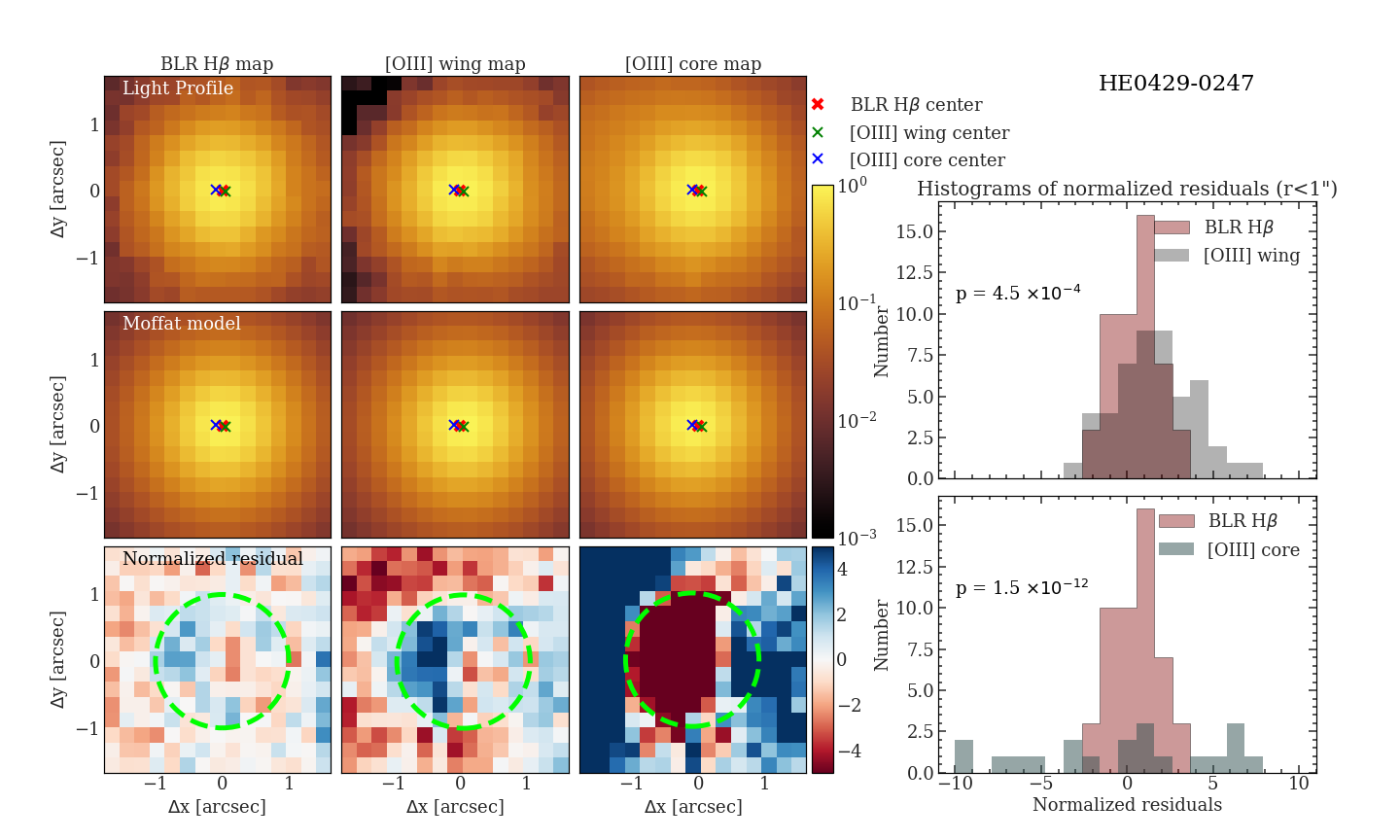}}
\end{figure*}

\begin{figure*}
   \resizebox{\hsize}{!}{\includegraphics{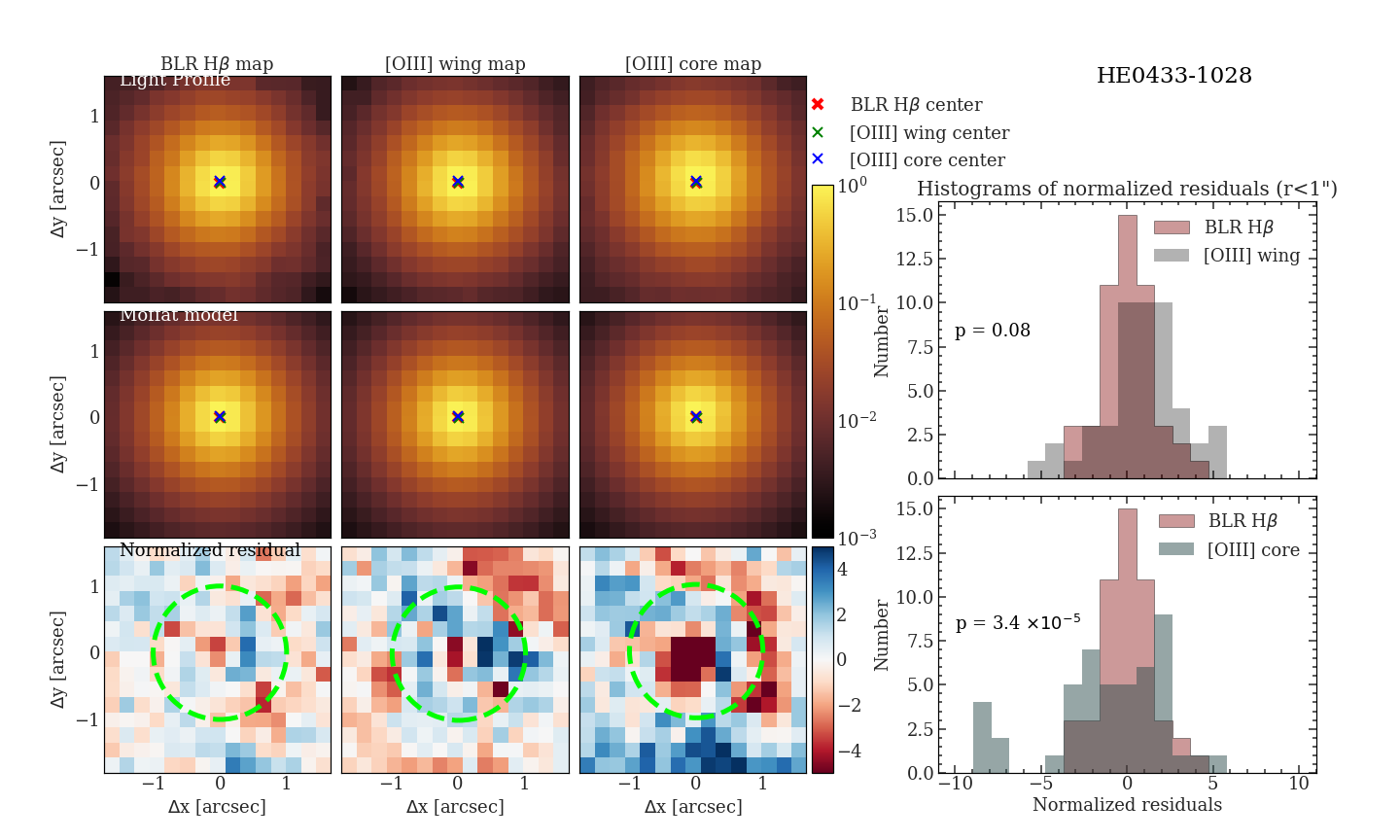}}
   \resizebox{\hsize}{!}{\includegraphics{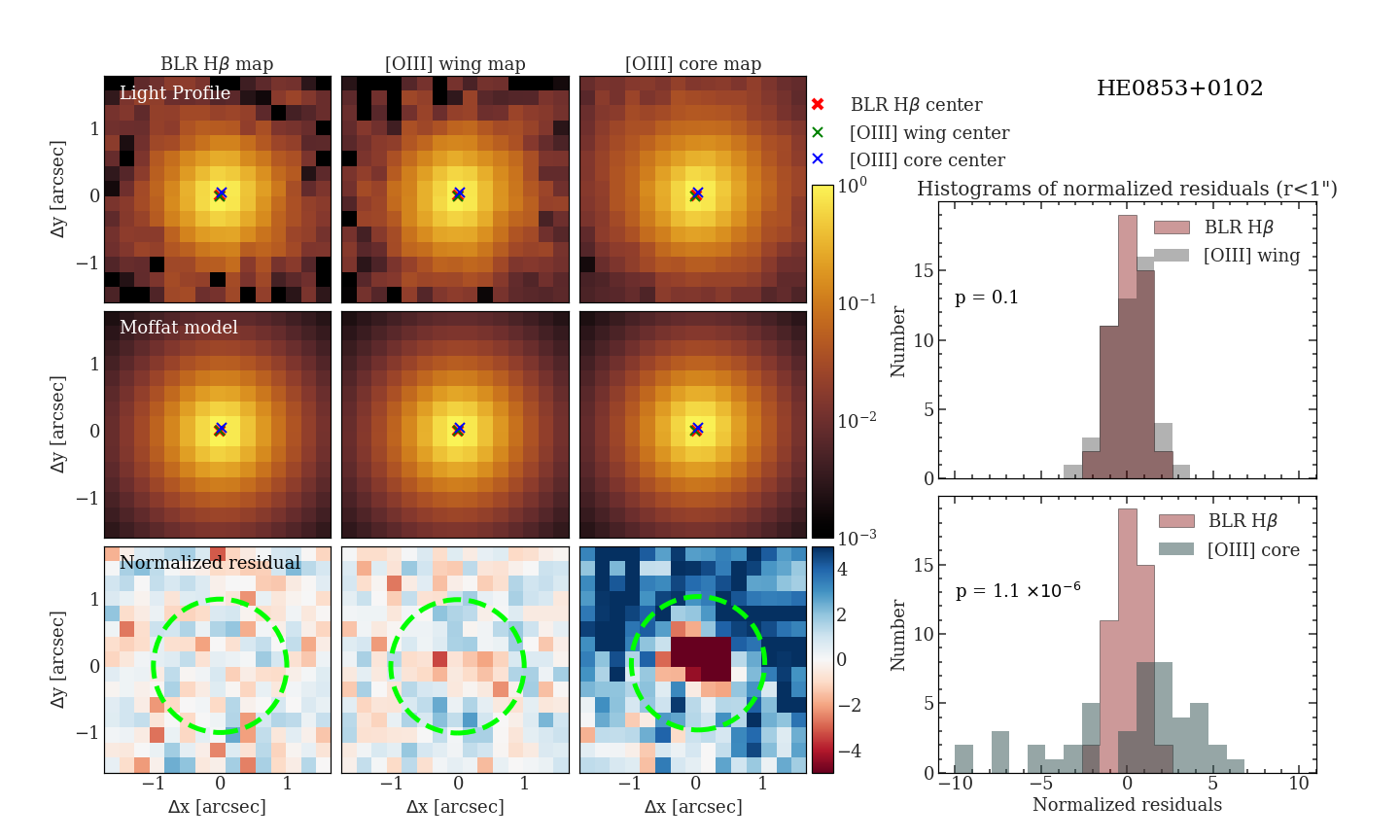}}
\end{figure*}

\begin{figure*}
   \resizebox{\hsize}{!}{\includegraphics{Maps/HE0934+0119_1_arcsec_maps.png}}
   \resizebox{\hsize}{!}{\includegraphics{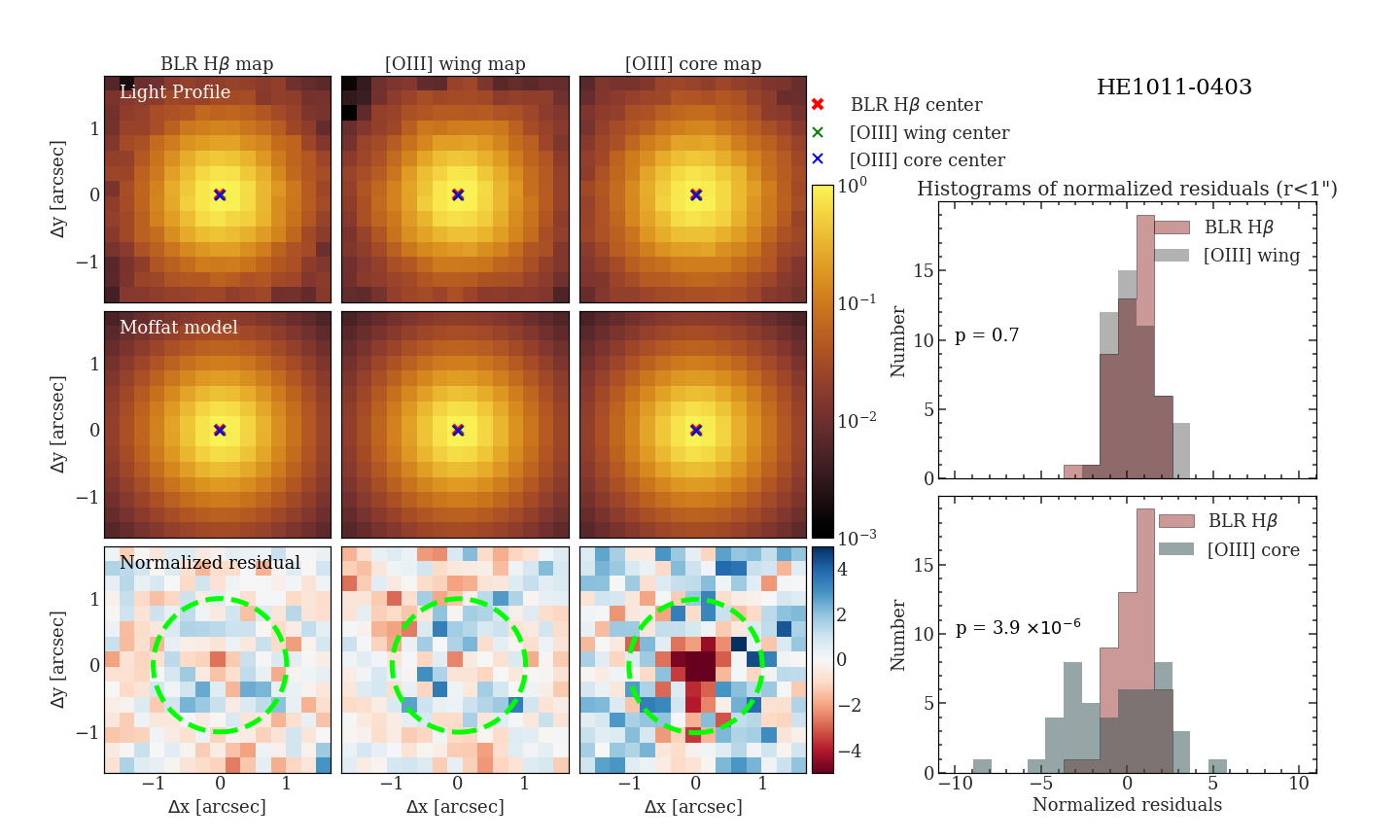}}
\end{figure*}

\begin{figure*}
   \resizebox{\hsize}{!}{\includegraphics{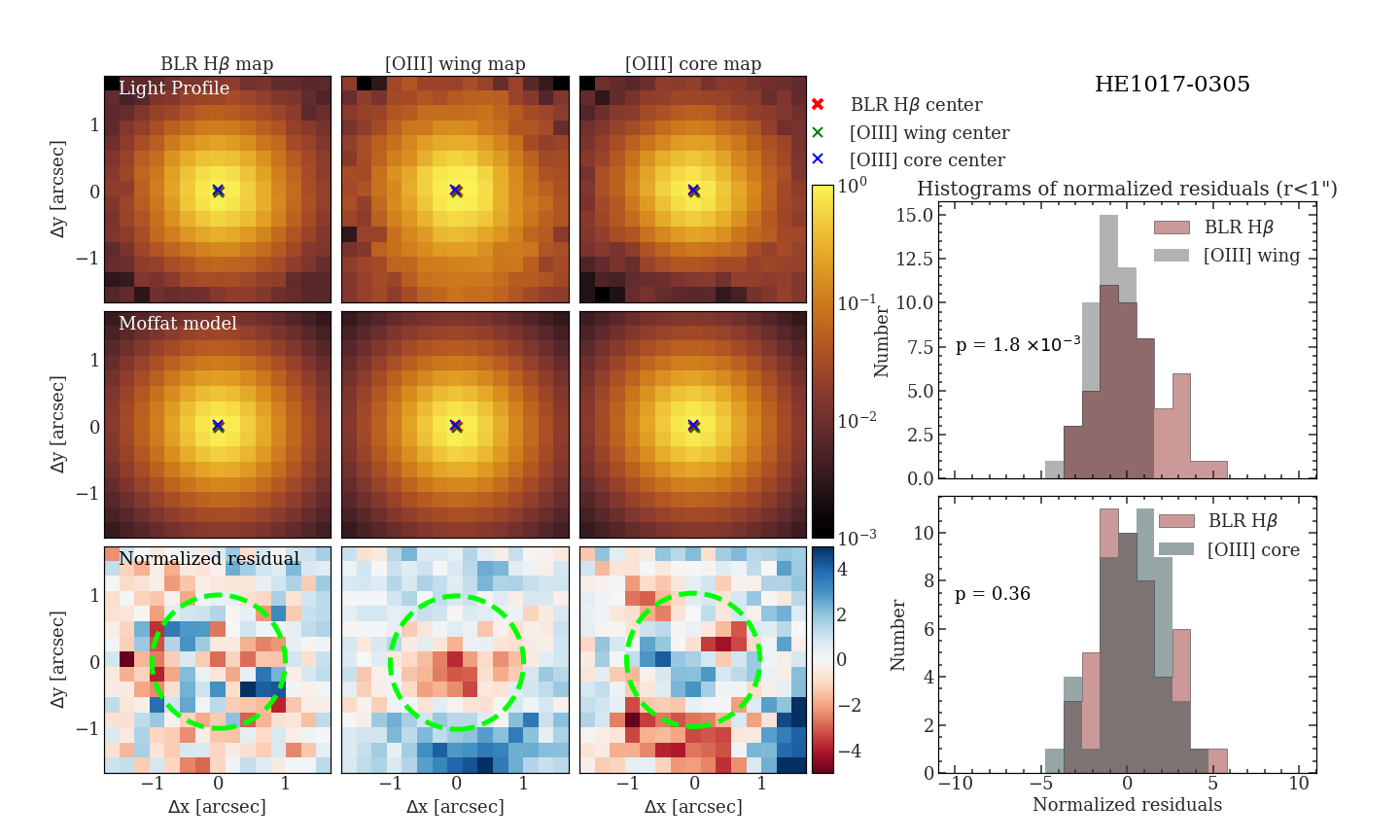}}
   \resizebox{\hsize}{!}{\includegraphics{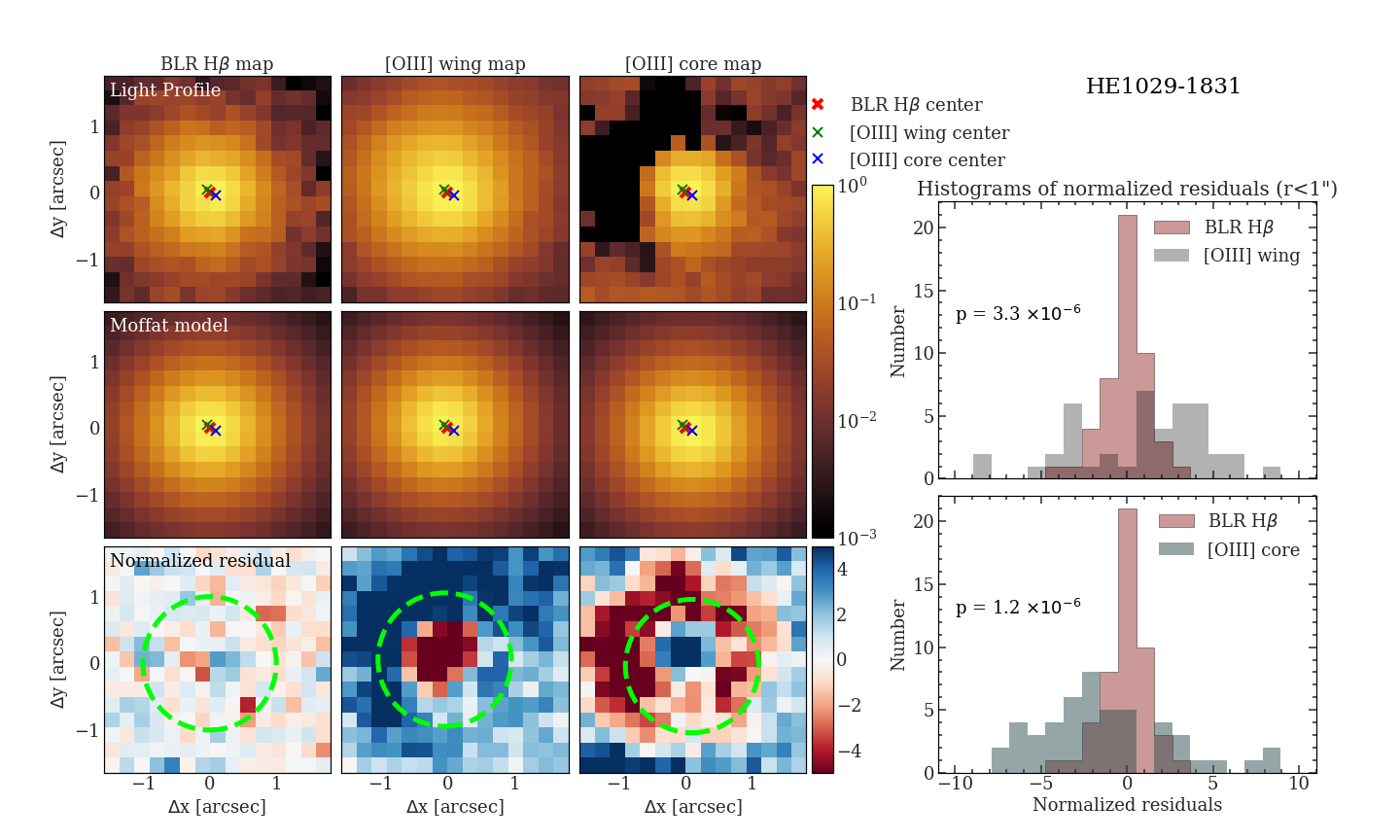}}
\end{figure*}

\begin{figure*}
   \resizebox{\hsize}{!}{\includegraphics{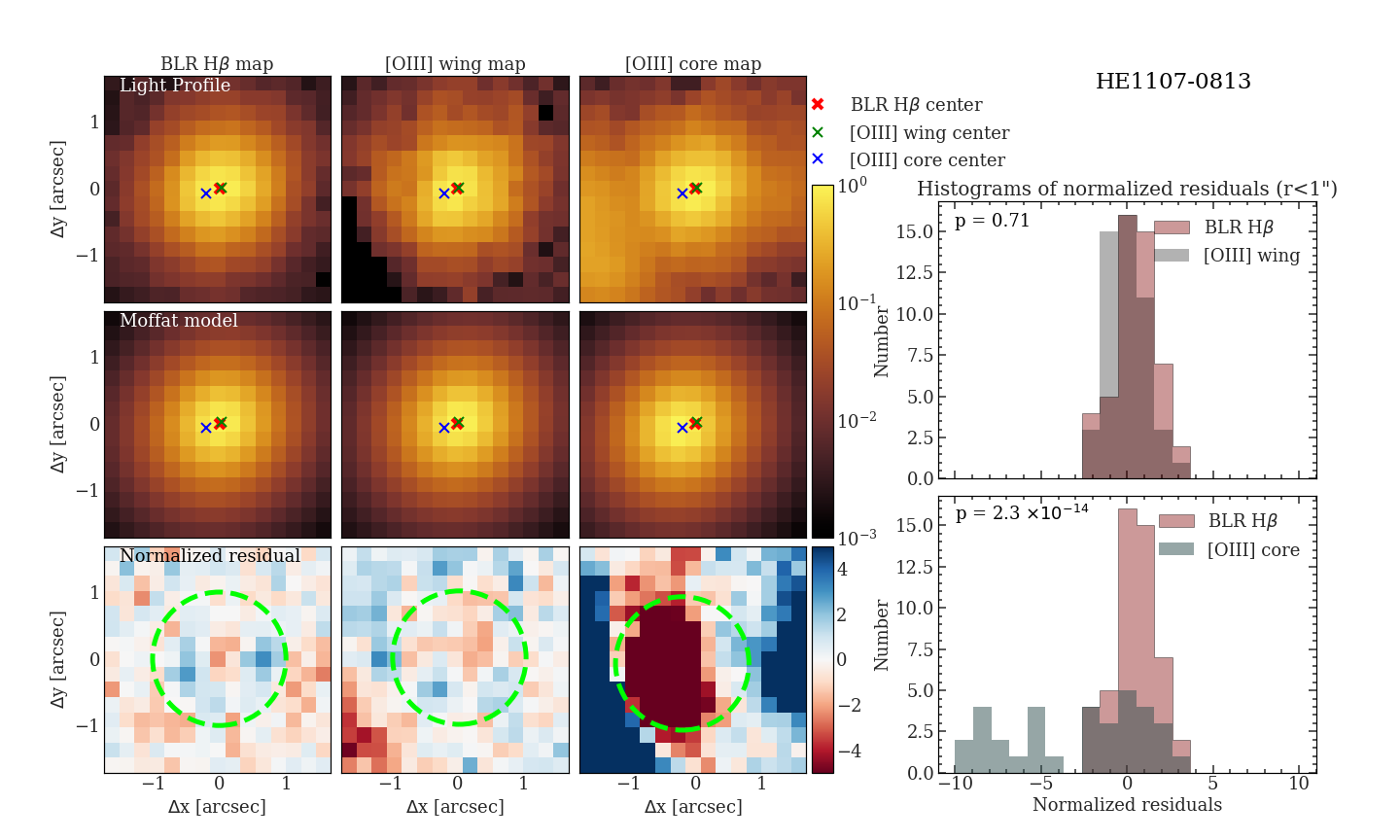}}
   \resizebox{\hsize}{!}{\includegraphics{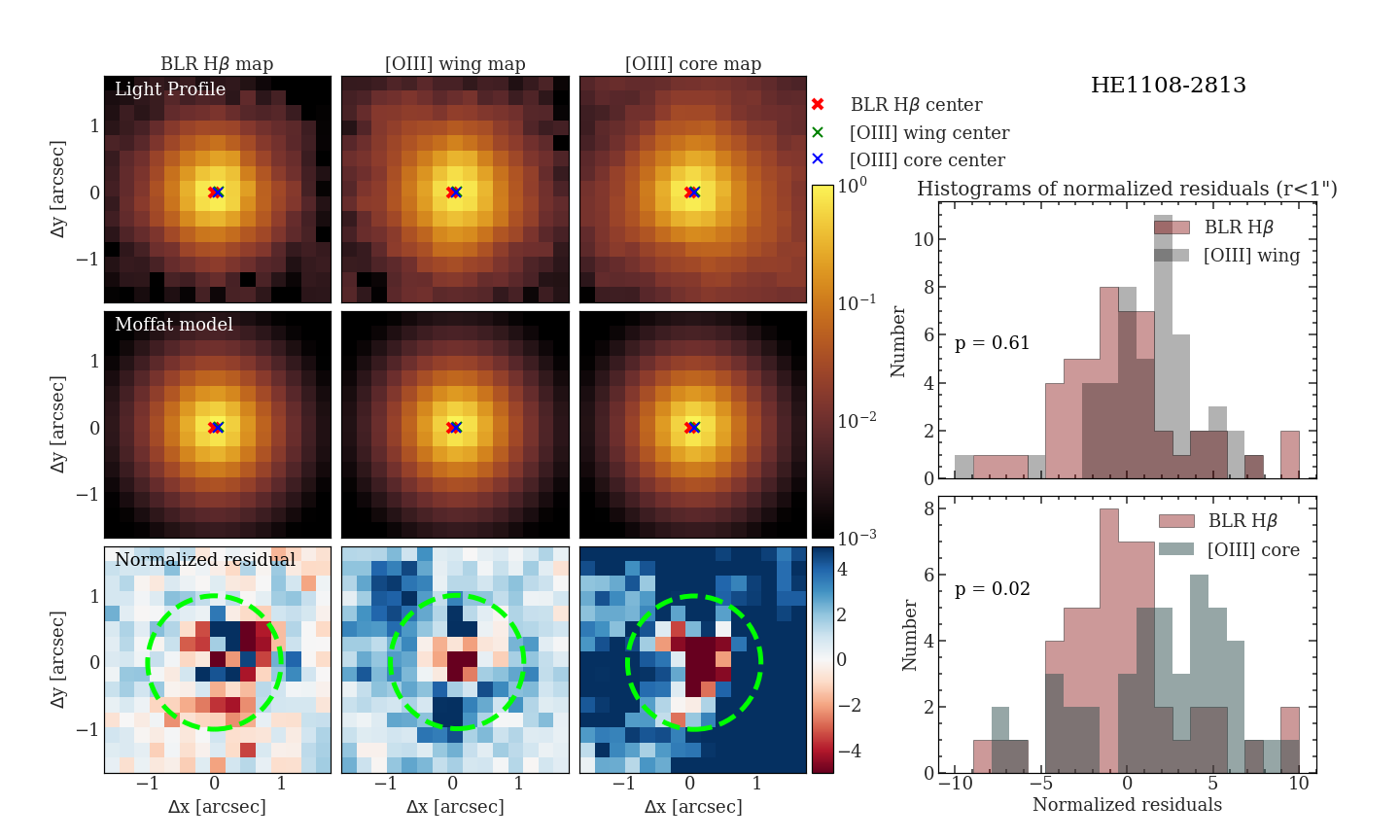}}
\end{figure*}

\begin{figure*}
   \resizebox{\hsize}{!}{\includegraphics{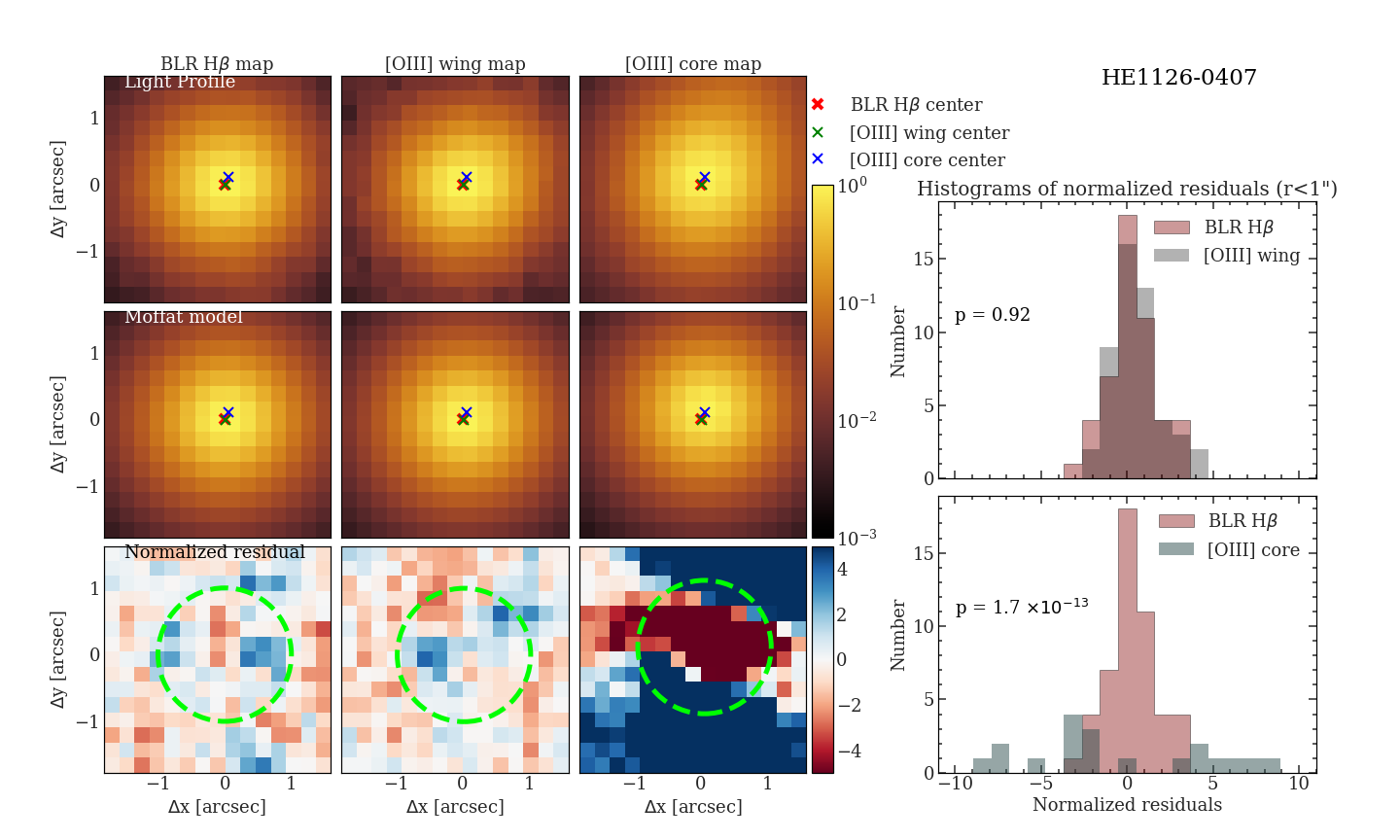}}
   \resizebox{\hsize}{!}{\includegraphics{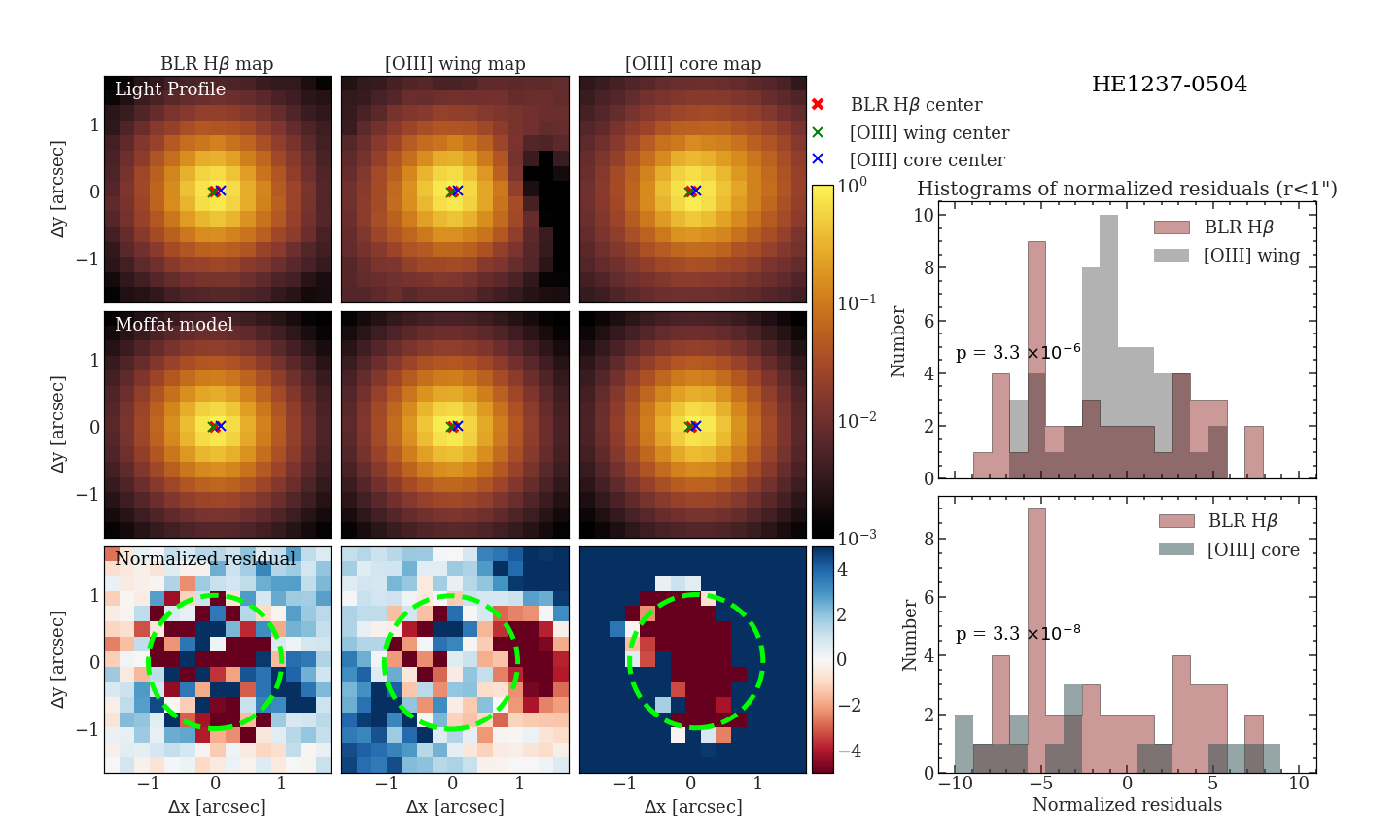}}
\end{figure*}

\begin{figure*}
   \resizebox{\hsize}{!}{\includegraphics{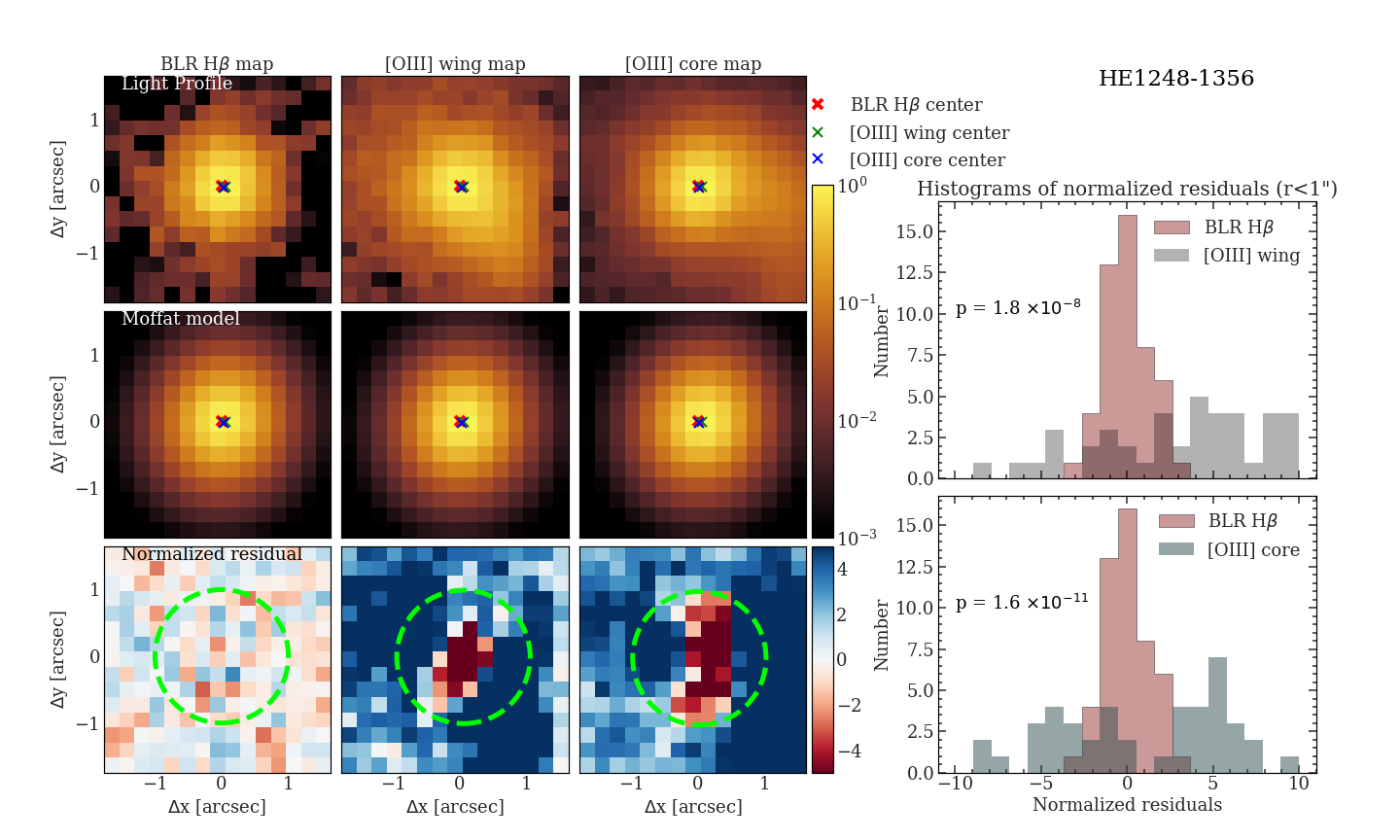}}
   \resizebox{\hsize}{!}{\includegraphics{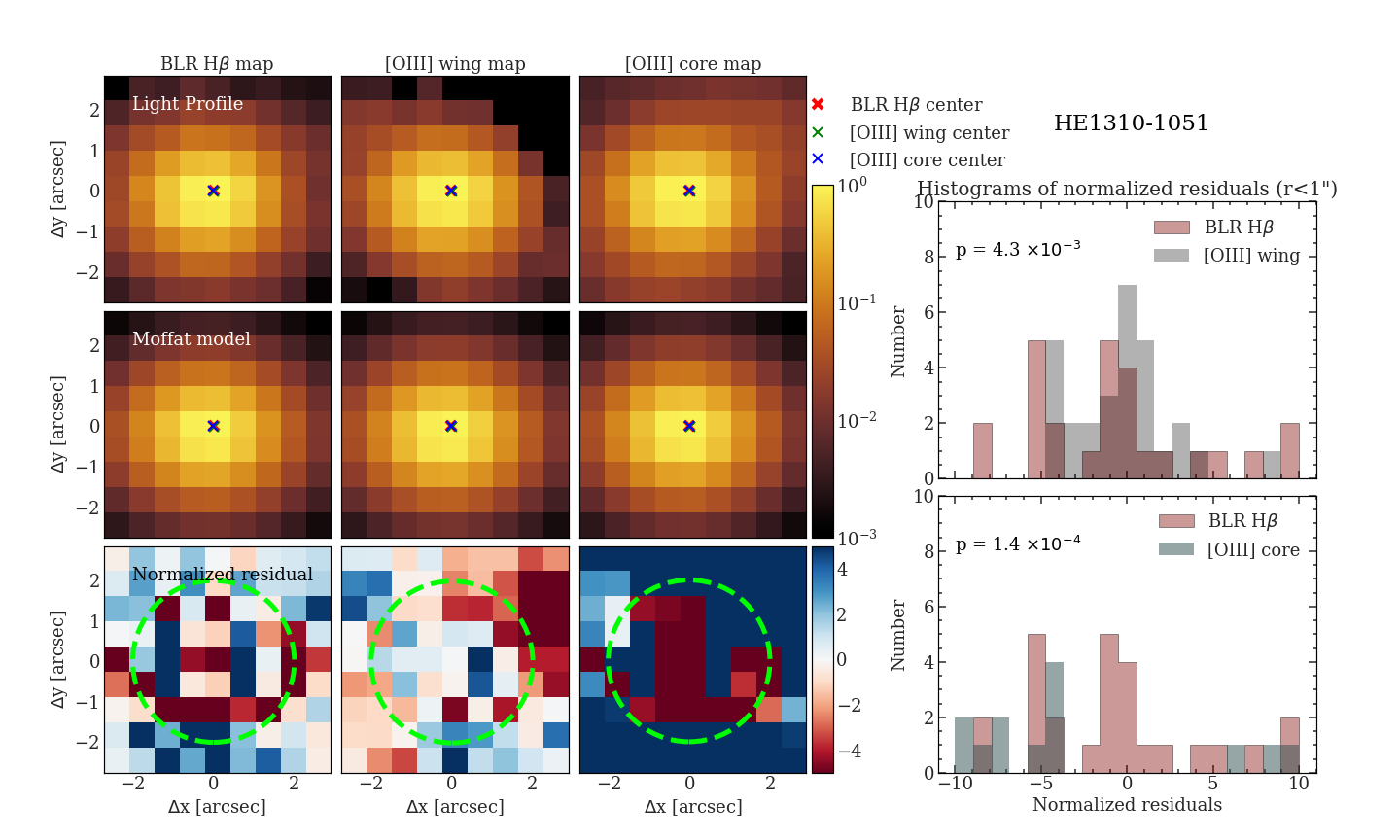}}
\end{figure*}

\begin{figure*}
   \resizebox{\hsize}{!}{\includegraphics{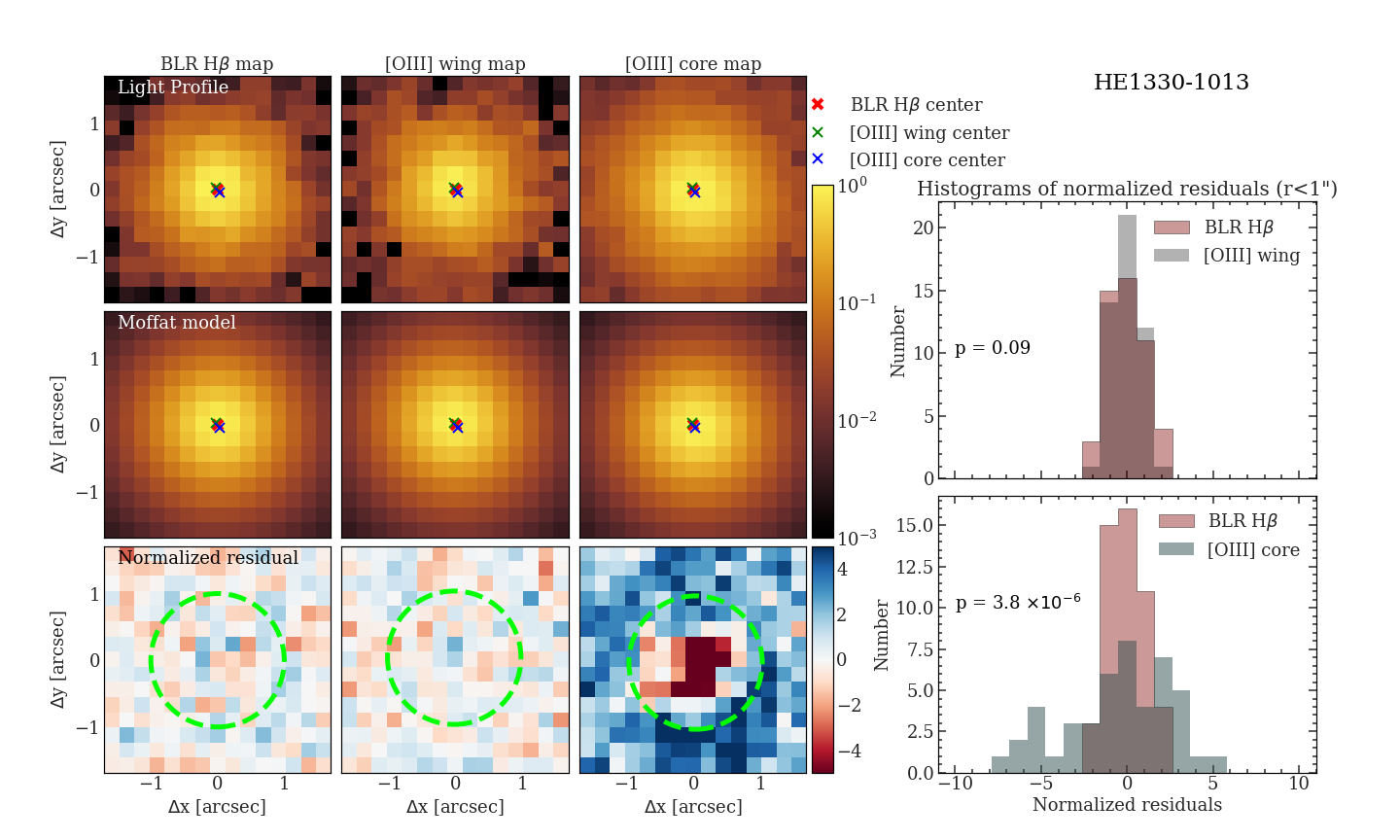}}
   \resizebox{\hsize}{!}{\includegraphics{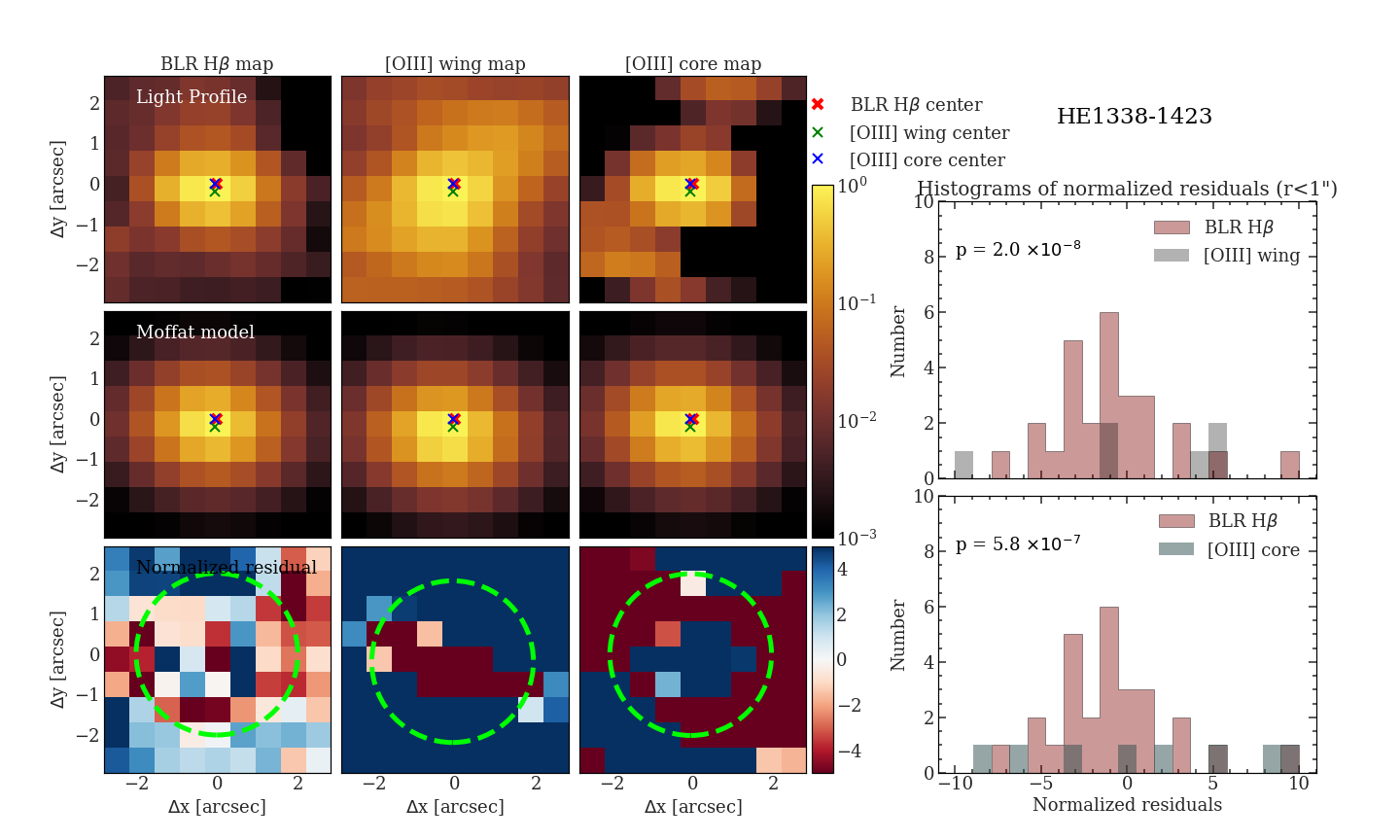}}
\end{figure*}

\begin{figure*}
   \resizebox{\hsize}{!}{\includegraphics{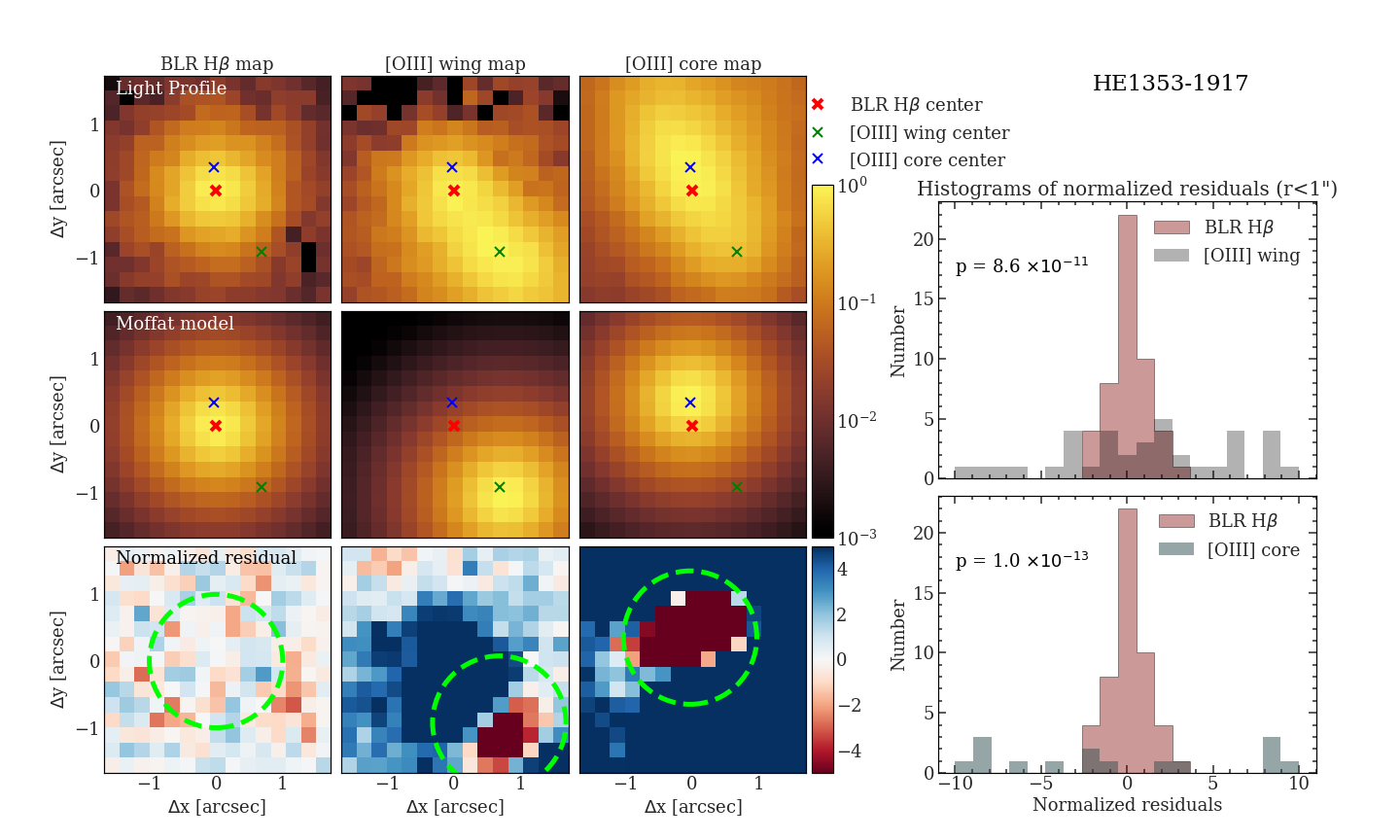}}
   \resizebox{\hsize}{!}{\includegraphics{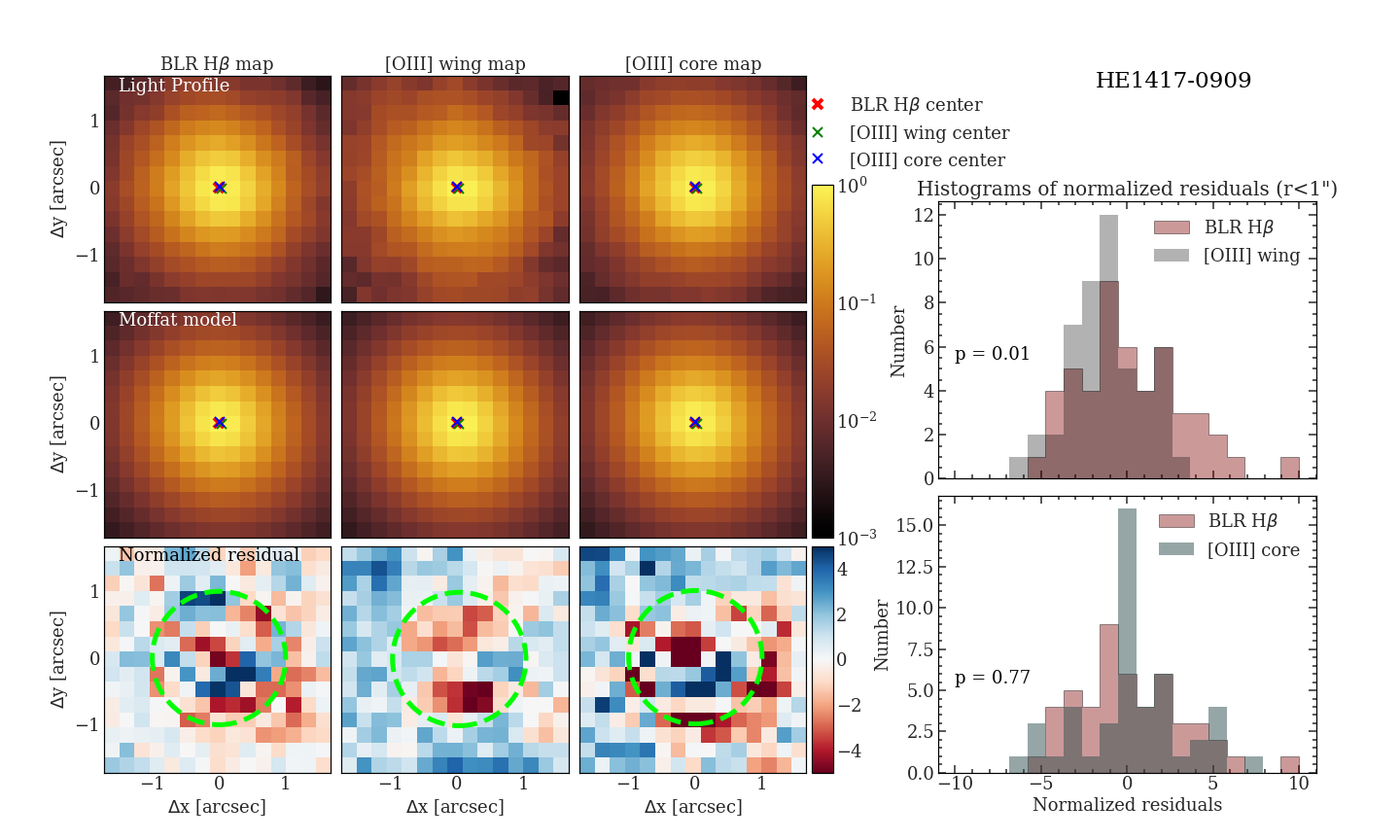}}
\end{figure*}

\begin{figure*}
   \resizebox{\hsize}{!}{\includegraphics{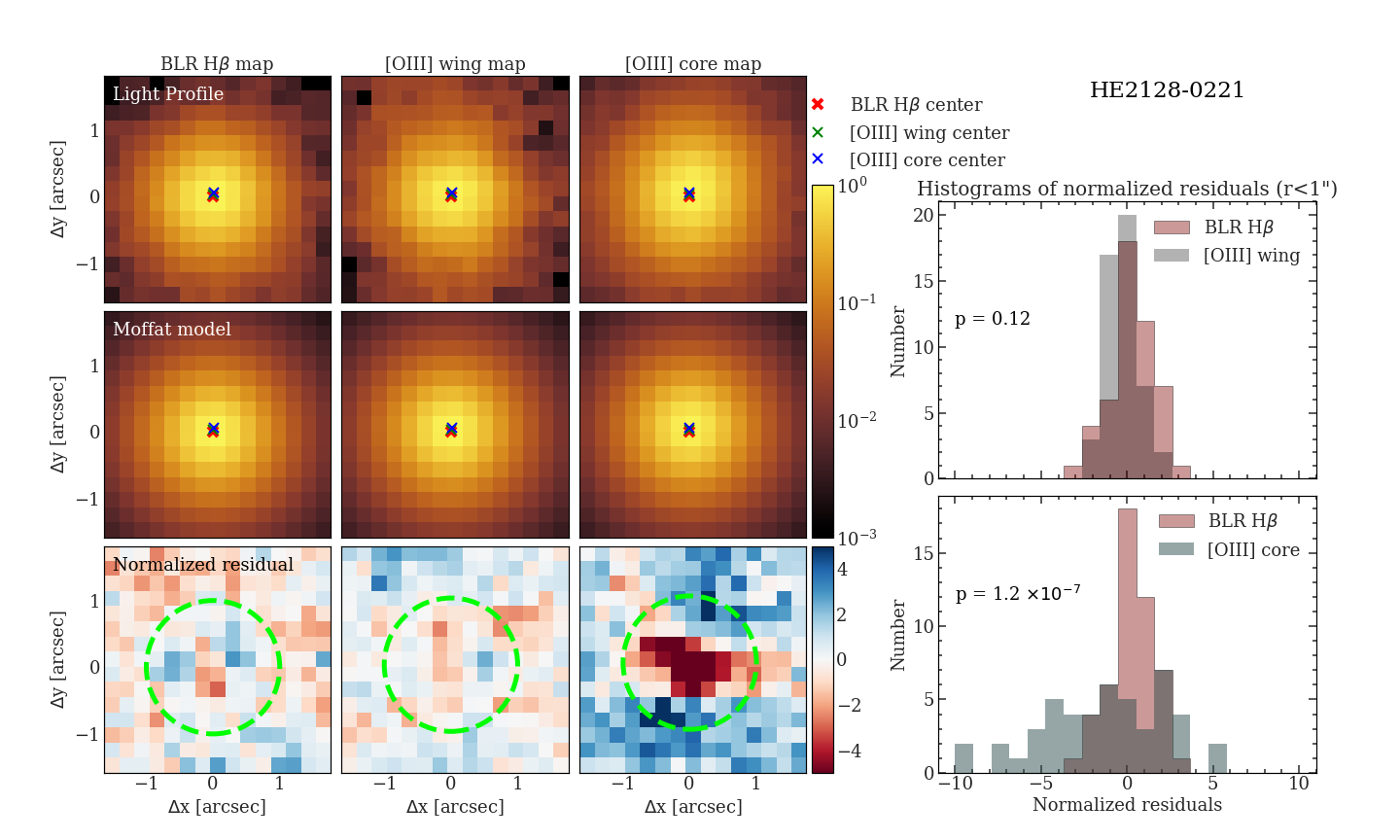}}
   \resizebox{\hsize}{!}{\includegraphics{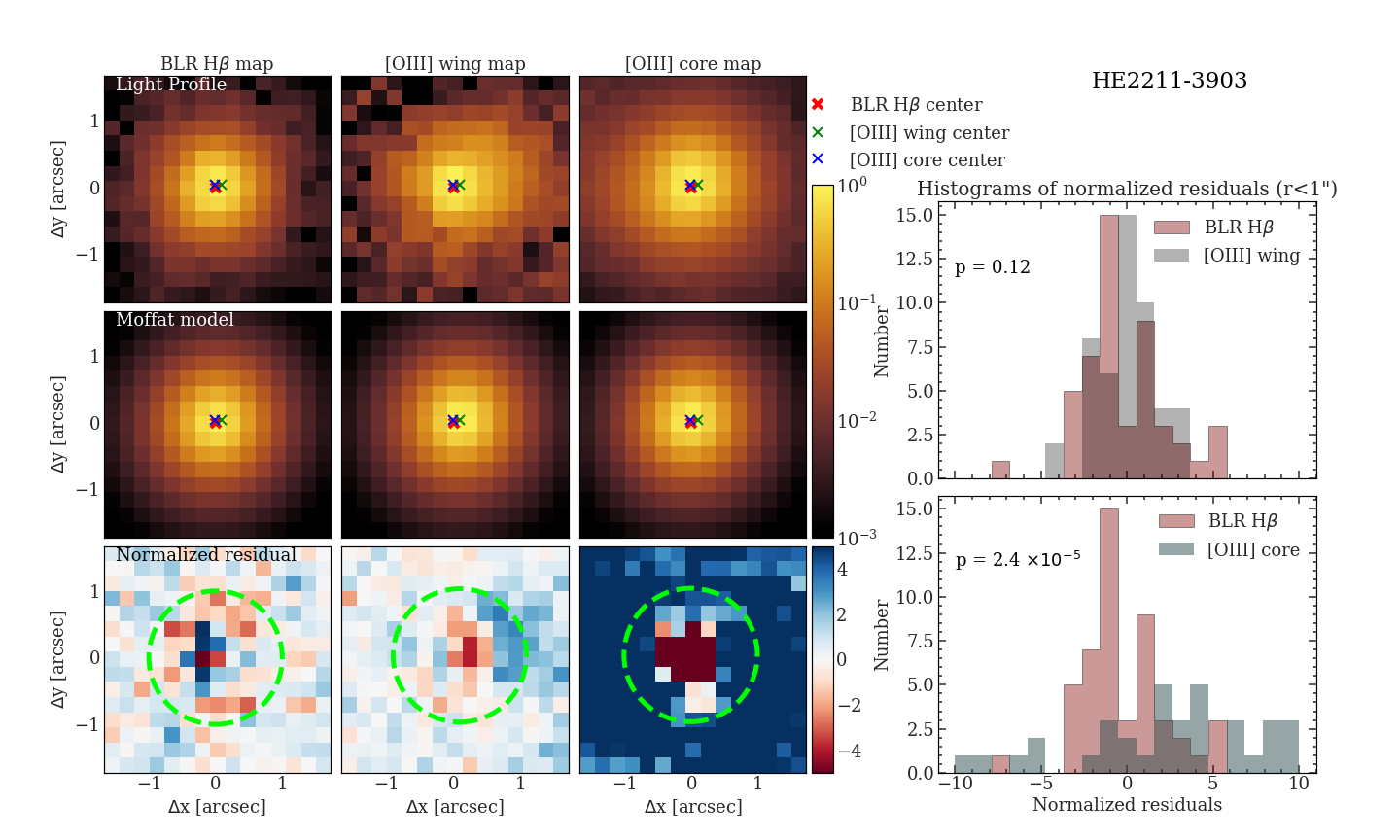}}
\end{figure*}

\begin{figure*}
   \resizebox{\hsize}{!}{\includegraphics{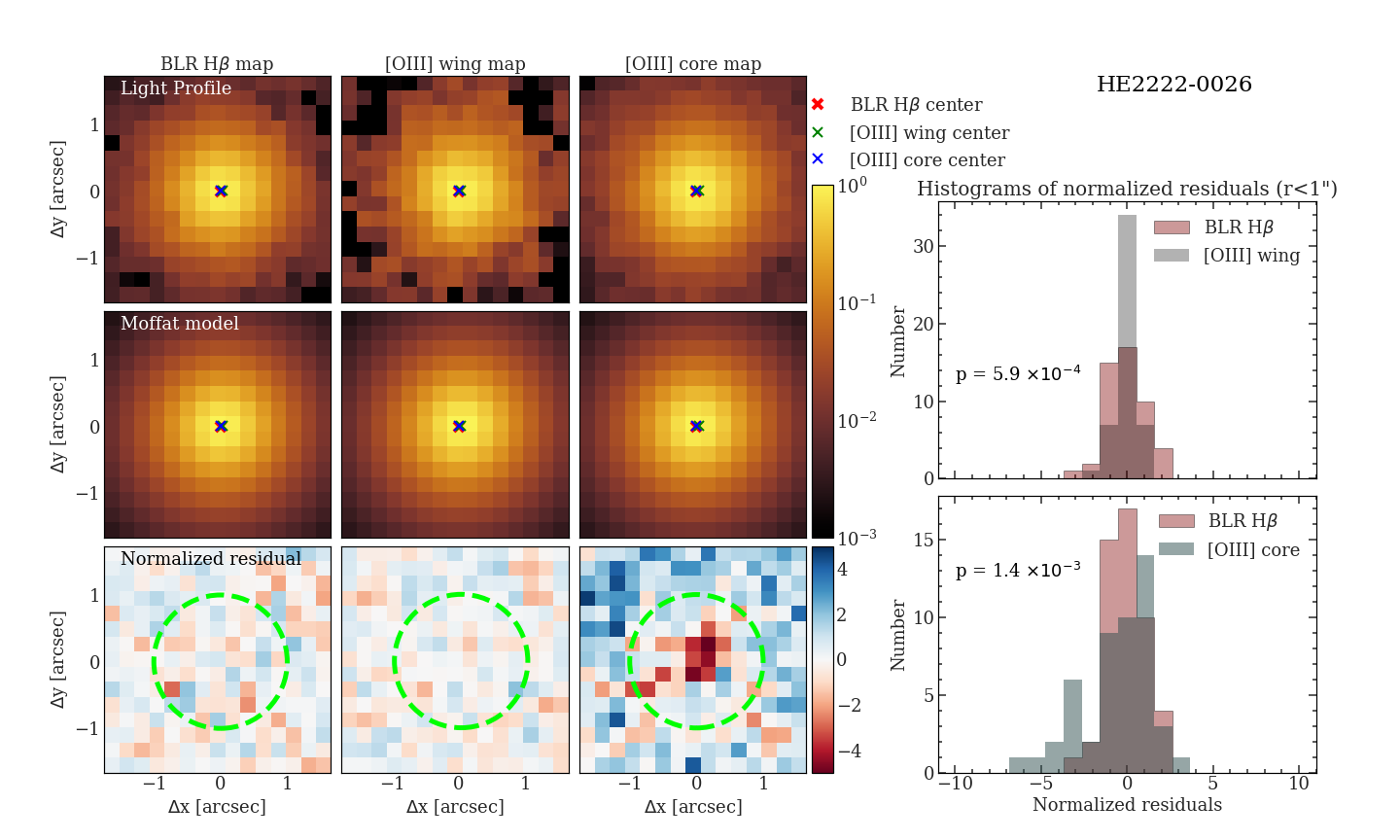}}
   \resizebox{\hsize}{!}{\includegraphics{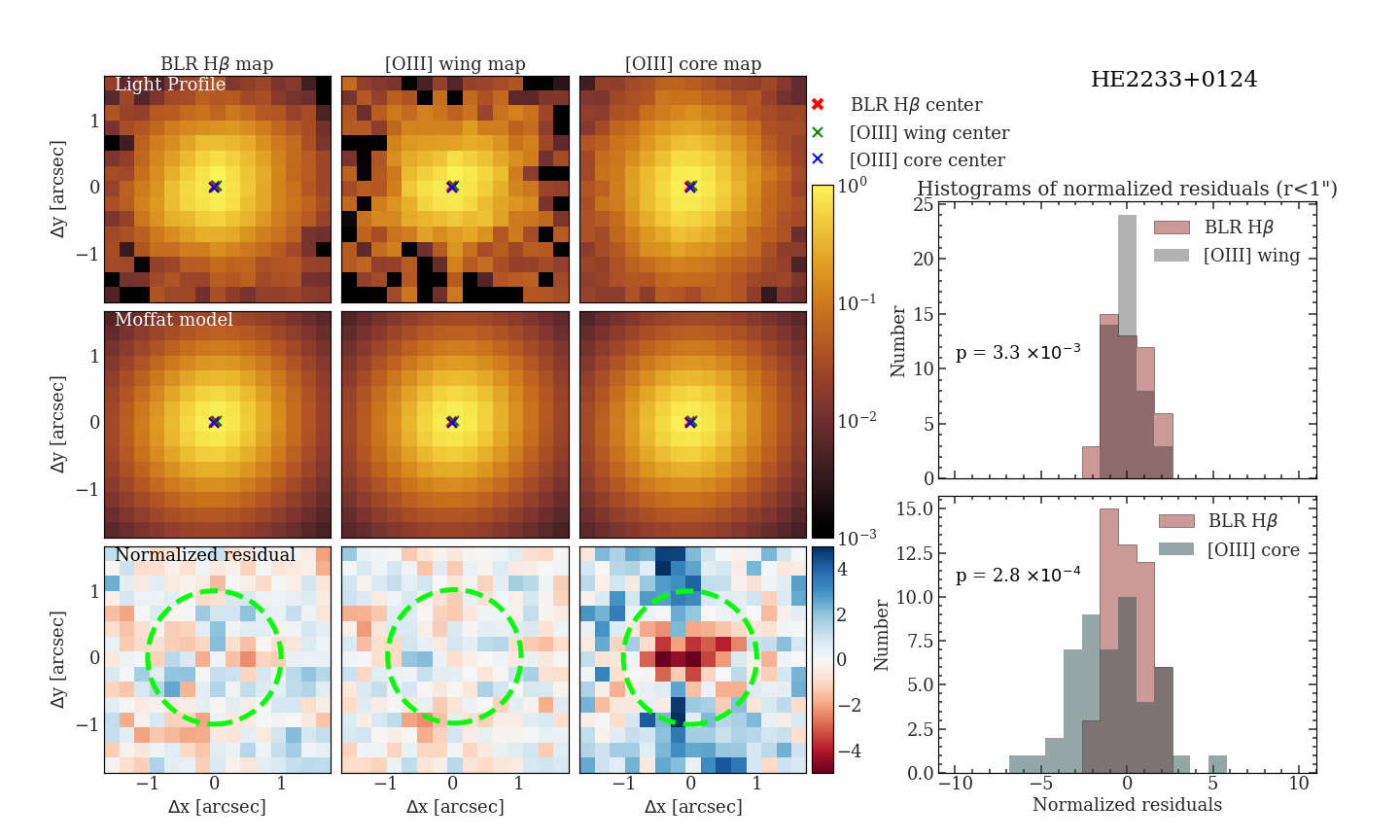}}
\end{figure*}

\begin{figure*}
   \resizebox{\hsize}{!}{\includegraphics{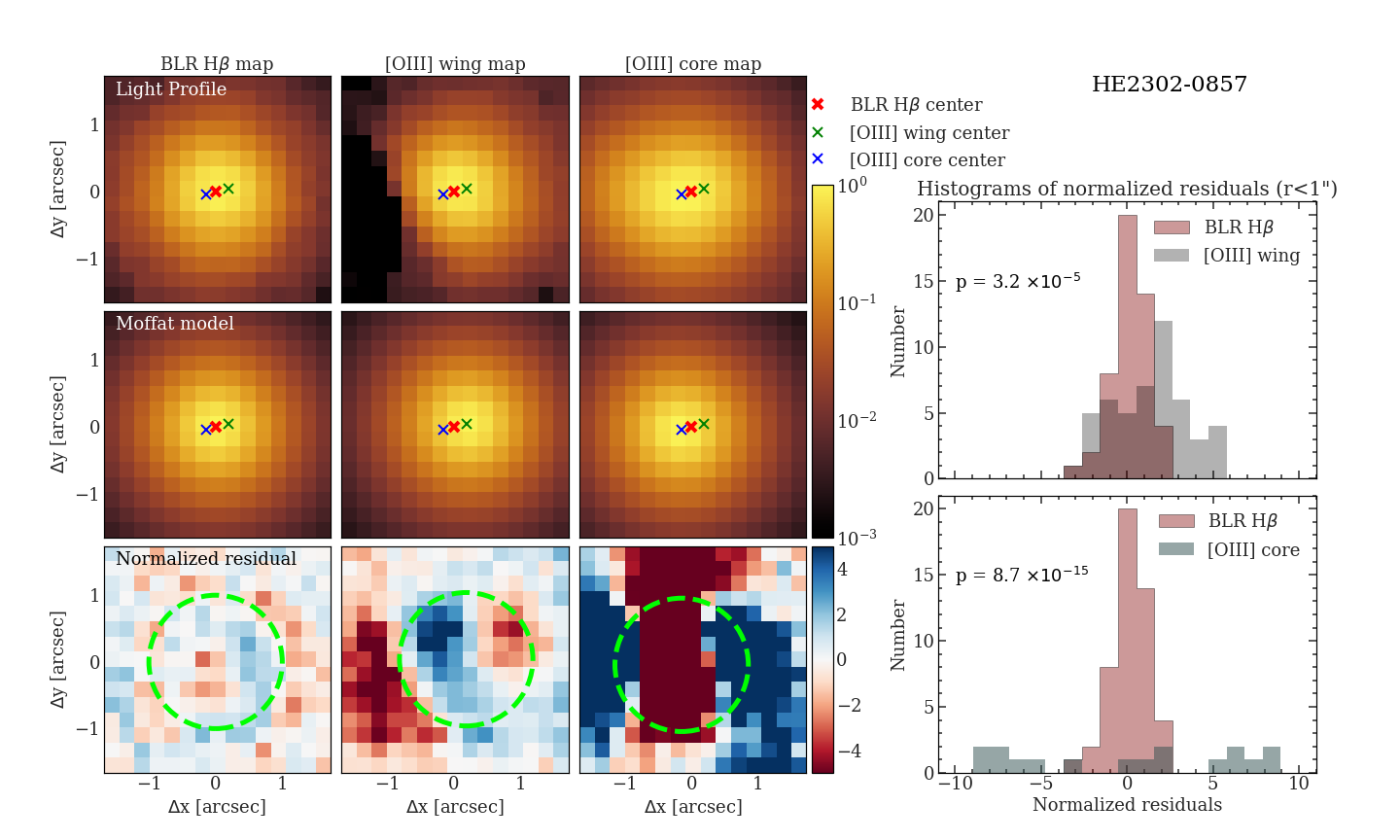}}
\end{figure*}

\pagebreak

\end{appendix}

\end{document}